\documentclass[aps,prd,amssymb,groupedaddress,nofootinbib,a4paper,notitlepage]{revtex4-1}
\pdfoutput=1
\usepackage{epsfig}

\usepackage[english]{babel}
\usepackage{amsfonts}
\usepackage{amssymb}
\usepackage{amsmath}
\usepackage{booktabs}
\usepackage{colortbl}
\usepackage{multirow}
\usepackage{bm}
\usepackage{hyperref}
\usepackage[labelfont={bf,scriptsize},textfont={scriptsize},justification=RaggedRight,format=hang]{caption,subfig}
\usepackage{natbib}

\def\fnl{f_\mathrm{NL}}
\def\gnl{g_\mathrm{NL}}
\def\gnlhat{\hat{g}_{\mathrm{NL}}}
\def\gnllocal{g_{\mathrm{NL}}^\text{loc}}
\def\taunl{\tau_\mathrm{NL}}
\def\taunlhat{\hat{\tau}_{\mathrm{NL}}}
\def\taunllocal{\tau_{\mathrm{NL}}^\text{loc}}

\def\fnlhat{\hat{f}_{\mathrm{NL}}}

\def\fnllocal{f_\mathrm{NL}^\text{loc}}

\def\bfnl{\kern2pt\overline{\kern-2ptf}_\mathrm{NL}}

\def\lmax{l_\text{max}}

\def\kmax{k_\text{max}}

\def\barQ{\kern2pt\overline{\kern-2pt\curl{Q}}}

\def\barR{\kern2pt\overline{\kern-2pt\curl{R}}}

\def\prd{Physical Review D}
\def\mnras{Monthly Notices of the Royal Astronomical Society}
\def\jcap{Journal of Cosmology and Astroparticle Physics}
\def\apj{Astrophysical Journal}
\def\apjs{Astrophysical Journal, Supplement}

\newcommand{\para}[1]{\par\vspace{2mm}\noindent\textbf{{#1}}}
\newcommand{\vect}[1]{\bm{\mathrm{{#1}}}}
\renewcommand{\d}{\mathrm{d}}
\newcommand{\e}[1]{\mathrm{e}^{{#1}}}

\renewcommand{\leq}{\leqslant}
\renewcommand{\geq}{\geqslant}

\newcommand{\bK}{\vect{K}}

\newcommand{\bk}{\vect{k}}
\newcommand{\bn}{\vect{n}}
\newcommand{\bx}{\vect{x}}

\newcommand{\npix}{n_{\text{pix}}}
\newcommand{\fsky}{f_{\text{sky}}}

\newcommand{\lmin}{l_{\text{min}}}

\newcommand{\ipleft}{\langle\kern-2.5pt\langle}
\newcommand{\ipright}{\rangle\kern-2.5pt\rangle}
\newcommand{\tripleft}{[\kern-1.5pt[}
\newcommand{\tripright}{]\kern-1.5pt]}

\newcommand{\baralpha}{\bar{\alpha}}

\newcommand{\nsrc}{n_{\text{src}}}
\newcommand{\Fsrc}{F_{\text{src}}}
\newcommand{\Tsrc}{T_{\text{src}}}

\newcommand{\Omegapix}{\Omega_{\text{pix}}}

\newcommand{\kB}{k_{\text{B}}}
\newcommand{\Tcmb}{T_{\text{CMB}}}

\newcolumntype{s}{>{\columncolor[gray]{0.9}$\displaystyle}c<{$}}
\newcolumntype{S}{>{$\displaystyle}c<{$}}

\begin{document}


\title{Constraining the WMAP9 bispectrum and trispectrum with needlets}

\author{Donough Regan}
\affiliation
{Astronomy Centre, University of Sussex, Brighton BN1 9QH, United Kingdom}

\author{Mateja Gosenca}
\affiliation
{Astronomy Centre, University of Sussex, Brighton BN1 9QH, United Kingdom}

\author{David Seery}
\affiliation
{Astronomy Centre, University of Sussex, Brighton BN1 9QH, United Kingdom}

\begin{abstract}
We develop a needlet approach to estimate the amplitude
of general (including non-separable) bispectra and trispectra
in the cosmic microwave background,
and apply this to the WMAP 9-year data.
We obtain estimates for the `orthogonal' bispectrum mode,
yielding results which are consistent with the WMAP 7-year
data.
We do not observe the frequency-dependence suggested by the WMAP team's
analysis of the 9-year data.
We present 1-$\sigma$ constraints on the `local' trispectrum
shape
$\gnl/10^5= -4.1\pm 2.3$, the `$c1$' equilateral model
$\gnl^{c_1}/10^6= -0.8\pm 2.9$, and the constant model
$\gnl^{\rm{const}}/10^6= -0.2\pm 1.8$,
together with a $95\%$ confidence-level
upper bound on the multifield local parameter $\taunl<22000$.
We estimate the bias on these parameters produced by point sources.
The techniques developed in this paper should prove useful
for other datasets such as Planck.
\end{abstract}

\maketitle


\section{Introduction}
\label{sec:introduction}
The study of primordial non-Gaussianity
is now a precision science,
making it increasingly important to develop general,
effective and efficient estimators (for a review see, for example, Refs.~\cite{0406398,2010AdAst2010E..71Y}).
Fergusson, Shellard and collaborators
developed a formalism using a
`modal' or `partial-wave' expansion~\cite{FLS1,RSF1,FLS2,FRS2}
which
enabled non-separable bi- and trispectrum shapes to be analysed,%
    \footnote{For the purposes of this paper,
    a function $f(k_1, k_2, \cdots, k_n)$ is \emph{separable}
    if it can be written as a sum of terms of the form
    $f_1(k_1)f_2(k_2) \cdots f_n(k_n)$.}
and applied this technology to a version of the KSW estimator~\cite{KSW}.
In Ref.~\cite{RMS2013} an alternative approach was developed which allowed
an arbitrary estimator to be coupled to the partial-wave decomposition.
This enables the benefits of a particular estimator to be exploited while
retaining the ability to detect $n$-point functions of arbitrary shape.
For example, wavelet-based estimators are
efficient detectors of point sources \cite{Curto:2009pv}.
Ref.~\cite{RMS2013} implemented such a wavelet-based estimator.

In this paper we make use of
a similar approach to couple the partial-wave
expansion to a \emph{needlet}-based estimator.
Needlets are a particular class of spherical wavelets which are designed
to be localized in both real-space and frequency,
and possess properties which make them attractive for
CMB analysis.
First, they do not require any tangent-plane approximation~\cite{Lan}.
Second,
unlike general wavelets, needlets are asymptotically uncorrelated
in frequency.
Therefore,
assuming Gaussianity,
the coefficients of a needlet decomposition
may be treated
as independent, identically-distributed random variables
at high frequency.
This makes them useful as diagnostics
of non-Gaussianity.

Needlets were first applied to CMB data by Pietrobon et al.~\cite{Pietrobon}
and subsequently used to study primordial non-Gaussianity by a number
of authors~\cite{Baldi,Baldi2,Lan,Pietrobon2008,Rudjord:2009mh,Pietrobon2009,Rudjord2010}.
In this paper we make use of \emph{Mexican needlets}, which were
discussed mathematically by Geller \&
Mayeli~\cite{2007arXiv0709.2452G,*2007arXiv0706.3642G,*2009arXiv0907.3164G}
and
first applied
to CMB analysis by Scodeller et al.~\cite{Scodeller2010}.
These can give a close approximation to the Spherical Mexican Hat wavelet,
but decorrelate more rapidly
with increasing frequency.
We will use them to develop a
nearly optimal estimator for both the three-
and four-point functions of the CMB,
and use it to obtain the first estimates
of
the bias on the four-point function due to point sources,
assuming a simple constant-flux model.
In addition to its application to primordial non-Gaussianity,
the four-point function estimator
can be adapted to study CMB lensing.
It could also be used to study perturbations generated
by cosmic strings, for which the trispectrum is expected to be
larger than the bispectrum~\cite{RS2009}.

\para{Summary.}
In~{\S\S}\ref{sec:bispneedlets}--\ref{sec:trispneedlets} we
summarize the use of partial-wave decompositions
of arbitrary bi- and trispectra.
We describe the construction of cubic and quartic needlet estimators
and explain how to calculate optimal constraints.
In~{\S\ref{sec:wmaptest}} we apply this methodology
to the 9-year WMAP data.
Our bispectrum analysis yields results which are consistent in frequency.
This disagrees with the analysis by the WMAP team, which suggested
mild tension between the V- and W-band constraints for the orthogonal mode.
We quote constraints for four trispectrum modes:
two local shapes ($\gnllocal$ and $\taunllocal$), the
constant shape,
and the `$c_1$' equilateral shape.
Our constraints for $\gnllocal$ are in agreement with the optimal constraints
given by Sekiguchi \& Sugiyama~\cite{SekSug2013} on the basis of WMAP9 data.
In addition we give constraints
on the bias induced by points sources for each of the trispectrum modes.
In~{\S\ref{sec:conclusions}} we present our conclusions.

\section{CMB bispectrum with needlets}
\label{sec:bispneedlets}

To fix notation, we briefly recall
some key steps in the decomposition of
an arbitrary bispectrum into `modes' or partial waves.
For more details we refer to the original literature~\cite{FLS1,RSF1,FLS2,FRS2}.

\para{CMB bispectrum.}
Consider the primordial gravitational potential $\Phi$,
which is related to the curvature perturbation on uniform density
hypersurfaces $\zeta$ by
$\Phi = 3 \zeta / 5$.
Denoting the spectrum and bispectrum of $\Phi$
by $P_{\Phi}(k)$ and
$B_{\Phi}(k_1,k_2,k_3)$,
and
using the basis of orthogonal polynomials $q_n(k)$ introduced
by Fergusson, Liguori \& Shellard~\cite{FLS1},
it is possible to decompose $B_\Phi$ into this basis by writing
\begin{equation}\label{eq:shapeloc1}
	S_{\Phi}^{\text{(loc)}}(k_1,k_2,k_3)
	\equiv
	\frac{B_{\Phi}(k_1,k_2,k_3)}{2 \big(
		P_{\Phi}(k_1) P_{\Phi}(k_2)
		+ P_{\Phi}(k_1) P_{\Phi}(k_3)
		+ P_{\Phi}(k_2) P_{\Phi}(k_3) \big)}	=
	\sum_{p r s} \alpha_{p r s}^{Q} q_{(p}(k_1)q_r(k_2)q_{s)}(k_3),
\end{equation}
where bracketed indices are symmetrized with weight unity.
The quantity $S_{\Phi}^{\text{(loc)}}$ is referred to as the
\emph{shape}.
Where $B_\Phi$ is given by the local model, the shape
is independent of the wavenumbers $k_i$
and corresponds to the amplitude $\fnl$.
For convenience we
represent unique triplets $(p,r,s)$ using a multi-index $n$
and define $Q_n \equiv q_{(p}(k_1) q_r(k_2) q_{s)}(k_3)$.
Then
the coefficients of the expansion $\alpha_n^{Q}$ are
to obtained by computing
\begin{equation}
\label{eq:alpha-n-Q}
\begin{split}
	\alpha_n^Q
	& = \sum_m \ipleft S_{\Phi}^{\text{(loc)}} , Q_m \ipright
	G_{m n}^{-1}
\end{split}
\end{equation}
where $G_{m n}= \ipleft Q_m , Q_n \ipright$,
and  $\ipleft f , g \ipright$ represents the inner produt
\begin{equation}
	\ipleft f, g \ipright
	\equiv
	\int_{\mathcal{V}} \d v \;
	f(k_1,k_2,k_3) g(k_1,k_2,k_3)\omega(k_1,k_2,k_3) .
	\label{eq:k-ip-def}
\end{equation}
In this equation
$\d v$ is an element of volume on
the integration domain $\mathcal{V}$
(defined by the triangle condition, $2 \max ( k_1,k_2,k_3 ) \leq k_1+k_2+k_3$),
and $\omega$ is a weight function which can be adjusted to suit our convenience.
To obtain a good approximation for $S_{\Phi}^{\text{loc}}$ we find that
it is necessary to use $\sim 100$ of the $Q_n$.

Eq.~\eqref{eq:shapeloc1} gives the CMB bispectrum
\begin{equation}\label{eq:primdecompcmb1}
	b_{l_1 l_2 l_3}
	=
	\sum_{n = (p,r,s)} 6 \alpha_n^Q
	\int \d x \, x^2 \;
	\tilde{q}^{(2)l_1}_{(p}(x) \tilde{q}^{(-1)l_2}_r(x) \tilde{q}^{(-1)l_3}_{s)}(x)
	\equiv
	\sum_n \alpha_n^Q b^{(n)}_{l_1 l_2 l_3}
\end{equation}
where the functions $\tilde{q}_p^{(2)l}$ and $\tilde{q}_p^{(-1)l}$
are defined by
\begin{equation}\label{eq:qplus2}
	\tilde{q}_p^{(2)l}(x)
	\equiv
	\frac{2}{\pi}
	\int \d k \; k^2 q_p\Big(\frac{k}{\kmax}\Big) \Delta_l(k) j_l(k x) \,,
	\quad \text{and} \quad
	\tilde{q}_p^{(-1)l}(x)
	\equiv
	\frac{2}{\pi}
	\int \d k \; k^2 P_{\Phi}(k) q_p\Big(\frac{k}{\kmax}\Big)
		\Delta_l(k) j_l(k x) \,,
\end{equation}
and $b^{(n)}_{l_1 l_2 l_3}$ is understood
to be defined by equation~\eqref{eq:primdecompcmb1}.
In these formulae, $j_l(kx)$ represents the spherical bessel function
of order $l$ and $\Delta_l (k)$ represents the CMB transfer function
used to map from primordial times to the surface of last scattering.
We compute $\Delta_l(k)$
by solving the collisional Boltzmann equations
using CAMB~\cite{9911177}.

Eq.~\eqref{eq:primdecompcmb1}
represents the CMB bispectrum $b_{l_1 l_2 l_3}$
using the coefficients $\alpha_n^Q$
which determine the primordial bispectrum $B_{\Phi}$,
and an explicit integration $\int \d x \, x^2 \cdots$
(the `line-of-sight' integral)
which appears in $b_{l_1 l_2 l_3}^{(n)}$.
Under some circumstances it may also be
possible to decompose this integral in terms of the
basis functions $Q_n$~\cite{RMS2013}.
Choosing a normalization for future convenience,
this yields
\begin{equation}\label{eq:latetime}
	s^{(n)}_{l_1 l_2 l_3}
	\equiv
	\frac{(2l_1+1)^{1/6}(2l_2+1)^{1/6}(2l_3+1)^{1/6}}{\sqrt{C_{l_1}C_{l_2}C_{l_3}}}
	b^{(n)}_{l_1l_2 l_3}
	=
	\sum_m \Gamma_{n m}Q_m(l_1,l_2,l_3) ,
\end{equation}
where the first equality is a definition of $s^{(n)}_{l_1 l_2 l_3}$.
For a fixed number of $Q_n$,
it need not happen that~\eqref{eq:latetime} is a good approximation.
Even in cases where
a sufficiently good approximation can be obtained
this may require more $Q_n$ than are needed for
$S_{\Phi}^{\text{loc}}$.
However, where it applies,
the advantage of~\eqref{eq:latetime}
is that the line-of-sight integral
is absorbed
in the transfer matrix $\Gamma_{nm}$
which need only be calculated once for each
choice of cosmology, represented by the transfer
function $\Delta_l(k)$.
An explicit expression for $\Gamma_{nm}$ was given in Ref.~\cite{RMS2013}.
We will return to the question
of whether Eq.~\eqref{eq:latetime}
is applicable
when discussing the
trispectrum in~{\S\ref{sec:trispneedlets}}.

In conclusion,
given the coefficients of the primordial decomposition $\alpha_n^Q$,
we may either
(\emph{a}) use Eq.~\eqref{eq:primdecompcmb1} to evaluate the CMB
bispectrum directly, or
(\emph{b}) use the transfer matrix to give the coefficients of the
CMB shape
$\baralpha_n^Q=\sum_m \Gamma_{n m}\alpha_m^Q$, from which
the CMB bispectrum can be reconstructed using
\begin{equation}\label{eq:latetime1}
	s_{l_1 l_2 l_3}
	\equiv
	\sum_n \alpha_n^Q s^{(n)}_{l_1 l_2 l_3}
	=
	\sum_n \baralpha_n^Q Q_n(l_1,l_2,l_3)\,.
\end{equation} 

\para{Simulating the CMB bispectrum.}
Now suppose we fix some primordial bispectrum $B(k_1, k_2, k_3)$
and include a component of this with amplitude $\fnl^B$ in
the gravitational potential $\Phi$,
so that
$B_\Phi(k_1, k_2, k_3) \supseteq \fnl^B B(k_1, k_2, k_3)$.
The symbol `$\supseteq$' indicates that $B_\Phi$ contains
this contribution among others.
Note that $\fnl^B$ is the amplitude of the
fixed bispectrum $B(k_1, k_2, k_3)$
and therefore does \emph{not} agree with the traditional $\fnl$ parameter
defined by Komatsu \& Spergel~\cite{KomatsuSpergel2001}
unless $B(k_1, k_2, k_3)$ is
the conventionally-normalized local bispectrum.
In this paper we always denote the Komatsu \& Spergel
parameter by $\fnllocal$.
Our objective is to estimate $\fnl^B$ from the data.

In order to compute the error associated with
our estimator
it will be necessary to
obtain its variance on an ensemble of maps
containing a bispectrum corresponding to $B_{\Phi}$,
and for that purpose we require
a suite of simulated maps with appropriate
statistical properties.
For a Gaussian map we could generate an
appropriate ensemble by
making
a multipole
decomposition $\delta T(\hat{\bn})/T
= \sum_{lm} a_{lm} Y_{lm}(\hat{\bn})$,
drawing the
coefficients $a_{lm}$
from a Gaussian distribution.
To account for the bispectrum $B_\Phi$
we must instead set
$a_{lm}=a^G_{lm}+ \fnl^B a^B_{lm}$,
with
$a^G_{lm}$ a dominant, Gaussian contribution
and $a^B_{lm}$ a non-Gaussian correction
chosen to reproduce
bispectrum $B(k_1, k_2, k_3)$.
A prescription for choosing
$a_{lm}^B$ was given by Fergusson,
Shellard \& Liguori~\cite{FLS1},
\begin{equation}
	a^B_{lm}
	\equiv \frac{1}{6}
	\sum_{l_2 m_2}\sum_{l_3 m_3} b_{l l_2 l_3}
	\mathcal{G}^{l l_2 l_3}_{m m_2 m_3}
	\frac{a^{G*}_{l_2 m_2}a^{G*}_{l_3 m_3} }{C_{l_2} C_{l_3}} ,
\end{equation}
where $a^{G*}_{lm}$ is the complex conjugate of $a^G_{l m}$.
In what follows we denote quantities which include the bispectrum
$B$ by the superscript `$B$'.
Using the expansion
of the CMB bispectrum $b_{l_1 l_2 l_3}$
in terms of the primordial coefficients $\alpha_n^Q$
given by Eq.~\eqref{eq:primdecompcmb1},
it follows that $a^B_{lm}$ can be written
\begin{align}\label{eq:latealmB}
    a_{lm}^B=\sum_n \alpha_n^Q
    \int \d x \; x^2
    \int \d\hat{\vect{n}} \; Y_{l m}(\hat{\vect{n}}) 
    \Big(
        \tilde{q}^{(2)l}_{(p}(x)
        \tilde{M}_r^{(-1)G}(x,\hat{\vect{n}})
        \tilde{M}_{s)}^{(-1)G}(x,\hat{\vect{n}})
    \Big) ,
\end{align}
where $\tilde{M}_p^{(-1)G}(x,\hat{\vect{n}})=\sum_{l m} \tilde{q}^{(-1)l}_p(x) a^G_{lm} Y_{lm}(\hat{\vect{n}}) / C_l$.
Alternatively we may use the CMB decomposition~\eqref{eq:latetime1}
with the result that
$a^B_{lm}=\sum_n \baralpha_n^Q \overline{a}^{B (n)}_{lm}$,
where
\begin{equation}\label{eq:almB}
    \overline{a}^{B (n)}_{lm}
	=
	\frac{1}{6}
	\frac{\sqrt{C_l}}{(2 l+1)^{1/6}}
	\int \d\hat{\vect{n}} \;
	Y_{ lm}(\hat{\vect{n}})
	q_{(p}\Big( \frac{l}{\lmax} \Big)
	M_r^G(\hat{\vect{n}}) M_{s)}^G(\hat{\vect{n}}) ,
\end{equation}
and
the weighted maps $M_p^G(\hat{\bn})$ are defined by
$M_p^G(\hat{\vect{n}})=\sum_{l m} q_p(l/\lmax)
a^G_{lm} Y_{lm}(\hat{\vect{n}}) / ((2 l+1)^{1/6}\sqrt{C_l})$.
We shall employ this latter expansion in this paper.

\begin{figure}
\centering
\vspace{0.25cm}
\hspace{0.1cm}
{
\includegraphics[width=0.44\linewidth]{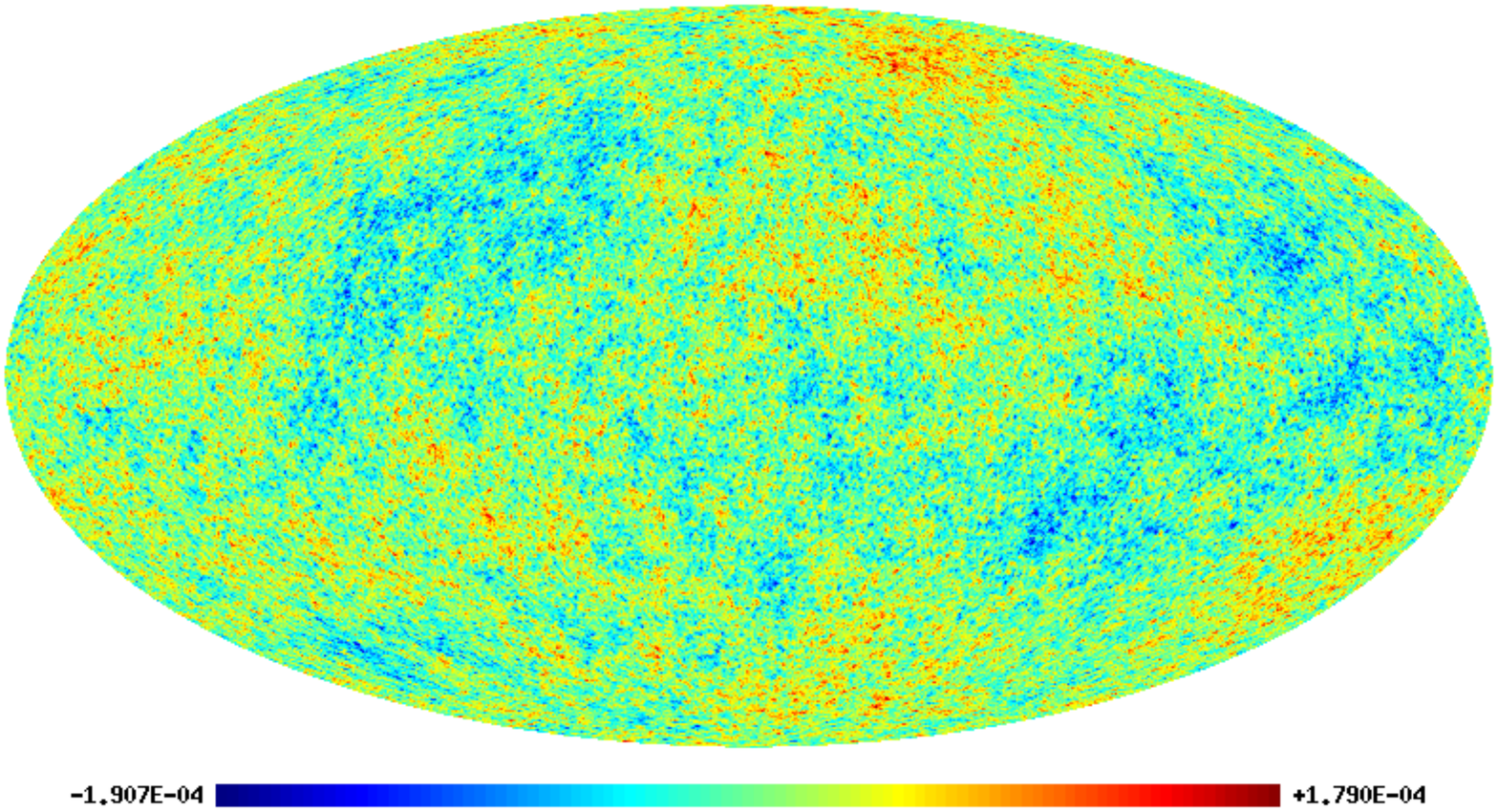}}
{
\includegraphics[width=0.44\linewidth]{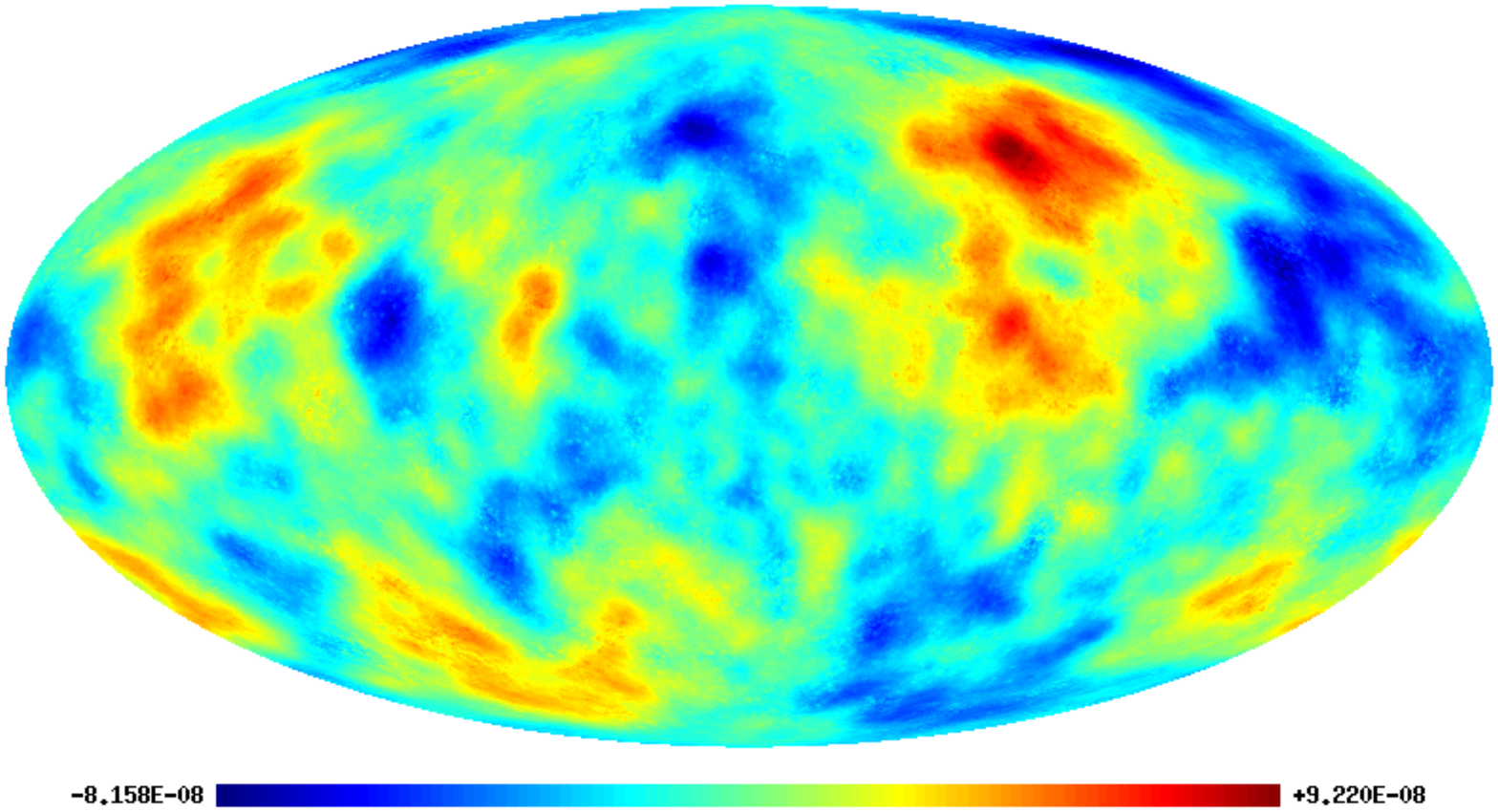}}
{
\includegraphics[width=0.44\linewidth]{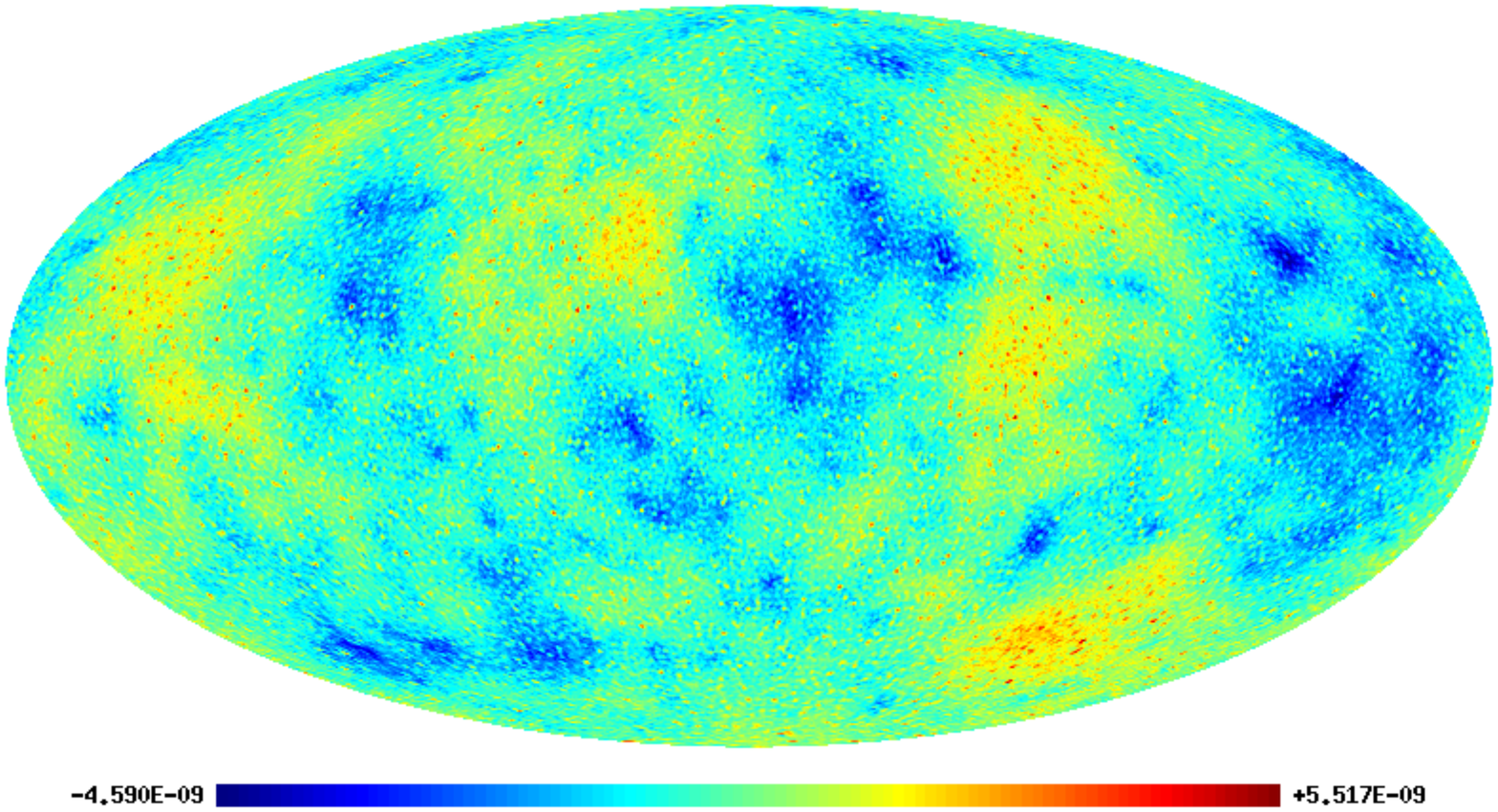}}
{
\includegraphics[width=0.44\linewidth]{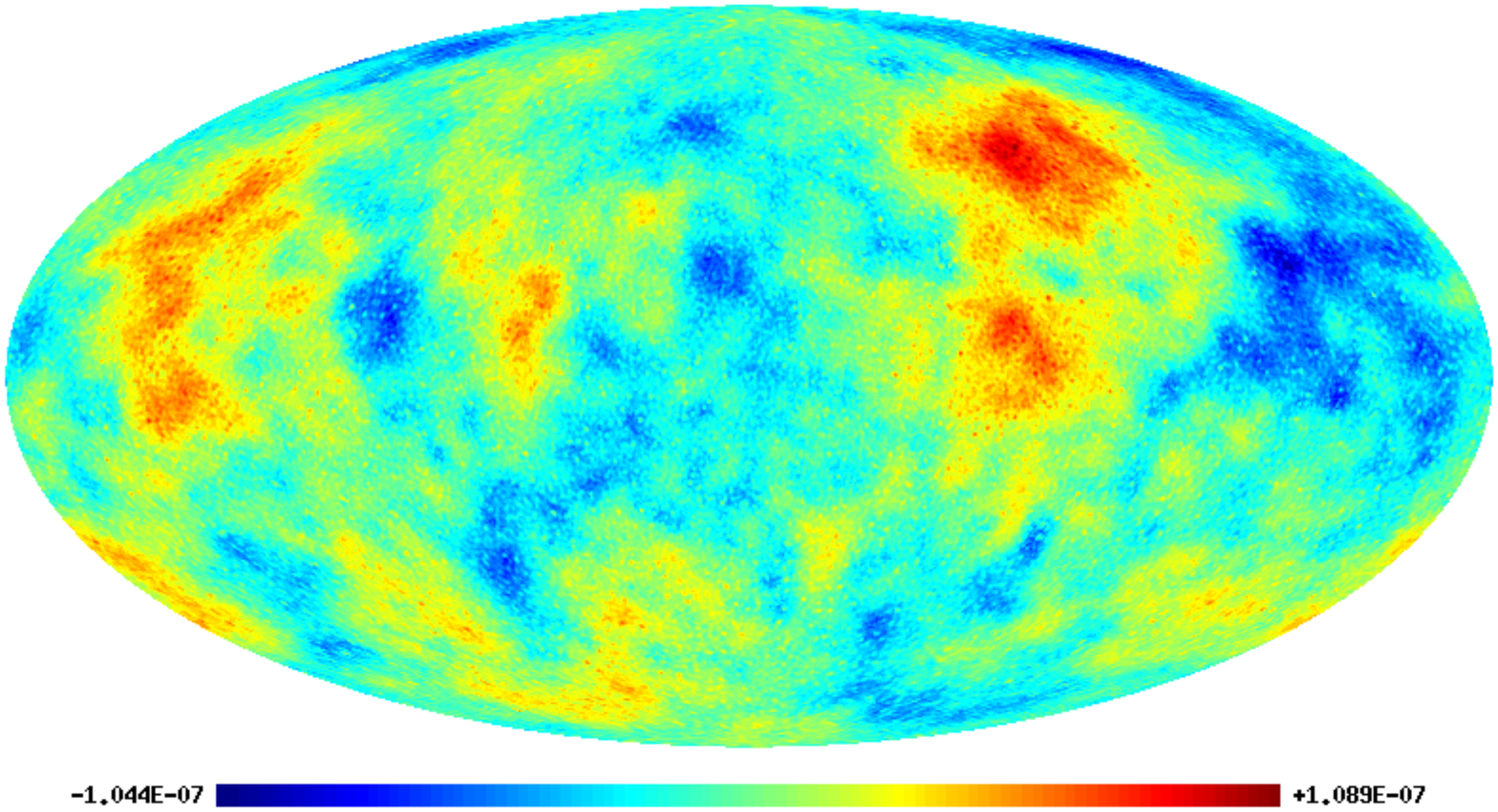}}
\caption{Represented clockwise from top left are a purely Gaussian simulation and the bispectrum simulations, computed using equation \eqref{eq:latealmB}, for the local, equilateral and flattened models, respectively.}
\label{fig:SimsMaps}
\end{figure}

\begin{figure}
\centering
\vspace{0.25cm}
\hspace{0.1cm}
{
\includegraphics[width=0.6\linewidth]{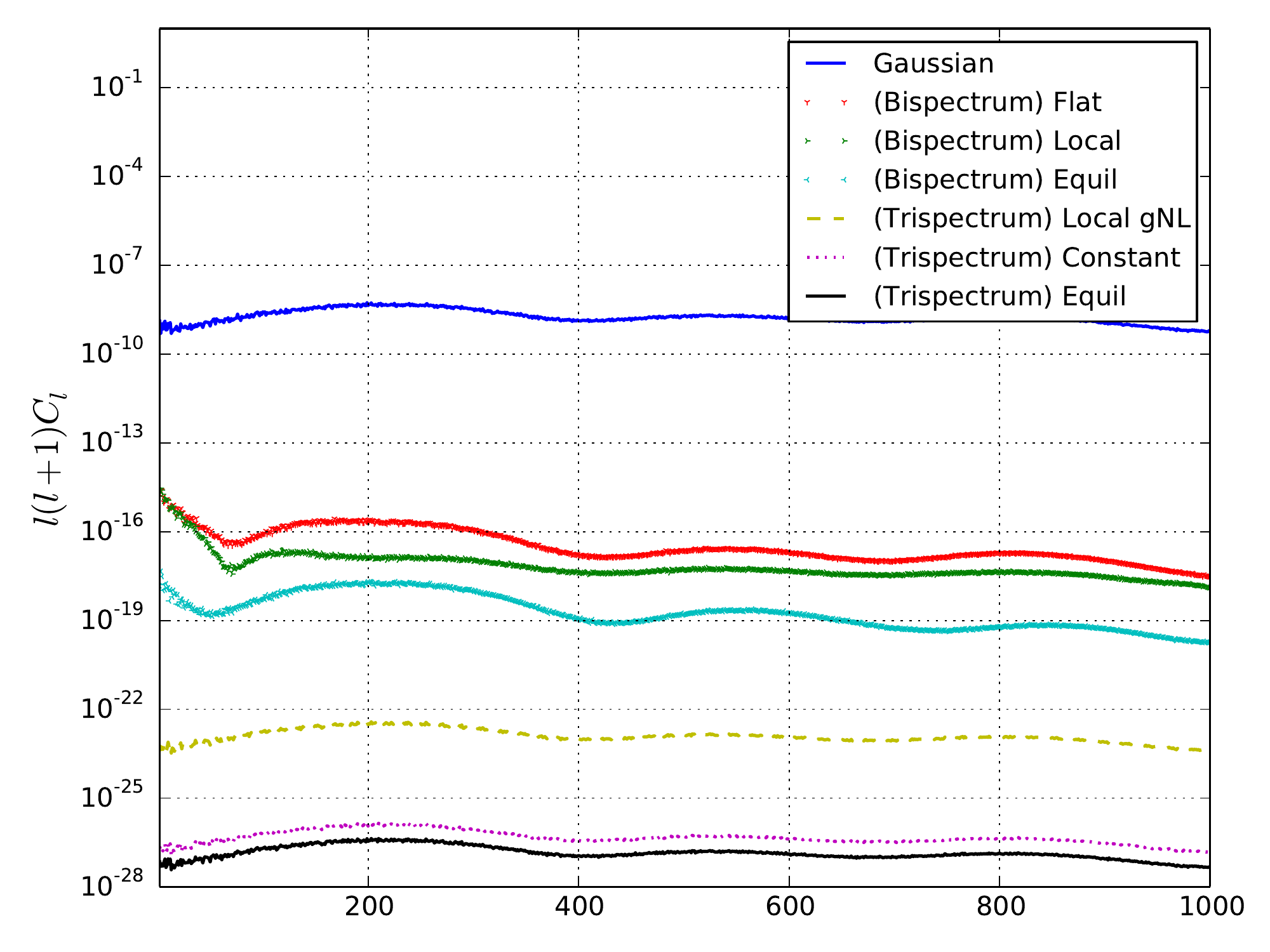}}
\caption{The power spectrum corresponding to the simulations of each of the models described in Fig.~\ref{fig:SimsMaps} is plotted. Also plotted are the power spectra computed for three (`diagonal-free') trispectrum models, as shall be described in \S\ref{subsec:trispecConstraints} and evaluated using equation~\eqref{eq:almT-diag-free}. It is clear that the bispectrum and trispectrum simulations are perturbative compared to the Gaussian part at all multipoles. Plotted from top-to-bottom are the Gaussian model, the flattened, local and equilateral bispectrum models, and the local ($g_{NL}$), constant and equilateral ($c_1$) trispectrum models, respectively.}
\label{fig:Cls_sims}
\end{figure}

In Fig.~\ref{fig:SimsMaps} we plot the simulated map for a Gaussian seed, and the local, equilateral and flattened models computed using this seed with the prescription described above. In Fig.~\ref{fig:Cls_sims} we plot the power spectrum for each map, verifying that the non-Gaussian contribution\footnote{Note that each of the bispectrum simulations is produced for $\fnl^B=1$.} is perturbative for all multipoles.

\para{Needlet bispectrum estimator.}
We now describe the construction of an estimator for $\fnl^B$ using needlets.

To define the family of needlets we will use,
consider a smooth weight function $b(t)$ of compact support and satisfying
the conditions
\begin{enumerate}
    \item $b(t)>0$ only if $t\in[\mu^{-1}, \mu]$ for some $\mu > 1$
    \item $\sum_{j=0}^{\infty}b^2(l/\mu^j)=1$, for $l=1,2,\dots$ .
\end{enumerate}
We pick a set of `scales' $j$ which characterize the needlets used in the analysis,
represents as powers of the basic scale $\mu = 1.53$.
For each $j$ we choose a set of $\npix$ points on the sphere, labelled $k$,
and denote these points $\xi_{jk}$.
Then, the needlet
functions $\psi_{jk}(\hat{\vect{n}})$ are defined by
\begin{equation}
\psi_{jk}(\hat{\bn})=\sqrt{\lambda_{jk}}\sum_{l=\mu^{j-1}}^{\mu^{j+1}}\sum_{m=-l}^l b\Big(\frac{l}{\mu^j}\Big)Y_{lm}(\hat{\bn})Y^*_{lm}(\xi_{jk}) ,
\end{equation}
where the $\lambda_{jk}$ are normalization coefficients
which are proportional the the pixel
area.%
    \footnote{In the case of the equal area pixel division
    used by HEALPix (http://healpix.jpl.nasa.gov), the coefficients $\lambda_{jk}$ may
    be given by an arbitrary constant value.}
Applied to a CMB temperature map $\Delta T(\hat{\vect{n}})/T$ this
yields needlet coefficients
\begin{equation}\label{eq:needletmap}
    \beta_{jk}
    =
    \int_{S^2} \d^2 \hat{\vect{n}} \;
    \frac{\Delta T(\hat{\bn})}{T}\psi_{jk}(\hat{\bn})
    =
    \sqrt{\lambda_{jk}}
    \sum_{l=\mu^{j-1}}^{\mu^{j+1}}
    \sum_{m=-l}^l
    b\left(\frac{l}{\mu^j}\right)
    a_{lm} Y_{lm}(\xi_{jk})
\end{equation}
It has been shown that
the mode--mode coupling introduced by masking and anisotropic noise
can be accounted for by subtracting,
for each scale $j$, the average over the $\npix$ pixels $k$~\cite{Donzelli2012}.
Defining the mean
for scale $j$ to be $\bar{\beta}_{j}=\sum_k \beta_{jk}/n_{\rm{pix}}$,
this gives
$\beta_{jk}\rightarrow \beta_{jk}-\bar{\beta}_{j}$.
In what follows we use these subtracted quantities.

In this paper we consider the Mexican needlets defined by
the weight function~\cite{Scodeller2010}
\begin{equation}
    b\left(\frac{l}{\mu^j}\right)=\left(\frac{l}{\mu^j}\right)^{2p}\exp\left(-\frac{l^2}{\mu^{2j}}\right) ,
\end{equation}
and choose the coefficients $\lambda_{jk}$ to be unity.
We set $p=1$,
for which the Mexican needlets
give a good approximation to the Spherical Mexican Hat wavelets
at high frequencies,
which corresponds here to large $j$.

The cubic needlet statistic is given by 
\begin{equation}
    N_{J}=\frac{1}{\sigma_{j_1}\sigma_{j_2}\sigma_{j_3}} \sum_{k=1}^{\npix}\beta_{j_1 k}\beta_{j_2 k}\beta_{j_3 k} ,
\end{equation}
where the triplet index $J$ represents $(j_1, j_2, j_3)$, and
$\sigma_j=(4\pi)^{-1}\sum_l (2l+1)C_l b(l/\mu^j)$.
The expectation value of $N_J$ for a given nonlinear map can be evaluated
by computing
\begin{equation}
\label{eq:NJ-ensemble-average}
    \langle N_J^{B}\rangle = \frac{1}{\sigma_{j_1}\sigma_{j_2}\sigma_{j_3}}
    \sum_{k=1}^{\npix}\langle\beta^G_{j_1 k}\beta^G_{j_2 k}\beta^{B}_{j_3 k}\rangle+\text{2 perms} ,
\end{equation}
where the superscripts `$G$' and `$B$'
indicate needlet maps
computed using,
respectively, the
Gaussian coefficients $a_{lm}^G$
and
non-Gaussian coefficients $a_{lm}^B$ which include the bispectrum $B$.
Using Eqs.~\eqref{eq:almB}, \eqref{eq:needletmap} and~\eqref{eq:NJ-ensemble-average}
we infer that
$\langle N^B_J \rangle$ can be written
\begin{align}
\langle N_J^{B}\rangle =\sum_n \baralpha_n^Q \langle N_{n J}^{B}\rangle ,
\end{align}
where the needlet maps for each mode $Q_n$, written $\langle N_{nJ}^B \rangle$,
are understood to be defined by this expression.
They allow for a change of basis to be performed between
needlets and partial waves.

The cubic needlet statistic enables us
to define an estimator for the amplitude of
the bispectrum $B$ which is present in the
measured CMB bispectrum.
Explicitly, we have
\begin{align}\label{eq:fnlB}
    \fnlhat^B= \frac{{\sum_{IJ}\langle N_I^{B} \rangle C_{IJ}^{-1} \hat{N}_J }}
    {{\sum_{IJ}\langle N_I^{B} \rangle C_{IJ}^{-1} \langle N_J^{B} \rangle}} , 
\end{align}
where the covariance matrix $C_{IJ}$ is defined by
$C_{IJ} \equiv
\langle N_I^G N_J^G\rangle
-
\langle N_I^G \rangle\langle N_J^G\rangle$,
and
\begin{equation}
    N_J^G = \frac{1}{\sigma_{j_1} \sigma_{j_2} \sigma_{j_3} \npix}
    \sum_{k=1}^{\npix} \beta^G_{j_1 k}  \beta^G_{j_2 k}  \beta^G_{j_3 k} .
\end{equation}
Finally,
$\hat{N}_J$ is the cubic needlet statistic evaluated from the data.
We invert the covariance matrix using principal component
analysis,
imposing a ratio of $10^{10}$ between the maximum and minimum eigenvalues
which are retained. This approach has been used in other studies of the bispectrum using wavelets \cite{Curto2011} and needlets \cite{Donzelli2012}.
The 1-$\sigma$ error bar
on $\fnlhat^B$
is
\begin{equation}
    \sigma(\fnlhat^B) = \Big( \sum_{IJ}\langle N_I^{B} \rangle C_{IJ}^{-1} \langle N_J^{B} \rangle \Big)^{-1/2} .
\end{equation}
Using the change-of-basis matrix
$\langle N_{nJ}^B \rangle$
we may rewrite Eq.~\eqref{eq:fnlB} in the form
\begin{align}\label{eq:fnlB2}
    \fnlhat^B=\frac{\sum_{n}\baralpha_n^Q \langle N_{nI}^{B} \rangle C_{IJ}^{-1} N_J }{\sum_{n m}\baralpha_n^Q\baralpha_m^Q\sum_{IJ}\langle N_{n I}^{B} \rangle C_{IJ}^{-1} \langle N_{m J}^{B} \rangle}=\frac{\sum_{n}\baralpha_n^Q \bar{\beta}_n^Q }{\sum_{n m}\baralpha_n^Q \gamma_{nm}\baralpha_m^Q }\,, 
\end{align}
with natural definitions
of $\bar{\beta}_n^Q$
and $\gamma_{n m}$
which may be deduced from this equation.
Performing the Cholesky decomposition
$\gamma_{n m}=\sum_r \lambda^{-1}_{n r}\lambda^{-1}_{m r}$,
and defining $\baralpha_r^R=\sum_n \lambda^{-1}_{n r} \baralpha_n^Q$
and $\bar{\beta}_r^R=\sum_n \lambda_{r n} \bar{\beta}_n^Q$,
we find that an ensemble of maps simulated with $\fnl^B=1$ satisfy
the consistency relation
\begin{align}
\langle \bar{\beta}_n^R\rangle=\baralpha_n^R\,.
\end{align}
Thus, the coefficients $\bar{\beta}_n^Q$ recovered from the
needlet maps may be used to reconstruct the underlying
bispectrum shape~\cite{RMS2013}. 

\section{CMB trispectrum with needlets}\label{sec:trispneedlets}

\para{Primordial trispectrum.}
The primordial trispectrum $T_{\Phi}$ is defined by
the connected four-point function of the primordial gravitational potential
\begin{equation}
    \label{eq:trispectrum-def}
    \langle\Phi(\bk_1)\Phi(\bk_2)\Phi(\bk_3)\Phi(\bk_4)\rangle_c
    =
    (2\pi)^3 \delta(\bk_1+\bk_2+\bk_3+\bk_4)
    T_{\Phi}(\bk_1,\bk_2,\bk_3,\bk_4) .
\end{equation}
In this paper we restrict attention
to trispectra which are `diagonal' in the sense that they can
be written
\begin{equation}
    T_{\Phi}(\bk_1, \bk_2, \bk_3, \bk_4)
    =
    p(k_1, k_2, k_3, k_4, K_{12})
    + p(k_1, k_2, k_3, k_4, K_{13})
    + p(k_1, k_2, k_3, k_4, K_{14}) ,
    \label{eq:diagonal-trispectrum}
\end{equation}
where $\bK_{ij} = \bk_i + \bk_j$
represents a diagonal of the quadrilateral formed
by the momenta $\bk_i$.
The zero-sum condition $\sum_i \bk_i = 0$ enforced by the
momentum-conservation $\delta$-function in~\eqref{eq:trispectrum-def}
means that it is unnecessary to include
the remaining combinations $\bK_{23}$, $\bK_{24}$ and
$\bK_{34}$.
The diagonal condition is not generic:
for example, it is not satisfied by
the microphysical component of the trispectrum which
is generated by
interactions near the epoch of horizon exit~\cite{Seery:2006vu,*Seery:2008ax,*Seery:2006js}.
However,
Eq.~\eqref{eq:diagonal-trispectrum} often does apply
for phenomenological shapes
generated with observable amplitude in certain models.

An interesting subclass of diagonal trispectra---%
including the `local' $\gnl$-shape,
the $c_1$ equilateral trispectrum~\cite{aChen}
and the constant trispectrum~\cite{FRS2}---
do not depend on the diagonals $\bK_{ij}$
but only the individual side-lengths $k_i$.
We describe these as `diagonal-free'.
In particular, the $\gnl$ shape satisfies
\begin{equation}\label{eq:trispgnl}
    T_{\Phi}^{\gnl}(k_1, k_2, k_3, k_4)
    =
    6\Big(
        P_{\Phi}(k_1)P_{\Phi}(k_2)P_{\Phi}(k_3)+\text{3 perms}
    \Big) .
\end{equation}
We define the shape function for diagonal-free trispectra by
\begin{equation}\label{eq:diagfree}
    S^{T}(k_1,k_2,k_3,k_4)
    =
    \frac{T_{\Phi}(k_1,k_2,k_3,k_4)}{T_{\Phi}^{\gnl}(k_1, k_2, k_3, k_4)} .
\end{equation}
Fergusson, Regan \& Shellard~\cite{FRS2} pointed out that
it is possible to decompose these diagonal-free trispectra
by analogy with Eq.~\eqref{eq:shapeloc1}.
Labelling unique 4-tuples $(n_1, n_2, n_3, n_4)$ by
a multi-index $n$,
in the same way that we used a multi-index to label
unique triplets in~{\S\ref{sec:bispneedlets}},
we write
\begin{align}\label{eq:trispdecompos}
    S^{T}(k_1,k_2,k_3,k_4)
    =
    \sum_{n} \alpha^{T Q}_n
        q^{T}_{( n_1}(k_1)
        q^{T}_{n_2}(k_2)
        q^{T}_{n_3}(k_3)
        q^{T}_{n_4)}(k_4) .
\end{align}
The basis functions $q^T_n$ are constructed so
that
\begin{equation}
    \int_{\mathcal{V}} \Big( \prod_{i=1}^4 \d x_i \Big) q^T_n (x_1) q^T_m(x_1)=\delta_{n m} ,
\end{equation}
where the integration domain $\mathcal{V}$
is defined by the condition $\sum_i x_i > 2\max \{ x_i\}$
and $0\leq x_i\leq 1$.
The expansion coefficients $\alpha_n^{TQ}$
are obtained by defining an inner product
analogous to~\eqref{eq:k-ip-def}
and using this to construct
coefficients analogous to~\eqref{eq:alpha-n-Q}.
For more details on the construction of the $q_n^T$,
the definition of this inner product and the calculation of the
expansion coefficients $\alpha_n^{TQ}$ we refer to Refs.~\cite{RSF1,FRS2}.


In this paper the only trispectrum
we will consider which is not
diagonal-free
is the local $\taunl$-shape given by
\begin{equation}\label{eq:ptaunl}
    p^{\taunl}(k_1, k_2, k_3, k_4; K)
    = \frac{25}{9}
    \big( P_{\Phi}(k_1)+P_{\Phi}(k_2) \big)
    \big( P_{\Phi}(k_3)+P_{\Phi}(k_4) \big)
    P_{\Phi}(K) .
\end{equation}
Because this cannot be decomposed using~\eqref{eq:trispdecompos},
we must deal with this model separately.
However, as will be evident from our treatment of the trispectrum
in what follows,
the formalism discussed in this paper is general and
applicable to arbitrary trispectra.

\para{CMB trispectrum: diagonal-free case.} 
The CMB trispectrum is given for diagonal-free trispectra by the connected four-point function of the spherical harmonics of the temperature map
\begin{equation}
    \langle a_{l_1 m_1}a_{l_2 m_2}a_{l_3 m_3}a_{l_4 m_4}\rangle_c
    =
    t_{l_1 l_2 l_3 l_4}
    \int \d^2\hat{\bn} \;
    Y_{l_1 m_1}(\hat{\bn})
    Y_{l_2 m_2}(\hat{\bn})
    Y_{l_3 m_3}(\hat{\bn})
    Y_{l_4 m_4}(\hat{\bn}) .
\end{equation}
where the `reduced' trispectrum $t_{l_1 l_2 l_3 l_4}$
corresponding to the diagonal-free decomposition~\eqref{eq:trispdecompos}
can be written~\cite{RSF1,FRS2}
\begin{equation}
    \label{eq:diagonal-free-t}
    t_{l_1 l_2 l_3 l_4}
    =
    \sum_n \alpha^{T Q}_n
    \int \d x \, x^2 \;
    \tilde{q}^{(T,2) l_1}_{(n_1}(x)
    \tilde{q}^{(T,-1) l_2}_{n_2}(x)
    \tilde{q}^{(T,-1) l_3}_{n_3}(x)
    \tilde{q}^{(T,-1) l_4}_{n_4)}(x) ,
\end{equation}
where $\tilde{q}^{(T,2)}_n$ and $\tilde{q}^{(T,-1)}_n$ are defined
as in Eq.~\eqref{eq:qplus2} with $q_n$ replaced by $q^{T}_n$.

We now proceed by analogy with the bispectrum,
defining a transfer matrix similar to $\Gamma_{mn}$
which accounts for the line-of-sight integral over the transfer function
$\Delta_l$,
and expressing the trispectrum in terms of
\emph{late-time} coefficients $\bar{\alpha}^{TQ}_n$
analogous to those of Eq.~\eqref{eq:latetime1}.
Therefore these results are limited to
diagonal-free trispectra.
The $\bar{\alpha}^{T Q}_n$, are chosen to satisfy
\begin{align}\label{eq:diagfree-late}
    s^T(l_1,l_2,l_3,l_4)\equiv
    \frac{t_{l_1 l_2 l_3 l_4}}{\sqrt{C_{l_1} C_{l_2} C_{l_3} C_{l_4} }}
    =
    \sum_n \bar{\alpha}_n^T Q_n^T(l_1,l_2,l_3,l_4) .
\end{align}
We define the inner-product
$\tripleft f, g \tripright$ by
\begin{equation}
    \tripleft f, g \tripright
    = \sum_{l_i} f(l_1,l_2,l_3,l_4)g(l_1,l_2,l_3,l_4)w(l_1,l_2,l_3,l_4) ,
\end{equation}
where the weight function $w$ satisfies
\begin{equation}
    w_{l_1 l_2 l_3 l_4} = \frac{1}{32\pi^2} \Big[
        \int_{-1}^1 \d\mu \prod_{i=1}^4 (2 l_i+1) P_{l_i}(\mu)
    \Big] .
\end{equation}
This choice is made so that the Fisher matrix
is equal to $\tripleft s^T, s^T \tripright / 24$,
which will appear in Eq.~\eqref{eq:FisherT} below.
With all these choices,
the late-time coefficients $\bar{\alpha}^{TQ}_n$ are given by
\begin{equation}
    \bar{\alpha}_n^T = \sum_{r m} \alpha^T_r
        \tripleft \tilde{Q}_r , Q^T_m \tripright H^{-1}_{mn}
        = \sum_r \alpha^T_r \Gamma^T_{rn},
\end{equation}
where $H_{mn} = \tripleft Q^T_m, Q^T_n \tripright$
and
\begin{equation}
    \tilde{Q}_n(l_1, l_2,l_3,l_4)
    =
    (C_{l_1} C_{l_2} C_{l_3} C_{l_4})^{-1/2}
    \int \d x \; x^2
    \tilde{q}^{(T,2) l_1}_{(n_1}(x)
    \tilde{q}^{(T,-1) l_2}_{n_2}(x)
    \tilde{q}^{(T,-1) l_3}_{n_3}(x)
    \tilde{q}^{(T,-1) l_4}_{n_4)}(x).
\end{equation}
The transfer matrix for the trispectrum is defined by
$\Gamma^T_{mn} = \sum_r \tripleft \tilde{Q}_m , Q^T_r \tripright H^{-1}_{rn}$.

\para{CMB trispectrum: diagonal case.}
These results do not apply for trispectra
which are not diagonal-free,
such as the local $\taunl$-shape~\eqref{eq:ptaunl}.
For a general diagonal trispectrum,
the analogue of Eq.~\eqref{eq:diagonal-free-t}
is
\begin{equation}\label{eq:gentrisp}
    \langle a_{l_1 m_1}a_{l_2 m_2}a_{l_3 m_3}a_{l_4 m_4}\rangle_c
    =
    \sum_{LM}(-1)^M \Big[
        p^{l_1 l_2}_{l_3 l_4}(L)
        \mathcal{G}^{l_1 l_2 L}_{m_1 m_2 M}
        \mathcal{G}^{l_3 l_4 L}_{m_3 m_4 -M}
        +
        (2\leftrightarrow 3)
        +
        (2\leftrightarrow 4)
    \Big] ,
\end{equation}
where $\mathcal{G}^{l_1 l_2 L}_{m_1 m_2 M}\equiv\int \d^2\hat{\bn} \; Y_{l_1 m_1}(\hat{\bn})Y_{l_2 m_2}(\hat{\bn})Y_{L M}(\hat{\bn})$
and $(2 \leftrightarrow 3)$, $(2 \leftrightarrow 4)$
represent the previous expression with the labels
$2$, $3$ and
$2$, $4$ exchanged, respectively.
For example,
the local $\taunl$-shape given in Eq.~\eqref{eq:ptaunl}
results in the CMB trispectrum
\begin{equation}\label{eq:taunlfull}
    p^{l_1 l_2}_{l_3 l_4}(L)
    =
    \int \d r_1 \, \d r_2 \; r_1^2 r_2^2
    F_L(r_1,r_2)
    \Big[
        \alpha_{l_1}(r_1) \beta_{l_2}(r_1)
        + \alpha_{l_2}(r_1) \beta_{l_1}(r_1)
    \Big]
    \Big[
        \alpha_{l_3}(r_3) \beta_{l_4}(r_2)
        + \alpha_{l_4}(r_3)\beta_{l_3}(r_2)
    \Big] ,
\end{equation}
where
\begin{subequations}
\begin{align}
    F_L(r_1,r_2)
    &
    =
    \int \d K \; K^2
    P_{\Phi}(K) j_L(Kr_1) j_L(K r_2) ,
    \\
    \label{eq:alpha-l-def}
    \alpha_{l}(x)
    &
    =
    \int \d k \; k^2
    \Delta_l(k) j_l(k x) ,
    \\
    \label{eq:beta-l-def}
    \beta_{l}(x)
    &
    =
    \int \d k \; k^2
    P_{\Phi}(k)\Delta_l(k) j_l(k x) .
\end{align}
\end{subequations}
Note that, despite the similarity of notation,
$\alpha_l$, $\beta_l$ as defined here are distinct from the
decomposition coefficients $\alpha_n^Q$, $\alpha_n^{TQ}$
and the needlet coefficients $\beta_{jk}$, $\bar{\beta}_j$.
The definitions~\eqref{eq:alpha-l-def}--\eqref{eq:beta-l-def}
are conventional.

It was shown by Pearson et al. that the following
approximation is accurate to within $\lesssim 2\%$ \cite{Pearson},
\begin{align}\label{eq:ptrisp}
    p^{l_1 l_2}_{l_3 l_4}(L)
    \approx
    C^{\zeta_*}_L (C_{l_1}+C_{l_2})(C_{l_3}+C_{l_4} ) ,
\end{align}
where $C^{\zeta_*}_L=(25/9)\int \d K \; K^2 P_{\Phi}(K) j_L(K r_*)^2$
is the angular power spectrum of the curvature perturbation $\zeta$,
and $r_*$ represents the distance to the last scattering surface.

\para{Simulating the CMB trispectrum.}
As in~{\S\ref{sec:bispneedlets}} we fix a choice of trispectrum
$T(\vect{k}_1, \vect{k}_2, \vect{k}_3, \vect{k}_3)$
and include it
in the trispectrum of the primordial gravitational potential
with amplitude $\gnl^T$,
so that
$T_\Phi(\vect{k}_1, \vect{k}_2, \vect{k}_3, \vect{k}_4)
\supset \gnl^T
T(\vect{k}_1, \vect{k}_2, \vect{k}_3, \vect{k}_4)$.
As for $\fnl^B$, it is important to be clear that
$\gnl^T$ does not coincide with the
traditional local
$\gnl$-parameter
\cite{Okamoto:2002ik,*Boubekeur:2005fj,*Sasaki:2006kq}
unless
$T$ is the conventionally-normalized
local trispectrum~\eqref{eq:trispgnl}.
In this paper, the traditional
local $\gnl$-parameter is always denoted $\gnllocal$.
Our task is to build an estimator for $\gnl^T$.

\begin{figure}
\centering
\vspace{0.25cm}
\hspace{0.1cm}
{
\includegraphics[width=0.3\linewidth]{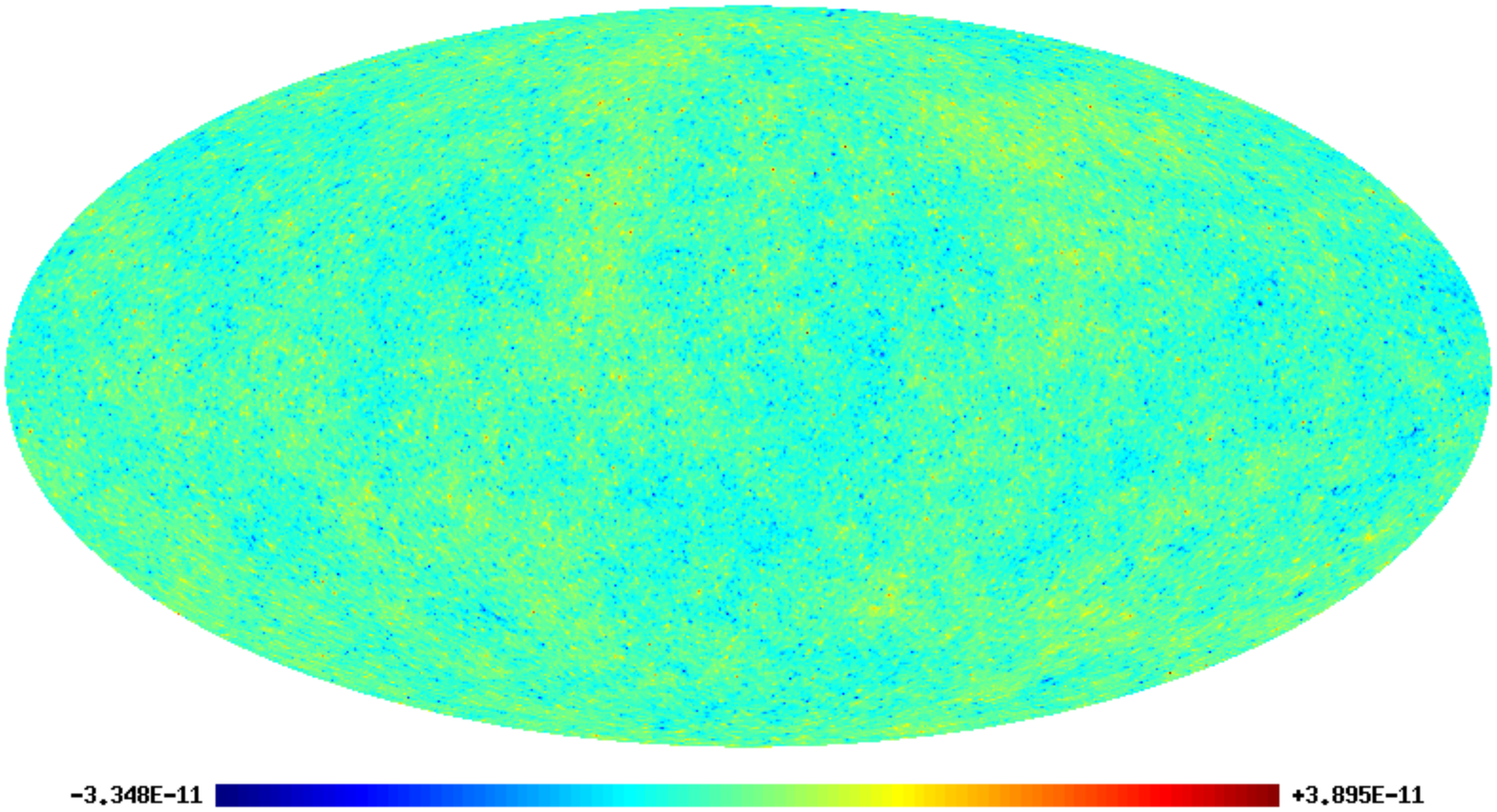}}
{
\includegraphics[width=0.3\linewidth]{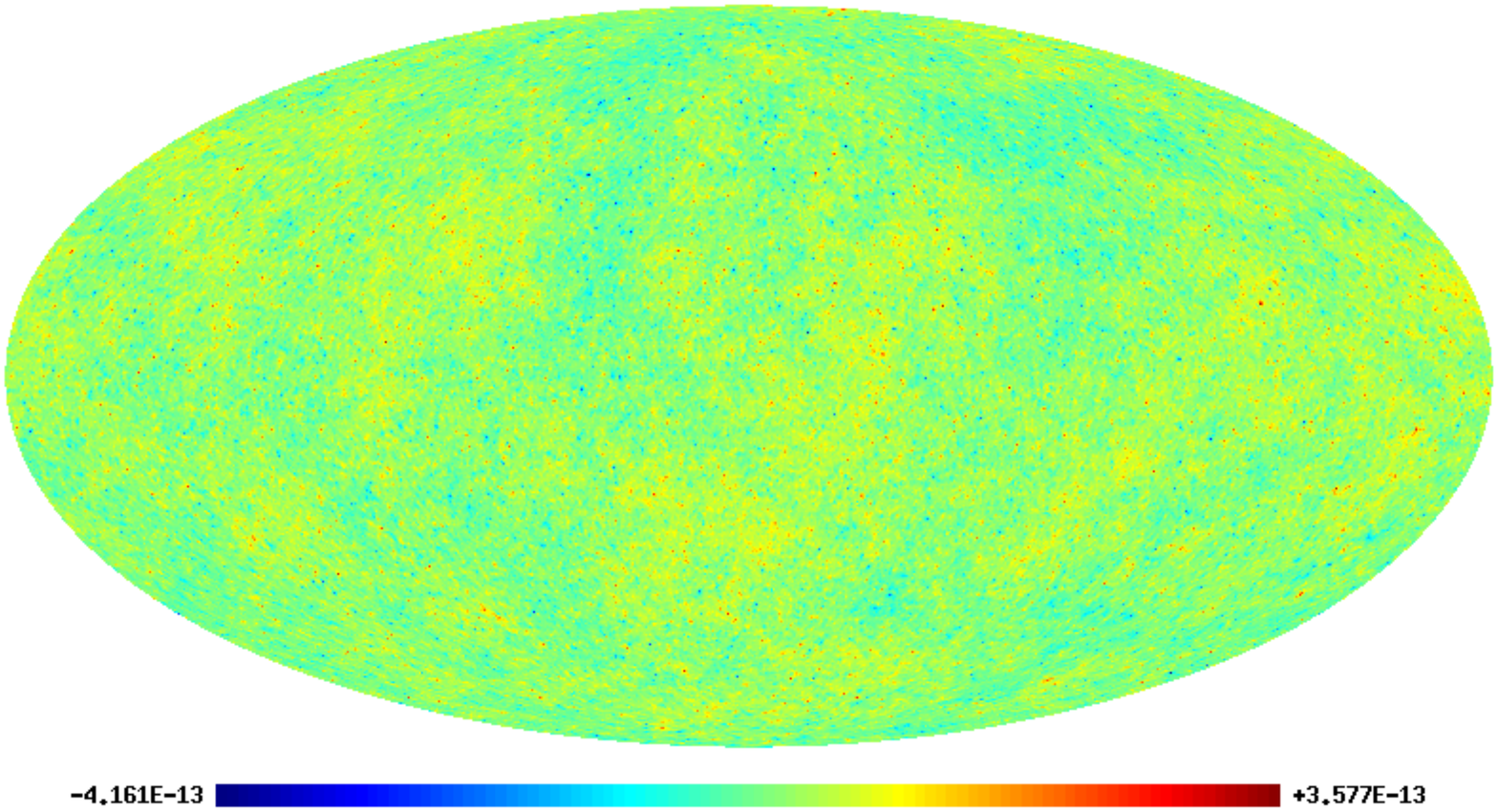}}
{
\includegraphics[width=0.3\linewidth]{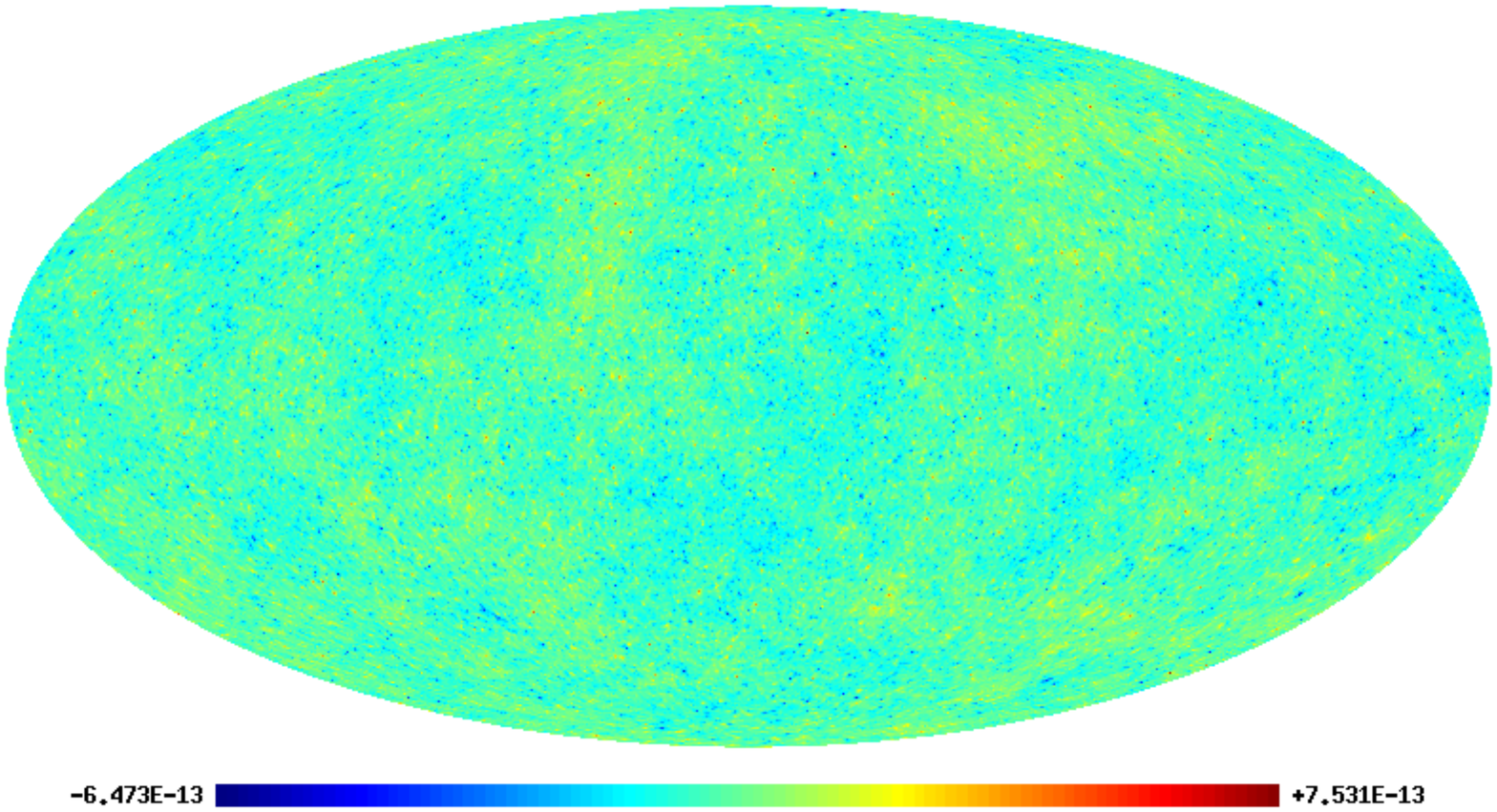}}
\caption{From left to right are the trispectrum simulations for the local ($g_{NL}$), equilateral ($c_1$) and constant trispectrum models, computed using equation \eqref{eq:almT-diag-free}.}
\label{fig:SimsMapsT}
\end{figure}

To estimate the covariance matrix
we must again
construct averages
over an ensemble of maps which
contain the trispectrum $T$.
For the analysis in this section,
and for the constraints reported
in~{\S\ref{sec:wmaptest}} below, we will
assume that there is no primordial bispectrum.
It follows that
we can generate appropriate maps
for the temperature anisotropy
by constructing multipole coefficients
$a_{lm} = a_{lm}^G + \gnl a_{lm}^T$,
were $a_{lm}^G$ continues to be the dominant Gaussian
contribution and $a_{lm}^T$ is a correction chosen
to reproduce the trispectrum $T$.
For a general trispectrum~\eqref{eq:gentrisp}
we have \cite{RSF1,FRS2}
\begin{align}
    \label{eq:almT-general}
    a_{lm}^T
    =
    \frac{1}{8}
    \sum_{LM}
    \sum_{l_i m_i}
    (-1)^M
    p^{l l_2}_{l_3 l_4}(L)
    \mathcal{G}^{l l_2 L}_{m m_2 M}
    \mathcal{G}^{l_3 l_4 L}_{m_3 m_4 -M}
    \frac{a_{l_2 m_2}^{G*}}{C_{l_2}}
    \frac{a_{l_3 m_3}^{G*}}{C_{l_3}}
    \frac{a_{l_4 m_4}^{G*}}{C_{l_4}}\ .
\end{align}
In the case of diagonal-free trispectra we may instead use the
expression~\cite{RSF1,FRS2}
\begin{equation}\label{eq:almT-diag-free}
\begin{split}
    a_{lm}^T
    & =
    \frac{1}{24}
    \sum_{l_i m_i}t_{l l_2 l_3 l_4}
    \int \d^2 \hat{\bn} \;
    Y_{lm}(\hat{\bn})
    Y_{l_2 m_2}(\hat{\bn})
    Y_{l_3 m_3}(\hat{\bn})
    Y_{l_4 m_4}(\hat{\bn})
    \frac{a_{l_2 m_2}^{G*}}{C_{l_2}}
    \frac{a_{l_3 m_3}^{G*}}{C_{l_3}}
    \frac{a_{l_4 m_4}^{G*}}{C_{l_4}}
    \\
    & =
    \frac{1}{24}
    \sum_n
    \alpha_n^{TQ}
    \int \d x \; x^2
    \int \d\hat{\vect{n}} \;
    Y_{l m}(\hat{\vect{n}})
    \Big[
        \tilde{q}^{(2)l}_{(n_1}(x)
        \tilde{M}_{n_2}^{(-1)(T,G)}(x,\hat{\vect{n}})
        \tilde{M}_{n_3}^{(-1)(T,G)}(x,\hat{\vect{n}})
        \tilde{M}_{n_4)}^{(-1)(T,G)}(x,\hat{\vect{n}})
    \Big] ,
\end{split}
\end{equation}
where
$\tilde{M}_p^{(-1)(T,G)}(x,\hat{\vect{n}})=\sum_{l m} \tilde{q}^{(-1,T)l}_p(x) a^G_{lm} Y_{lm}(\hat{\vect{n}})/C_l$.
Alternatively, using a similar approach to that described for the bispectrum, we may utilise the
late-time CMB trispectrum expansion given by equation~\eqref{eq:diagfree-late}, and write
\begin{equation}\label{eq:almT_late}
a_{lm}^T = \sum_{n} \bar{\alpha}_n^{TQ} \frac{\sqrt{C_l}}{24} \int \d\hat{\bn}
\; Y_{l m}(\hat{\bn})q_{(n_1}\big( \frac{l}{\lmax} \big)
\bar{M}^{G}_{n_2}(\hat{\bn})\bar{M}^{G}_{n_3}(\hat{\bn})\bar{M}^{G}_{n_4)}(\hat{\bn}) ,
\end{equation}
where $\bar{M}^{G}_{n_2}(\hat{\bn})=\sum_{lm} q_{n_2}(l/\lmax)a_{lm}^G Y_{l m}(\hat{\bn})/\sqrt{C_l}$.
Using equation~\eqref{eq:almT-diag-free}, in Fig.~\ref{fig:SimsMapsT} we plot the simulated maps for the local ($\gnl$), equilateral ($c_1$) and constant trispectrum models, which will be described in~\S\ref{subsec:trispecConstraints}. The corresponding power spectra are plotted in Fig.~\ref{fig:Cls_sims} emphasising the perturbative nature of the trispectra compared to the Gaussian seed.

However we choose to obtain the $a^T_{lm}$,
we define
the quartic needlet statistic by
\begin{equation}
    N_J^{T}
    =
    \frac{1}{\sigma_{j_1}\sigma_{j_2}\sigma_{j_3}\sigma_{j_4}}
    \sum_{k=1}^{\npix}
    \beta^G_{j_1 k}\beta^G_{j_2 k}\beta^{G}_{j_3 k}\beta^{T}_{j_4 k}
    +
    \text{3 perms} ,
\end{equation}
where $J$ represents the unique 4-tuple $(j_1, j_2, j_3, j_4)$
and
the $\beta_{jk}$ are the needlet coefficients
defined in Eq.~\eqref{eq:needletmap}.
A superscript `$G$' indicates
that these coefficients are to be computed
using a Gaussian map,
and the superscript `$T$'
indicates that they should be computed
from a map which includes the trispectrum
correction $a_{lm}^T$.
We use Eq.~\eqref{eq:almT_late} to write
the expectation value of
$N_J^T$ over an ensemble of maps
\begin{align}\label{eq:NJT_late}
    \langle N_J^{T}\rangle =\sum_n \bar{\alpha}_n^{TQ} \langle N_{n J}^{T}\rangle .
\end{align}
As above, the change-of-basis matrix
$\langle N_{nJ}^T \rangle$
is defined by this expression.

In the case of the $\taunl$ trispectrum we may utilize equations \eqref{eq:ptrisp} and \eqref{eq:almT-general} in order to write
\begin{align}\label{eq:taunlNJ}
\langle N_J^{T,\taunl}\rangle=\sum_L C_L^{\zeta_*}\langle N_{L J}^{T,\taunl}\rangle\equiv \sum_L C_L^{\zeta_*}\sum_M 4\int \d^2\hat{\bn} \; Y_{LM}(\hat{\bn})\tilde{\beta}_{(j_1}(\hat{\bn})\tilde{\beta}_{j_2}(\hat{\bn}) \int \d^2\hat{\bn} \; Y_{LM}^*(\hat{\bn})\tilde{\beta}_{j_3}(\hat{\bn})\tilde{\beta}_{j_4)}(\hat{\bn})\,,
\end{align}
where we express $\tilde{\beta}_{j k}$ as $\tilde{\beta}_j(\hat{\bn})$,
with tilde representing a needlet map, \eqref{eq:needletmap},
with appropriate weighting determined by equation~\eqref{eq:ptrisp}.
The beam and mask properties must also be accounted for via a transformation of these
spherical harmonics, as will be described in \S\ref{sec:wmaptest}.

\para{Needlet trispectrum estimator.}
In
Ref.~\cite{RSF1},
the Edgeworth expansion was used to derive the optimal trispectrum estimator.
This was
\begin{equation}
\begin{split}
    \mathcal{E}^{\text{opt}}
    = &
    \sum_{l_i m_i, l_i' m_i'}
    \big \langle
        \prod_{i=1}^4 a_{l_i m_i}
    \big \rangle_c
    C^{-1}_{l_1 m_1, l_1' m_1'}
    C^{-1}_{l_2 m_2, l_2' m_2'}
    C^{-1}_{l_3 m_3, l_3' m_3'}
    C^{-1}_{l_4 m_4, l_4' m_4'}
    \\
    & \times
    \Big[
        \big(
            \prod_{i=1}^4 a_{l_i' m_i'}
        \big)
        - \big(
            C_{l_1' m_1',l_2' m_2'}
            a_{l_3' m_3'}
            a_{l_4' m_4'}
            + \text{5 perms}
        \big)
        + \big(
            C_{l_1' m_1',l_2' m_2'}
            C_{l_3' m_3',l_4' m_4'}
            + \text{2 perms}
        \big)
    \Big] ,
\end{split}
\end{equation}
where $C_{l_1 m_1,l_2 m_2} = \langle a_{l_1 m_1}a_{l_2 m_2}\rangle$.
Approximating the inverse covariance matrix as diagonal,
i.e. $C^{-1}_{l_1 m_1,l_2 m_2} = \delta_{l_1,l_2}\delta_{m_1,-m_2}(-1)^{m_1} C_{l_1}^{-1}$,
it follows that the Fisher matrix
$F\equiv \langle\mathcal{E}^{\rm{opt}}\rangle/24$
roughly satisfies
\begin{equation}\label{eq:FisherT}
    F
    \approx
    \frac{\fsky}{24}
    \sum_{l_i m_i}
    \frac{1}{C_{l_1}C_{l_2}C_{l_3}C_{l_4}}
    \big\langle
        \prod_{i=1}^4 a_{l_i m_i}
    \big\rangle_c^2
    ,
\end{equation}
where $\fsky$ represents the sky fraction
covered by the map
and $C_{l}$ is the total power spectrum,
including beam effects and noise contributions. It was this approximate expression that
was used in Refs.~\cite{RSF1,FRS2}.
A similar estimator
may be derived for needlets,
with the inverse covariance matrix used to optimize the signal to noise.
The estimator may be used to give an estimate for the
trispectrum amplitude,
$\gnlhat^T = \mathcal{E}/\langle\mathcal{E}\rangle$.
We find
\begin{equation}\label{eq:gnlT}
    \gnlhat^T =
    \frac{\sum_{IJ} \langle N_I^T \rangle C_{IJ}^{-1} \hat{\mathcal{N}}_J}
        {\sum_{IJ}\langle N_I^{T} \rangle C_{IJ}^{-1} \langle N_J^{T} \rangle}
    ,
\end{equation}
where $\hat{\mathcal{N}}_J$ is to be obtained from the data,
\begin{equation}
    \hat{\mathcal{N}}_J =
    \hat{N}_J -
    \Bigg(
        \frac{\sum_k \langle
            \beta^G_{j_1 k} \beta^G_{j_2 k}
        \rangle
        \hat{\beta}_{j_3 k} \hat{\beta}_{j_4 k}}
        {\npix \prod_{i=1}^4 \sigma_{j_i}}
        + \text{5 perms}
    \Bigg)
    +
    \frac{\sum_k \langle
            \beta^G_{j_1 k} \beta^G_{j_2 k} \beta^G_{j_3 k} \beta^G_{j_4 k}
        \rangle}
            {\npix \prod_{i=1}^4 \sigma_{j_i}} ,
\end{equation}
with the covariance matrix
defined by $C_{IJ} = \langle N_I^G N_J^G \rangle
- \langle N_I^G \rangle \langle N_J^G \rangle$.
The 1-$\sigma$ error bar is
$\Delta \gnlhat^T = \langle\mathcal{E}\rangle^{-1/2}$.

For the special case of diagonal-free trispectra,
where the decomposition~\eqref{eq:diagfree-late} applies,
we may write the trispectrum needlet map in the form~\eqref{eq:NJT_late}.
Then
the needlet estimator becomes
\begin{equation}\label{eq:gnlT2}
    \gnlhat
    =
    \frac{\sum_{n} \bar{\alpha}_n^{TQ} \langle N_{nI}^{T} \rangle C_{IJ}^{-1} \hat{\mathcal{N}}_J}
        {\sum_{n m}\bar{\alpha}_n^{TQ} \gamma_{nm}^T \bar{\alpha}_m^{TQ}}=\frac{\sum_{n}\bar{\alpha}_n^{TQ}\bar{\beta}_n^{TQ} }{\sum_{n m}\bar{\alpha}_n^{TQ} \gamma^{T}_{nm}\bar{\alpha}_n^{TQ} }\,, 
\end{equation}
where $\gamma_{nm}^T = \sum_{IJ}\langle N_{n I}^{T} \rangle C_{IJ}^{-1} \langle N_{m J}^{T} \rangle$
represents the covariance matrix
projected into modal space and
is notationally the same as
the matrix $\gamma_{nm}$
defined in Eq.~\eqref{eq:fnlB2},
except that it should be computed
using maps including the trispectrum
contribution $T$ rather than a bispectrum
contribution from $B$.
In equation \eqref{eq:gnlT2} we have defined
$\bar{\beta}_n^{TQ} \equiv \langle N_{nI}^T \rangle C_{IJ}^{-1} \hat{\mathcal{N}}_J$.

In the case of the local $\taunl$ trispectrum, using equation~\eqref{eq:taunlNJ} the estimator may be written in the form
\begin{align}\label{eq:taunlT}
\hat{\tau}_{\rm{NL}}=\frac{\sum_L C_L^{\zeta_*}\sum_{IJ}\langle N_{L I}^{T,\taunl}\rangle C_{IJ}^{-1}\mathcal{N}_J }{\sum_{LL'}\sum_{IJ}C_L^{\zeta_*} \langle N_{L I}^{T,\taunl}\rangle C_{IJ}^{-1}\langle N_{L' J}^{T,\taunl}\rangle C_{L'}^{\zeta_*}}=\frac{\sum_L C_L^{\zeta_*} D_L}{\sum_{LL'}C_L^{\zeta_*} G_{L L'}C_{L'}^{\zeta_*}}\,,
\end{align}
with the $1\sigma$ error bar is given by $1/\sqrt{3\sum_{LL'}C_L^{\zeta_*} G_{L L'}C_{L'}^{\zeta_*}}$, and
where the quantities $D_L$ and $G_{L L'}$ may be inferred from the second and third equalities.
The factor of $3$ arises due to the three instances of
$p^{l_1 l_2}_{l_3 l_4}(L)$ in the CMB trispectrum~\eqref{eq:gentrisp}.
In the application of this estimator we restrict the range of $L$ to $1\leq L\leq L_{\rm{max}} =50$. 

\section{Application to 9-year WMAP data}\label{sec:wmaptest}

In this section we apply the formalism described in
{\S\S\ref{sec:bispneedlets}--\ref{sec:trispneedlets}}
to the foreground-cleaned, coadded $V+W$
maps from the 9-year WMAP data release \cite{WMAP9}.
We work up to $\lmax = 1000$.
The data are supplied in HEALPix format with
a resolution of $6.9$ arcmin and
$N_{\text{side}} = 512$,
together with the necessary beam and noise properties
to perform realistic simulations.
In this analysis we use cosmological parameters
corresponding to those of the WMAP9
fiducial cosmology,
given in Table~\ref{table:wmap9params}.
\begin{table}[htp]
\begin{center}
	\heavyrulewidth=.08em
	\lightrulewidth=.05em
	\cmidrulewidth=.03em
	\belowrulesep=.65ex
	\belowbottomsep=0pt
	\aboverulesep=.4ex
	\abovetopsep=0pt
	\cmidrulesep=\doublerulesep
	\cmidrulekern=.5em
	\defaultaddspace=.5em
	\renewcommand{\arraystretch}{1.8}

	\begin{tabular}{SSSSSS}
		
		\toprule
	
	    \Omega_b h^2 &
	    \Omega_c h^2 &
	    \Omega_\Lambda &
	    \tau &
	    A_{\Phi} &
	    n_s \\

		0.02256 &
		0.11142 &
		0.7185 &
		0.0851 &
	    1.705 \times 10^{-8} &
		0.971 \\

		\bottomrule
	\end{tabular}
\end{center}
\caption{Parameters for the WMAP9 fiducial cosmology \cite{WMAP9}.
The optical depth is measured by $\tau$.
The parameters $A_{\Phi}$ and $n_s$
parametrize the primordial power spectrum, with
$P_{\Phi}(k) = A_{\Phi} (k / k_\star )^{n_s - 1}$.
The pivot scale $k_\star$ is chosen to be
$k_\star = 0.002 h \; \text{Mpc}^{-1}$.
\label{table:wmap9params}}
\end{table}


\para{Simulated maps.}
We simulate the Gaussian spherical harmonics amplitudes $a_{lm}^G$
with a variance given by the angular power spectrum $C_l$.
The non-Gaussian amplitudes $a_{lm}^B$ and $a_{lm}^T$
are simulated according to the prescriptions
outlined in~{\S\S\ref{sec:bispneedlets}--\ref{sec:trispneedlets}},
in particular
Eqs.~\eqref{eq:almB} and~\eqref{eq:almT-general}--\eqref{eq:almT-diag-free}.
We incorporate the effect of the WMAP beam $b_l$ and noise $n_{lm}$ for each
channel
$X \in \{ V_1, V_2, W_1, W_2, W_3, W_4 \}$
by making that transformation
\begin{equation}\label{eq:beamadd}
	a_{lm} \rightarrow \tilde{a}_{lm} = b_l a_{lm} + n_{l m} .
\end{equation}
For each data channel, $i$, we model $n_{lm}$ as white noise with variance per pixel
given by
$N_i(\hat{\bn}) \equiv \sigma_{0,i}/\sqrt{N_{i,\text{obs}}}$
where $\sigma_{0,i}$ is the sensitivity per data channel,
and $N_{i,\rm{obs}}(\hat{\bn})$ represents the corresponding number
of observations per pixel.
The simulations and data maps for each channel, $M_i(\hat{\bn})$,
are coadded optimally with inverse noise weighting per pixel, i.e.
\begin{equation}\label{eq:mapadd}
    M(\hat{\bn})
    =
    \frac{\sum_{i\in X}M_{i}(\hat{\bn})N_{i}(\hat{\bn})^{-2} }{\sum_{i\in X}N_{i}(\hat{\bn})^{-2}} ,
\end{equation}
where $X$ is the set of channels defined above~\eqref{eq:beamadd}.

We apply a suitable mask
and remove the monopole and dipole using HEALPix.
The map is then re-decomposed into spherical harmonics,
and needlet maps are evaluated by convolving with
the needlet function $\psi_{jk}$ as in~\eqref{eq:needletmap}.
For the bispectrum we choose
weight functions by setting
$\mu=1.53$
and $j = 0, 3, 4, 5, \ldots, 15, 16$
giving fifteen distinct scales.
Therefore there are ${15 + 2 \choose 3} = 680$ distinct
cubic statistics $N_J$.
For the trispectrum
it is not necessary to use all scales
because they provide little extra information,
so we thin the range
and choose
$j = 0, 4, 6, 8,9, \ldots 15$.
We will show later that this thinning does
not impair the optimality of the estimator.
These choices give $11$ distinct scales,
and therefore ${11 + 3 \choose 4} = 1001$
distinct quartic statistics.

As described in~{\S\ref{sec:bispneedlets}},
we subtract the mean from each
wavelet coefficient,
setting
$\beta_{jk} \rightarrow \beta_{jk} - \bar{\beta}_j$,
and evaluate the covariance matrix $C_{IJ}$
for the bi- and trispectrum estimators.
This requires computation of the
ensemble averages
$\langle N_I N_J \rangle$ and
$\langle N_I^T N_J^T \rangle$,
together with the corresponding one-point statistics
$\langle N_I \rangle$ and $\langle N_I^T \rangle$.
For the bispectrum estimator
we use a suite of $60,000$ simulations.
For the trispectrum estimator
we use a suite of $300,000$ simulations
because we find that more samples are required
to achieve convergence.
The $\taunl$ model requires special treatment,
and it is necessary to compute
its associated one-point statistic
$\langle N_I^{\taunl} \rangle$,
using~\eqref{eq:taunlNJ}.
In each case
we evaluate the inverse covariance matrix
$C_{IJ}^{-1}$ using principal component
analysis,
keeping only eigenvalues up to a factor
$10^{11}$ smaller than the largest eigenvalue.%
    \footnote{We have verified that our
    results are independent of the precise
    cut which is chosen.}
We evaluate the change-of-basis
matrices
$\langle N_{nI} \rangle$ and $\langle N_{nI}^T \rangle$
using $1,000$ simulations.
Finally, the trispectrum estimators~\eqref{eq:gnlT}
and~\eqref{eq:taunlT}
require the two-point expectation value
$\langle \beta^G_{j_1 k} \beta^G_{j_2 k} \rangle$,
which we obtain using $50,000$ Gaussian simulations.

At the end of this process
we are able to estimate the observables
$\fnl^B$, $\gnl^T$ and $\taunl$.
For $\taunl$ we may immediately apply Eq.~\eqref{eq:taunlT},
whereas $\fnl^B$ and $\gnl^T$
first require a suitable decomposition of the primordial
bi- and tri-spectra $B$ and $T$.
As explained in~{\S\S\ref{sec:bispneedlets}--\ref{sec:trispneedlets}},
for the bi- and tri-spectrum we fold the transfer function
$\Delta_l$ and the line-of-sight integral into
the respective transfer matrices $\Gamma_{nm}$ and $\Gamma^T_{nm}$ for computation of a 
decomposition of the CMB shapes.


\subsection{Validation procedure}
To verify that both the cubic and quartic estimators are unbiased
we apply them to
$2,000$ Gaussian simulations.
The mean of each recovered cubic and quartic coefficent,
$\bar{\beta}_n^R$ and $\bar{\beta}_n^{RT}$, 
respectively, are verified to be consistent with zero within two standard errors of the mean.
For the local bispectrum mode
and local $\gnl$-mode trispectrum we find
\begin{subequations}
\begin{align}
    \langle \fnlhat^{\text{loc}} \rangle
    & = 0.1\pm 23.1 \\
    \langle \gnlhat^{\text{loc}} \rangle
    & = [0.03\pm 2.33]\times 10^5 .
\end{align}
\end{subequations}
The error bar for $\langle \gnlhat^\text{loc} \rangle$
establishes that the procedure described in this paper
is close to optimal,
because in Ref.~\cite{SekSug2013}
the optimal error bar was shown to
be $2.2\times 10^5$.

This establishes that the estimator correctly gives zero
when applied to Gaussian maps, but
it is also necessary to check
that it recovers the correct non-Gaussian amplitude
when applied to non-Gaussian maps.
We test this using 200 simulations of the local-mode
bispectrum and $\fnllocal = 100$,
constructed using the method described by Hanson et al.~\cite{Hanson:2009kg}.
We recover the result
$\langle \fnlhat^{\text{loc}} \rangle = 100.3 \pm 0.34$,
where the error bar represents the standard error of the mean.

\begin{figure}
\centering
\vspace{0.25cm}
\hspace{0.1cm}
{
\includegraphics[width=0.44\linewidth]{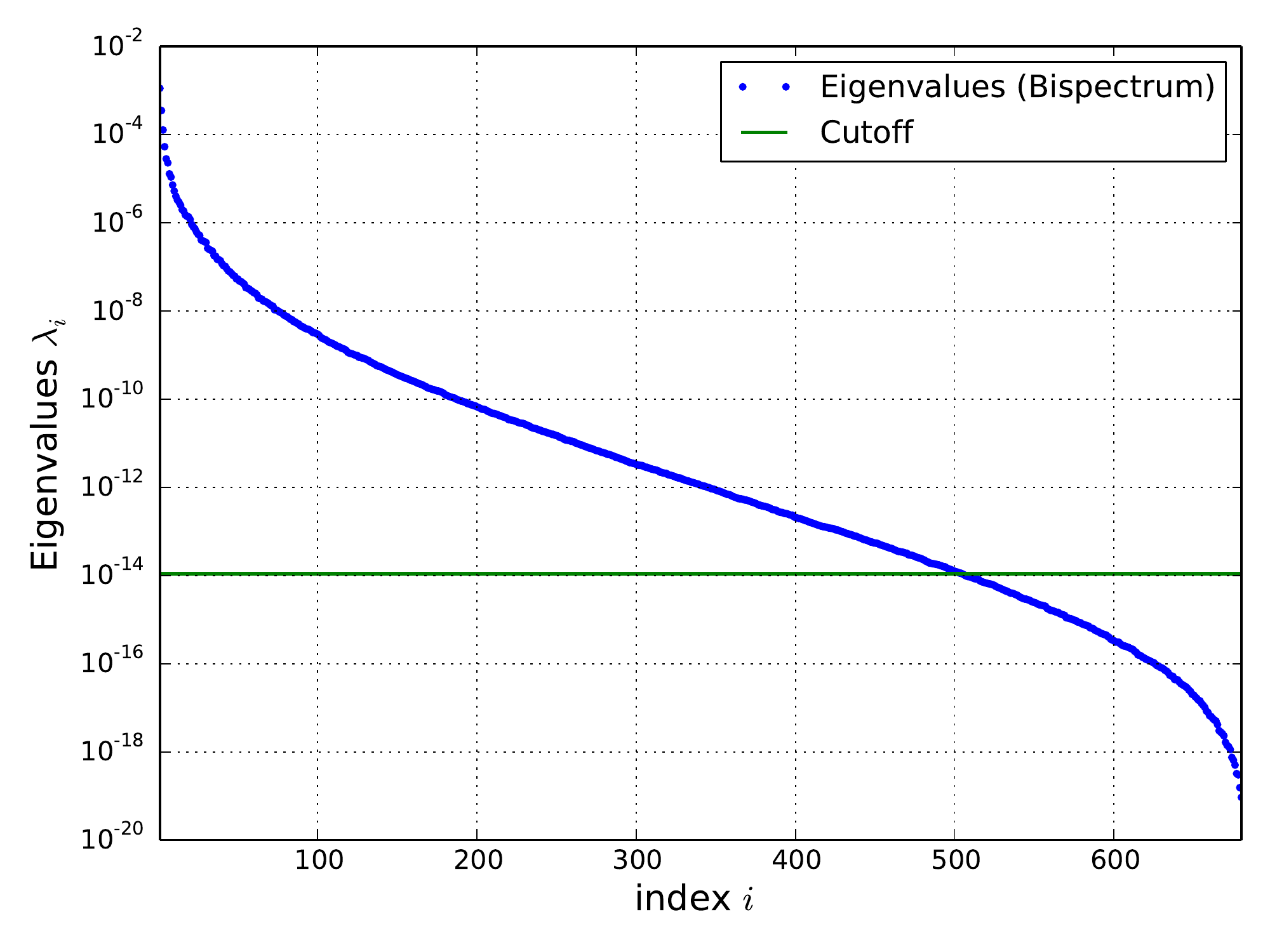}}
{
\includegraphics[width=0.44\linewidth]{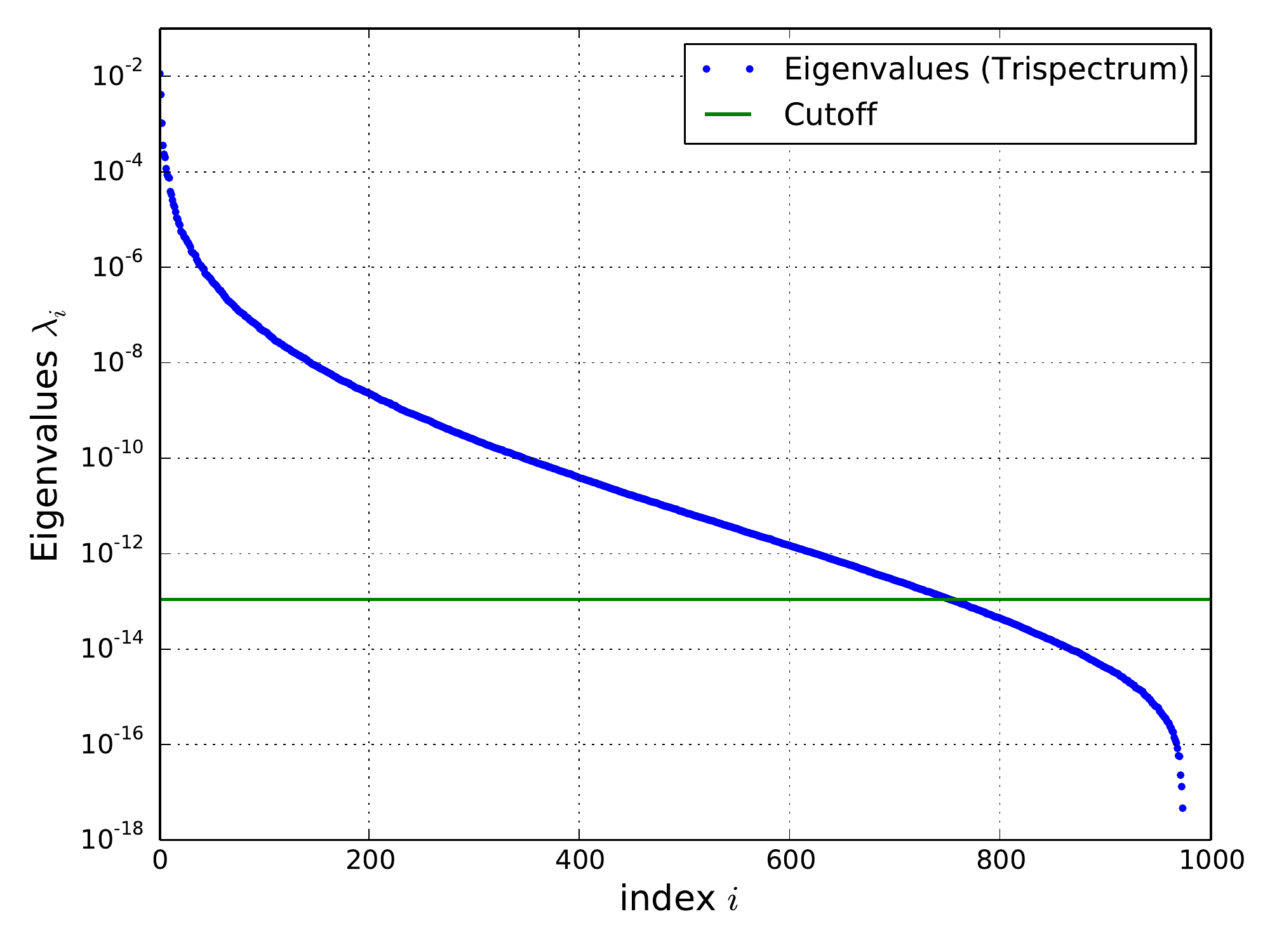}}
\caption{Plotted are the principal eigenvalues of the covariance matrix for the cubic (left) and quartic (right) needlet statistics. The cutoff value of $10^{-11}$ times the maximum eigenvalue applied in the inversion of the covariance matrix is also indicated in both cases.}
\label{fig:eigensvals}
\end{figure}

In Fig.~\ref{fig:eigensvals}
we plot the eigenvalues for both the bispectrum (cubic statistics)
and trispectrum (quartic statistics),
in order to assess the number of elements included in the analysis
after applying principal-component analysis to the covariance matrix.
The indicated cutoff value is $10^{-11}$ times the maximum eigenvalue.

\subsection{Point source Simulations}\label{subsec:ptsource}
To estimate the effect of unresolved point sources we adopt the
constant-flux model described in Ref.~\cite{WMAP5}.
We assume that
a population of sources,
each with constant flux
$\Fsrc$,
and number density $\nsrc$ per steradian,
contaminate each pixel with
a frequency-dependent temperature increment
$\delta \Tsrc(\hat{\bn}) =
\Fsrc g(x) \epsilon(\hat{\bn}) / \Omegapix$,
where $\Omegapix = 1 / \npix$ is the solid angle per pixel
and $x = h\nu / \kB \Tcmb = \nu / (56.8 \, \text{GHz})$.
The function $g(x)$ satisfies
\begin{equation}
    g(x) = \frac{\sinh^2 (x/2)}{x^4} \frac{1}{24.8 \; \text{Jy} \, \text{K}^{-1} \, \text{sr}^{-1}}
\end{equation}
and represents the conversion factor between brightness and
temperature~\cite{TegEfsth96}.
Finally, $\epsilon(\hat{\bn})$ is a Poisson-distributed
random variable with mean
$\nsrc \Omegapix$.
This Poisson distribution of sources introduces
a non-Gaussian signature.
This constant-flux model satisfactorily reproduces the
power spectrum and bispectrum of point sources
measured by the WMAP team~\cite{WMAP5,Nolta08},
given the values $\Fsrc = 0.5 \; \text{Jy}$
and a source density of $\nsrc = 85 \; \text{sr}^{-1}$.

To estimate the influence of point sources on the estimators
described above, we perform a separate set of $1,000$ simulations
in which we construct a set of Gaussian maps
using~\eqref{eq:beamadd} and~\eqref{eq:mapadd}.
These maps are
modified by adding a point-source contamination
according to the prescription above.
We compute the cubic and quartic needlet statistics
both with and without point sources.
Taking the difference for each realization
gives an estimate of the bias on
$\fnl^B$ or $\gnl^T$
for each primordial shape of interest.
We give numerical results in
Table~\ref{table:bispresults}.

\subsection{Bispectrum Constraints}\label{subsec:bispecConstraints}

In this section
we tabulate the constraints and estimated point source contamination for the
local, DBI, equilateral, constant, orthogonal and flattened bispectrum models.
Constraints for these models have previously been published
by
Fergusson, Liguori \& Shellard
based on the KSW estimator~\cite{FLS2}
and using a wavelet-based estimator in Ref.~\cite{RMS2013}.

\begin{figure}
\centering
\vspace{0.25cm}
\hspace{0.1cm}
{
\includegraphics[width=0.34\linewidth]{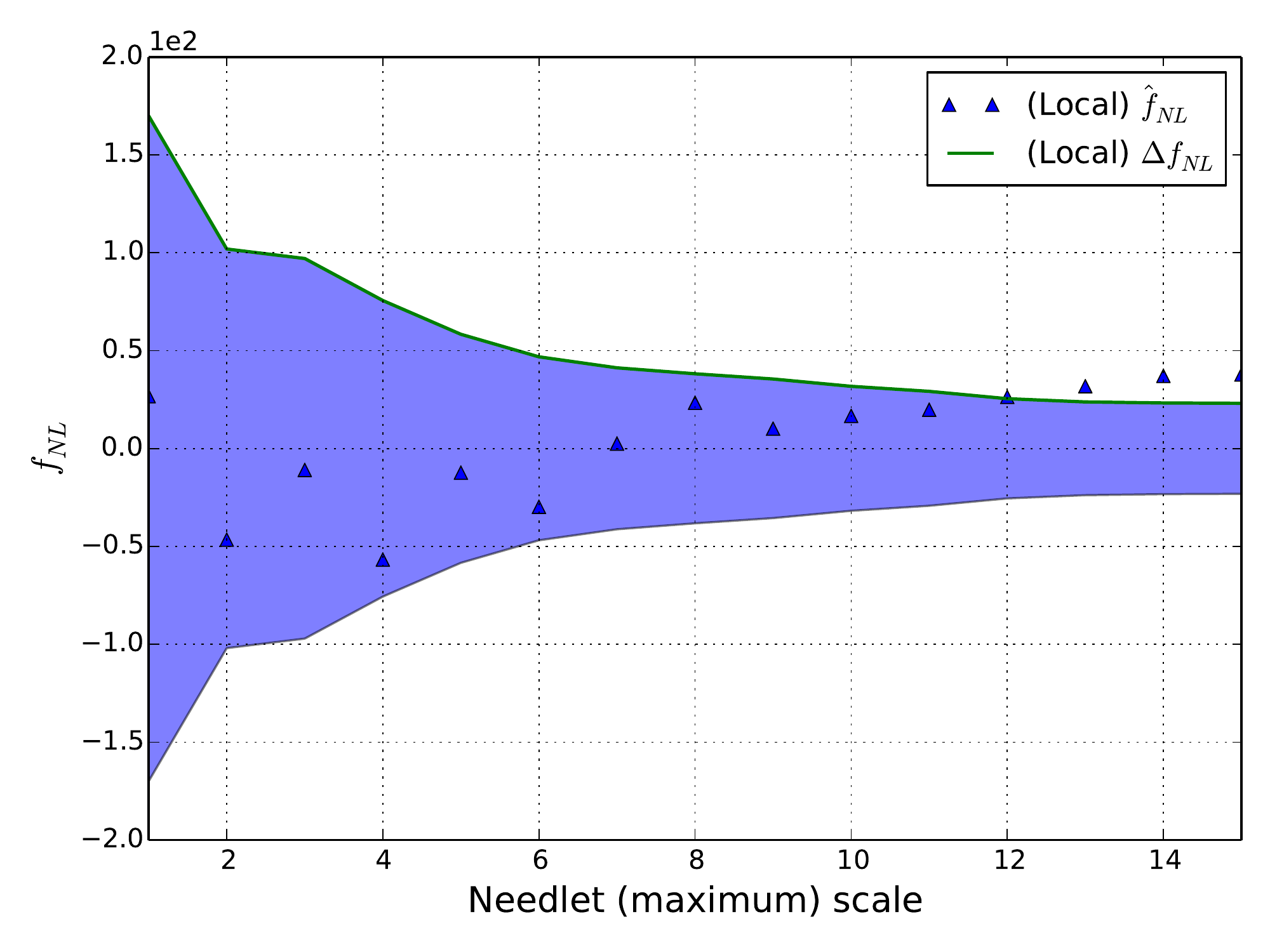}}
{
\includegraphics[width=0.34\linewidth]{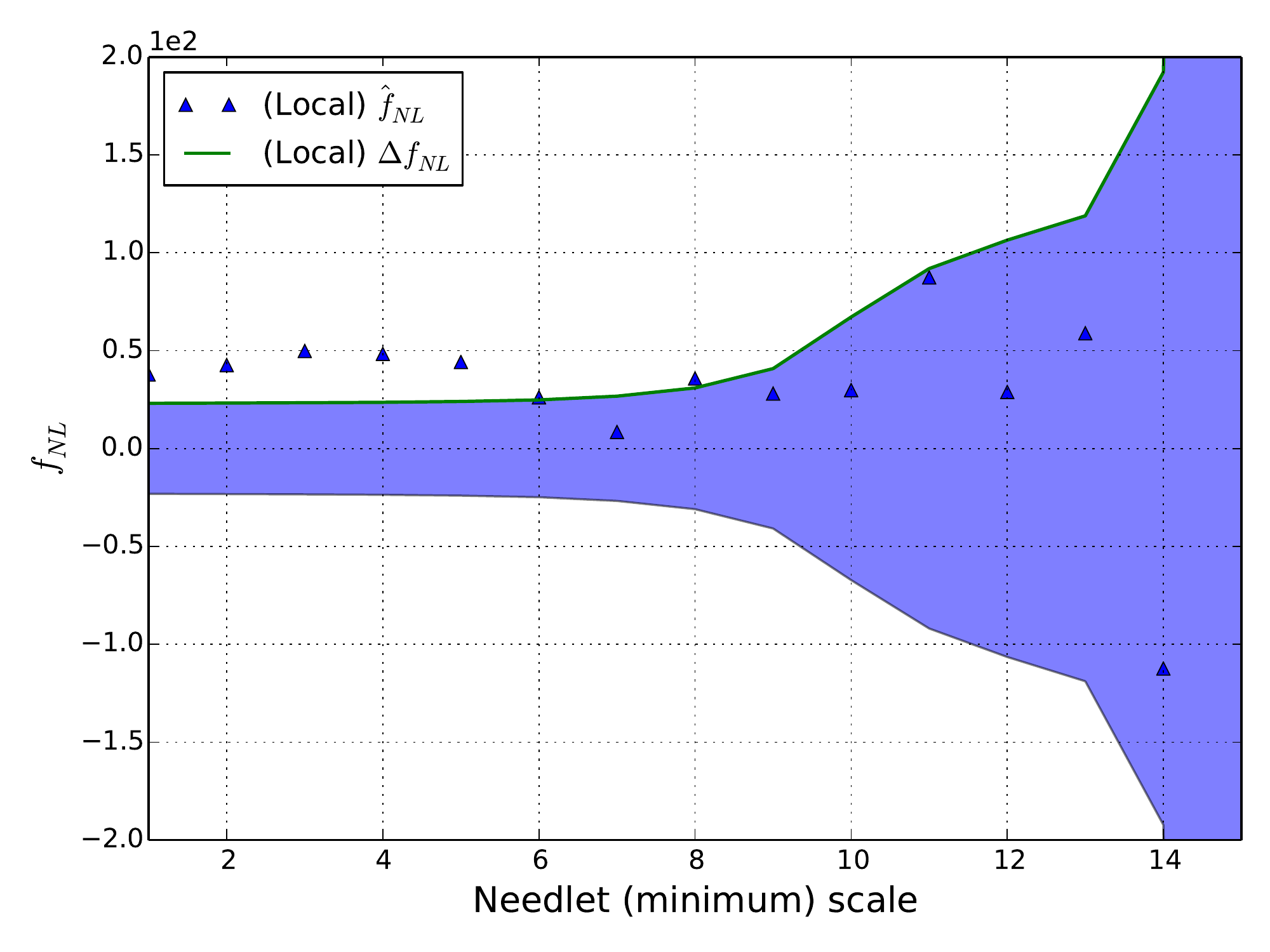}}\\
{
\includegraphics[width=0.34\linewidth]{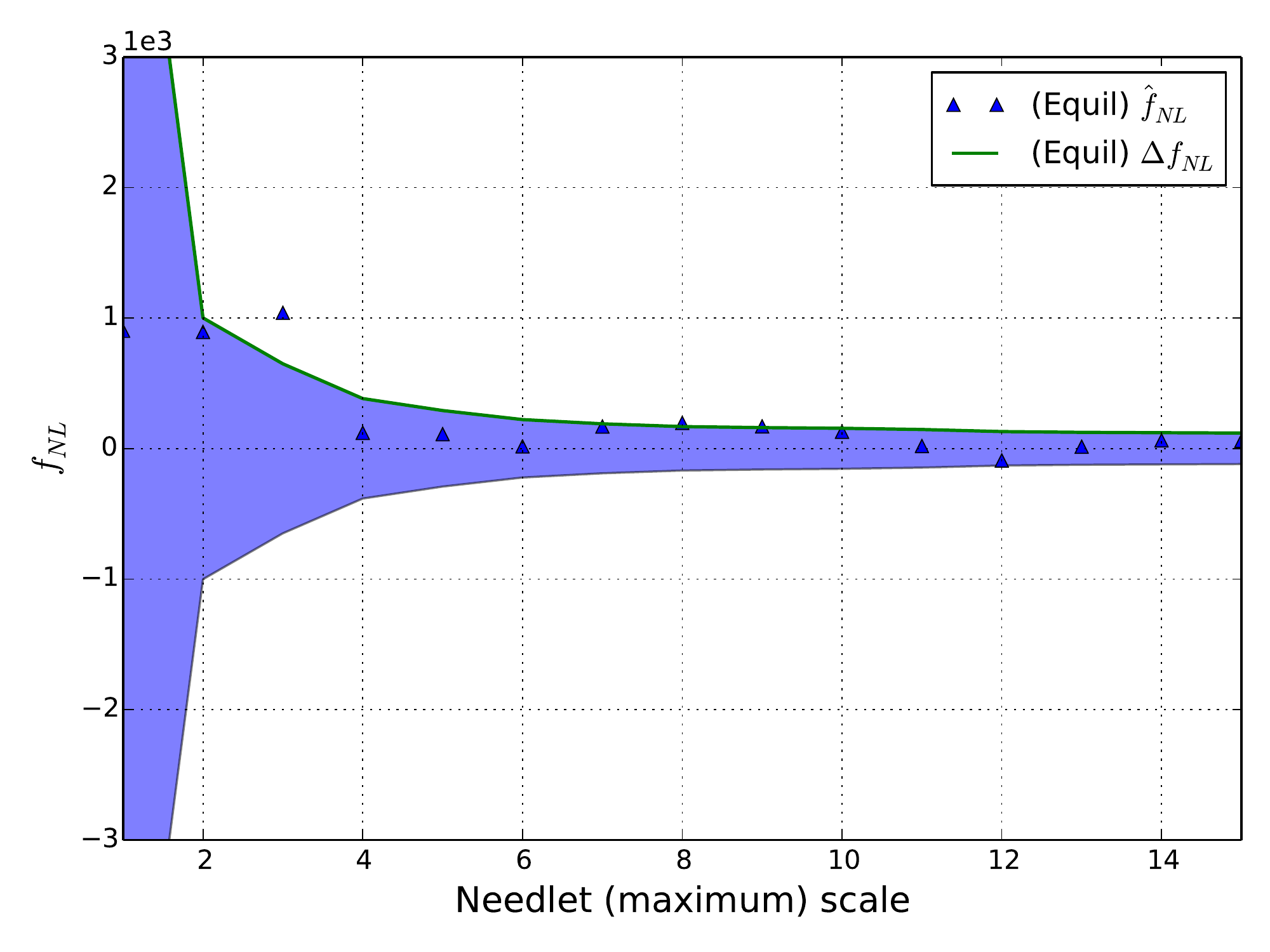}}
{
\includegraphics[width=0.34\linewidth]{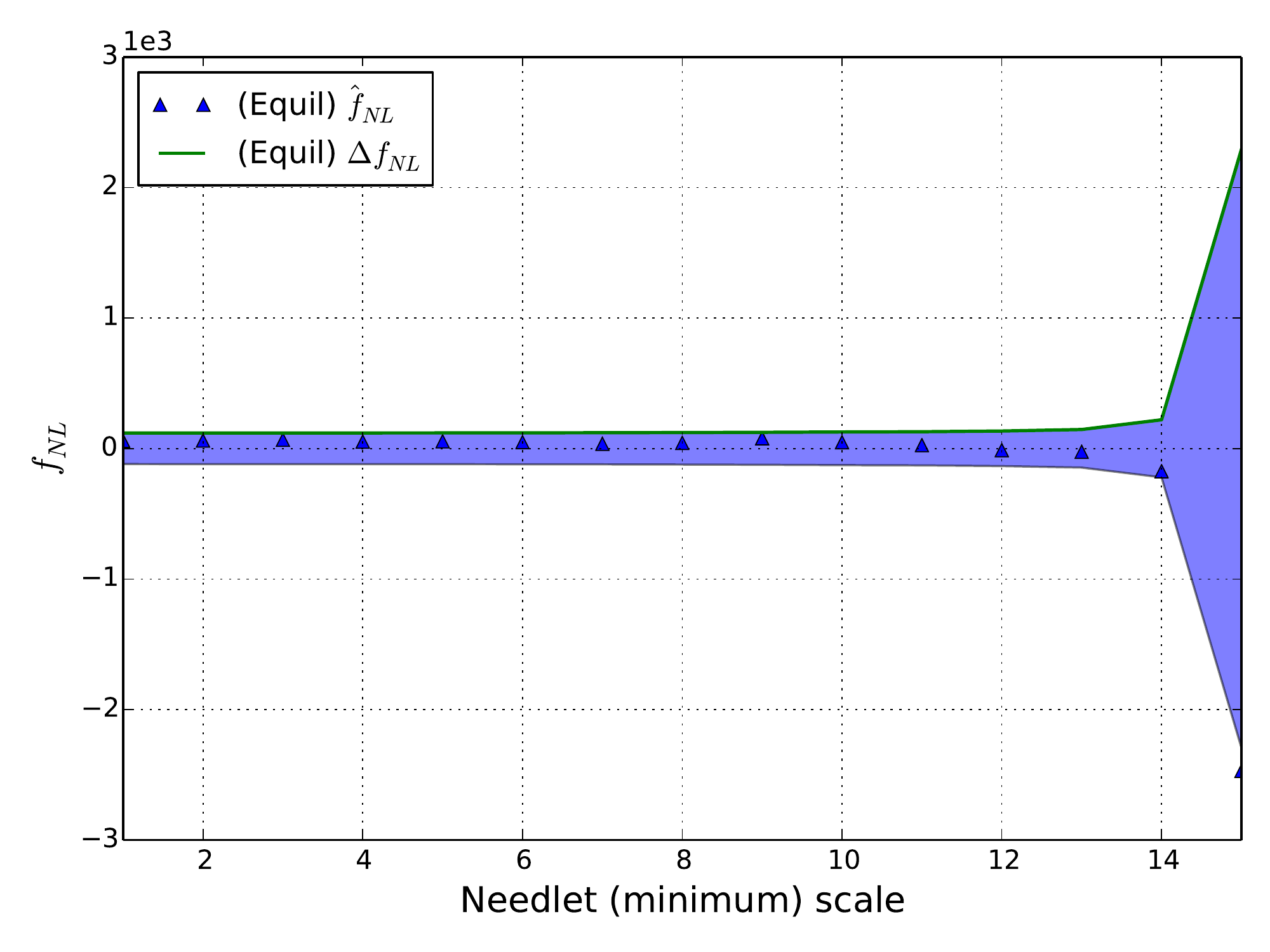}}\\
{
\includegraphics[width=0.34\linewidth]{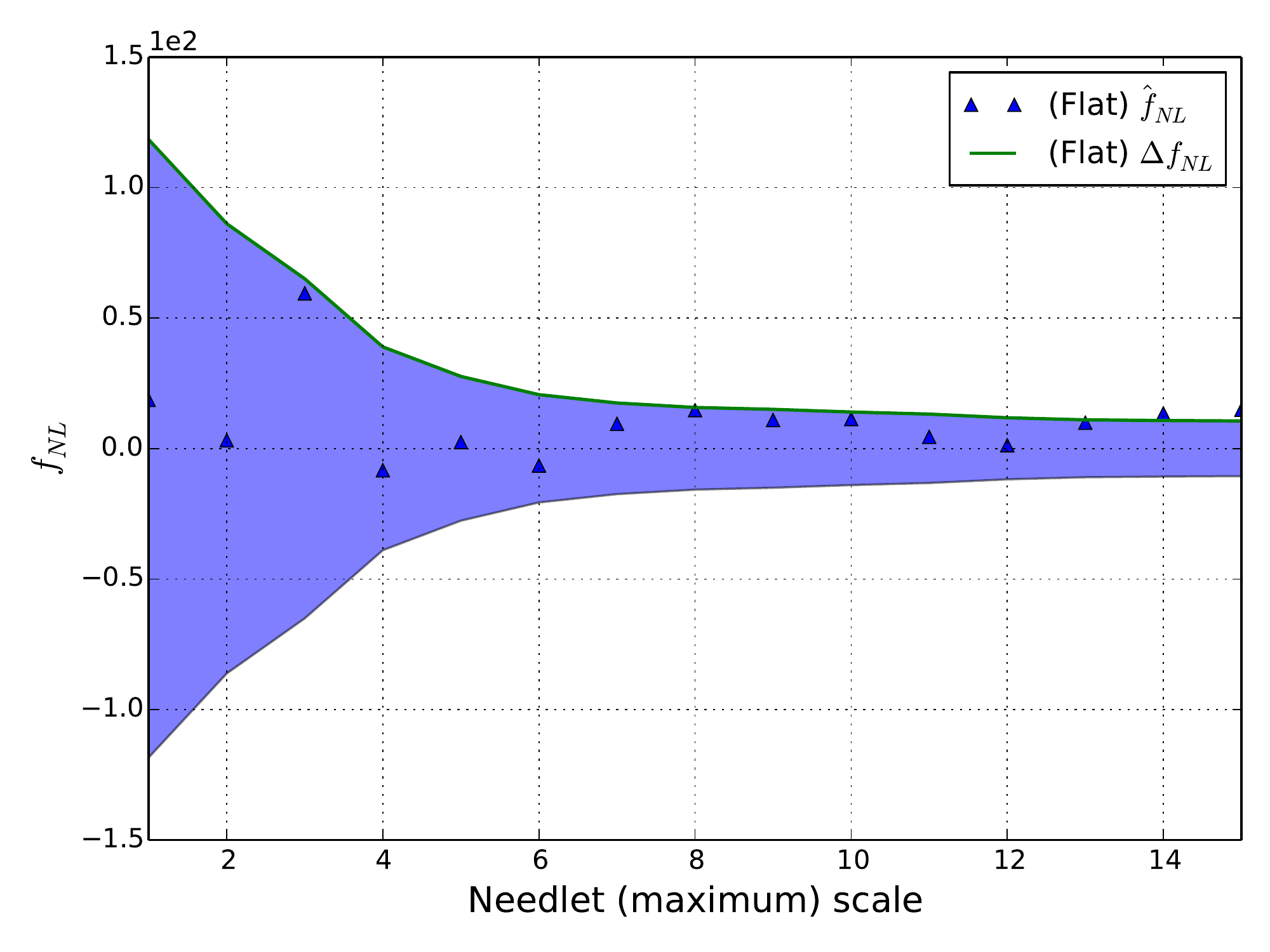}}
{
\includegraphics[width=0.34\linewidth]{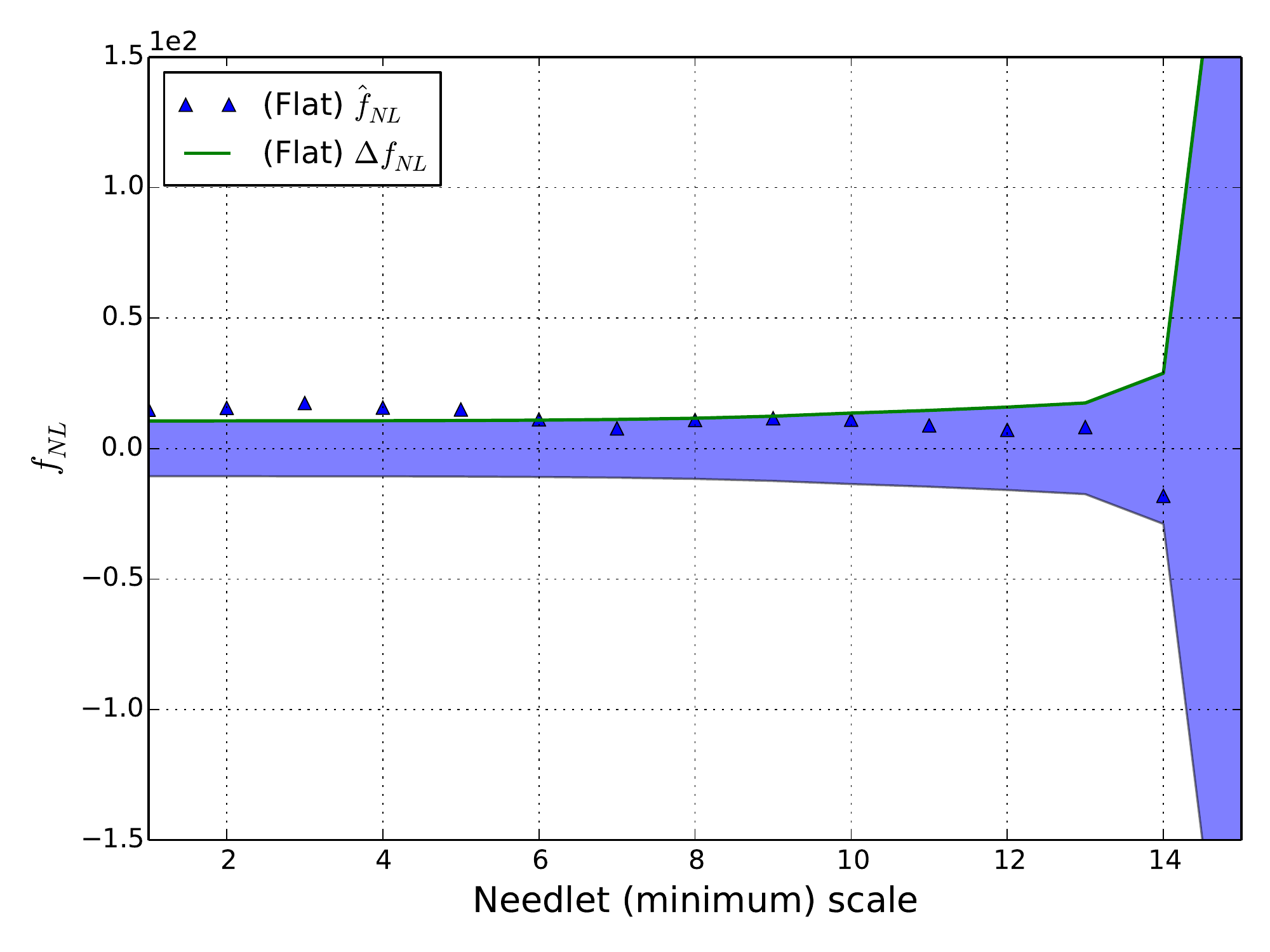}}
\caption{Plotted on the left column is the dependency of the best fit value - and error bar - of $\fnl$ on the maximum needlet scale. The triangle glyph represents the best fit value, while shading is used to represent the region $\fnl\in [-\Delta \fnl, \Delta \fnl]$. Represented are the constraints using all measured needlet data up to the maximum scale indicated. On the right hand column we represent the dependency on the minimum needlet scale, with all needlet data used beyond this scale. From top to bottom we plot these quantities for the local, equilateral and flattened models respectively.}
\label{fig:NeedletScale}
\end{figure}

\begin{itemize}
\item \textbf{Local model.}
A Taylor expansion around a Gaussian gravitational potential
$\phi_G$
defines both the local-mode bispectrum and local $\gnl$-mode trispectrum.
It accurately represents the type of non-Gaussianity generated
by evolution on superhorizon scales
in multiple-field inflationary models~\cite{LythRod2005,Seery:2006js}
or the curvaton model~\cite{EnqvistSloth2001,LythWands2001,Moroi2001}.
We write~\cite{Salopek}
\begin{equation}\label{eq:localmodel}
	\Phi(\bx)=\Phi_G(\bx)+\fnllocal(\Phi_G(\bx)^2-\langle \Phi_G(\bx)^2\rangle)+\gnllocal\Phi_G(\bx)^3+\dots\,.
\end{equation}
The resulting primordial bispectrum is given by
\begin{equation}
    B_{\Phi}(k_1,k_2,k_3)=2\fnllocal (P_{\Phi}(k_1) P_{\Phi}(k_2)
		+ P_{\Phi}(k_1) P_{\Phi}(k_3)
		+ P_{\Phi}(k_2) P_{\Phi}(k_3)) .
\end{equation}
Our constraints on $\fnllocal$ and the bias
$\Delta\fnllocal$
due to point
sources are
\begin{equation}\label{eq:local-bisp-result}
	\fnllocal = 38.6\pm 23.1
	\quad \text{and} \quad
	\Delta \fnllocal = 9.6\pm 4.1.
\end{equation}
This result is consistent with the needlet-based constraint
$\fnl^{\text{loc}} =37.5\pm 21.8$
reported by
Donzelli et al. \cite{Donzelli2012},
and with the wavelet-based
constraint $\fnl^{\text{loc}} =38.4\pm 23.6$
reported in
Ref.~\cite{RMS2013}.

The lower error bar in the case of
Donzelli et al.~\cite{Donzelli2012} may be partly attributed to the
larger number of needlet scales used in that work, as well as the use of a linear term correction. We have chosen to ignore the linear correction term in this work, because Refs.~\cite{Curto2011,RMS2013} demonstrated (in the case of wavelets) that it leads to only a $\lesssim 2\%$ correction; the mean scale subtraction largely accounting for anisotropies due to the mask and noise. In addition the remaining bispectrum models achieve optimality and therefore we persist with only $15$ needlet scales.
However, we note that to achieve optimality in the case of the local model, we
may require more scales. While this issue is somewhat parenthetic to the aims of this paper, we present an investigation of the dependency of the results on the maximum and minimum needlet scale in Fig.~\ref{fig:NeedletScale}.

The estimate for the bias due to point sources
given in~\eqref{eq:local-bisp-result}
for the contribution of point sources is
larger than predicted using wavelets,
which give $\Delta \fnllocal = 3.1\pm 3.7$~\cite{RMS2013}.

\item \textbf{DBI and equilateral models.}
Under certain circumstances, non-standard kinetic
terms may lead to strong self-interactions between
modes as they leave the horizon~\cite{AlishahihaSilversteinTong2004,ChenetAl2007,0605045}.
An example is DBI inflation, for which the equilateral model provides
an accurate separable approximation.
The bispectra are
\begin{align}
	B_{\Phi}^{\text{DBI}}
	& =
	\frac{1}{(k_1 k_2 k_3)^3 (\sum_i k_i)^2}
	\bigg(
		\sum_i k_i^5
		+ \sum_{i\neq j}(2 k_i^4 k_j-3 k_i^3 k_j^2)
		+ \sum_{i\neq j\neq l}(k_i^3 k_j k_l - 4 k_i^2 k_j^2 k_l)
	\bigg) ,
	\\
	B_{\Phi}^{\text{eq}}
	& =
	6\bigg(
		- \Big[ P_{\Phi}(k_1)P_{\Phi}(k_2)+\text{2 perms} \Big]
		- 2 \Big[ P_{\Phi}(k_1)P_{\Phi}(k_2)P_{\Phi}(k_3) \Big]^{2/3}
		+ \Big[
			P_{\Phi}^{1/3}(k_1)P_{\Phi}^{2/3}(k_2)P_{\Phi}(k_3) + \text{5 perms}
		\Big]
	\bigg) .
\end{align}
The needlet-based estimator
gives
\begin{align}
	\fnl^{\text{DBI}}  &= 65.5\pm 99.4
	&
	\Delta\fnl^{\text{DBI}} & = 9.5\pm 14.1\,,\nonumber
	\\
	\fnl^{\text{eq}}  &= 64.5\pm 117.3
	&
	\Delta\fnl^{\text{eq}} & = 11.0\pm 19.3\,.
\end{align}

\item \textbf{Constant model.} The constant model
gives a primordial bispectrum
corresponding to
\begin{equation}
    B^{\text{const}}_{\Phi}(k_1,k_2,k_3)=6(P_{\Phi}(k_1)P_{\Phi}(k_2) P_{\Phi}(k_3))^{2/3} .
\end{equation}
The CMB bispectrum is entirely due to the transfer function;
see Ref.~\cite{ChenWang2009} for a possible microphysical realization.
The needlet estimator gives
\begin{equation}
	\fnl^{\text{const}}= 60.8\pm 62.7
	\quad\qquad
	\Delta\fnl^{\text{const}}= 9.3\pm 9.9.
\end{equation}
\item \textbf{Orthogonal model.} The orthogonal model is given by
a linear combination of the equilateral and constant models,
$B^{\rm{orthog}}_{\Phi}=3B^{\rm{eq}}_{\Phi}-2B^{\rm{const}}_{\Phi}$.
We obtain
\begin{equation}
    \label{eq:fnl-orthog}
	\fnl^{\text{orthog}}=-175.0\pm 101.8 
	\quad\qquad
	\Delta\fnl^{\text{orthog}}= -24.3\pm 16.7.
\end{equation}
For comparison, the constraint from 7-year WMAP data using the wavelet-based
estimator of Ref.~\cite{RMS2013} was
$\fnl^{\text{orthog}} = -173.2 \pm 101.4$.
By comparison, the WMAP team report
$\fnl^{\text{orthog}} = -245 \pm 100$ from the 9-year data \cite{WMAP9}.
The constraint given in~\eqref{eq:fnl-orthog} uses 9-year data
and is consistent with the 7-year result.
It is substantially less significant than the result obtained by the
WMAP team. Below, we investigate this further by providing
a frequency-band analysis.

\item \textbf{Flattened model.}
A `flattened' configuration may be produced by a nontrivial initial state,
including a non-Bunch--Davies vacuum.
For such states, long-lived excitations with nearly zero energy
can be
formed by a combination of positive- and negative-energy modes.
These generate strong correlations.
Physical models realizing this effect are discussed, for example,
in Refs.~\cite{HolmanTolley2008,Ashoorioon}.
The bispectrum is
\begin{equation}
	B_{\Phi}^{\text{flat}}
	=
	\frac{6}{39(k_1 k_2 k_3)^2}
	\Bigg[
		\left(
			\frac{k_1^2+k_2^2-k_3^2}{k_2 k_3} + \text{2 perms}
		\right)
		+ 12
		+ 8 \left(
			\frac{k_1 k_2+ k_1 k_3 -k_2 k_3}{(k_2+k_3-k_1)^2} + \text{2 perms}
		\right)
	\Bigg]\,.
	\label{eq:flattened-bispectrum}
\end{equation}
In order to handle the divergence we set the bispectrum to zero for $k_1 + k_2 - k_3 < Z \sum_i k_i$
(or its permutations) with $Z=0.03$ and employ a low pass (Gaussian) filter
in order to smoothen the shape near the edges as in Ref.~\cite{FLS1}.
The resulting constraints are
\begin{equation}
	\fnl^{\text{flat}} = 14.9\pm 10.5
	\qquad
	\Delta \fnl^{\text{flat}} = 2.5\pm 1.7 .
\end{equation}
\end{itemize}
\begin{table}[htp]
\begin{center}
	\heavyrulewidth=.08em
	\lightrulewidth=.05em
	\cmidrulewidth=.03em
	\belowrulesep=.65ex
	\belowbottomsep=0pt
	\aboverulesep=.4ex
	\abovetopsep=0pt
	\cmidrulesep=\doublerulesep
	\cmidrulekern=.5em
	\defaultaddspace=.5em
	\renewcommand{\arraystretch}{1.8}

	\begin{tabular}{cSSSSSS}
		
		\toprule

        &&&& \multicolumn{3}{c}{Point source contamination} \\
        \cmidrule{5-7}
		\rowcolor[gray]{0.9} \textbf{Shape} & \multicolumn{1}{c}{\textbf{V-band}}  & \multicolumn{1}{c}{\textbf{W-band}} & \multicolumn{1}{c}{\textbf{V+W}}& \multicolumn{1}{c}{\textbf{V-band}}  & \multicolumn{1}{c}{\textbf{W-band}}& \multicolumn{1}{c}{\textbf{V+W}}\\
		Local & 44.0\pm 26.2& 30.2\pm 26.6& 38.6\pm23.1& 12.0\pm 5.1& 1.6\pm 2.4& \,\,\, 9.6\pm 4.1\\
		\rowcolor[gray]{0.9} DBI & 53.3\pm 106.4& 86.4\pm 107.2& 65.5\pm99.4&31.9 \pm 18.5& 3.5\pm 9.1& \,\,\, 9.5\pm 16.1\\
		Equilateral & 54.2\pm 125.6& 88.6\pm 126.7&64.5 \pm 117.3& 38.3\pm 22.2& 4.4\pm 10.9& \,\,\, 11.0\pm 19.3\\
		\rowcolor[gray]{0.9} Constant & 54.1\pm 67.5& 85.1\pm 67.6&60.8 \pm 62.7& 22.8\pm 11.6& 2.6\pm5.7& \,\,\, 9.3\pm 9.9\\
		Orthogonal & -154.7\pm 114.9& -182.8\pm 114.0& -175.0\pm 101.8& -36.0\pm 20.1&-3.9 \pm 9.6& \,\,\, -24.3\pm 16.7\\
		\rowcolor[gray]{0.9} Flat &  13.1\pm 11.6& 15.6\pm 11.6& 14.9\pm 10.5& 4.7\pm 2.1& 0.6\pm 1.0& \,\,\, 2.5\pm 1.7\\

		\bottomrule
	\end{tabular}
\end{center}
\caption{Constraints on the various bispectrum shapes computed for,
respectively, the  V-band, W-band and coadded data.
The last three columns give estimates for the point source
contamination in each case.
\label{table:bispresults}}
\end{table}
\para{Frequency dependence.}
Instead of using the entire dataset, it is possible to obtain constraints using only V- or W-band data
and corresponding simulations of the maps and covariance matrix.
The bias due to point sources can be taken into account as described above.
We tabulate our results in Tables~\ref{table:bispresults} and~\ref{table:bispresults2}.

We find that point-source contamination in the V-band is more significant than in the W-band.
In comparison to alternative estimators, such as Spherical Mexican Hat wavelets,
needlets show more sensitivity to point-source contamination of the local shape.
The frequency-band analysis shows that the V- and W-band constraints are consistent
for each model (within 1-$\sigma$), including for the orthogonal shape
for which the WMAP team obtained discrepant results from the 9-year data \cite{WMAP9}
($\fnl^{\text{orthog}} = -245.5 \pm 99.6$ from the coadded map,
$\fnl^{\text{orthog}} = -125.9 \pm 112.7$ from V-band only, and
$\fnl^{\text{orthog}} = -320.2 \pm 112.1$ from W-band only).
The needlet-based analysis given here produces much weaker frequency dependence.


\begin{table}[htp]
\begin{center}
	\heavyrulewidth=.08em
	\lightrulewidth=.05em
	\cmidrulewidth=.03em
	\belowrulesep=.65ex
	\belowbottomsep=0pt
	\aboverulesep=.4ex
	\abovetopsep=0pt
	\cmidrulesep=\doublerulesep
	\cmidrulekern=.5em
	\defaultaddspace=.5em
	\renewcommand{\arraystretch}{1.8}

	\begin{tabular}{cSSS}
		
		\toprule
	
		\rowcolor[gray]{0.9} \textbf{Shape} & \multicolumn{1}{c}{\textbf{V-band}}  & \multicolumn{1}{c}{\textbf{W-band}}& \multicolumn{1}{c}{\textbf{V+W}}\\
		Local &  32.0\pm 26.2& 28.6\pm 26.6& 29.0\pm23.1\\
		\rowcolor[gray]{0.9} DBI &  21.4\pm 106.4& 82.9\pm 107.2& 56.0\pm99.4\\
		Equilateral &  15.9\pm 125.6& 84.2\pm 126.7&53.5 \pm 117.3\\
		\rowcolor[gray]{0.9} Constant &  31.3\pm 67.5& 82.5\pm 67.6&51.5 \pm 62.7\\
		Orthogonal & -118.7\pm 114.9& -178.9\pm 114.0& -150.7\pm 101.8\\
		\rowcolor[gray]{0.9} Flat & 8.4\pm 11.6& 15.0\pm 11.6& 12.4\pm 10.5\\

		\bottomrule
	\end{tabular}
\end{center}
\caption{Constraints on bispectrum models corrected for the effect of point sources (see Table~\ref{table:bispresults}). \label{table:bispresults2}}
\end{table}

\para{Point Source Model Investigation.} One may be
concerned that the constant-flux point source model described in~\S\ref{subsec:ptsource} is too simplistic. Therefore, we also implement a more realistic point source model which provides a better match to observations at each flux value. We use the analytic fit \cite{Argueso:2006gu} to the de Zotti et al. \cite{DeZotti:2004mn} observations, with the proper distribution of number counts $dn/dS=A S^{-\alpha}$, where $A=22.1\pm 1.5$ and $\alpha=2.32\pm 0.06$ with the best fit value chosen for our simulations. Extending the constant-flux model we integrate over fluxes using this model with $S\in [10^{-3},1]\,{\rm Jy}$. In Table~\ref{table:pointsource2} we list the corresponding estimates for the contamination due to point sources. Comparison to Table~\ref{table:bispresults} reveals the consistency of the results obtained using both point source models.  In the context of searches for primordial non-Gaussianity beyond Planck, more detailed source modelling may be necessary. Indeed, the extra intensity of the point source contamination in the V-band suggests the impact of radio-sources.\footnote{We thank an anonymous referee for drawing our attention to this detail.} A study of different families of point sources was performed in Ref.~\cite{Curto:2013hi} detailing the potentially strong non-Gaussian deviations due to unresolved point sources for both high $(>225 {\rm GHz})$ and low $(<100 {\rm GHz})$ frequency data.

\begin{table}[htp]
\begin{center}
	\heavyrulewidth=.08em
	\lightrulewidth=.05em
	\cmidrulewidth=.03em
	\belowrulesep=.65ex
	\belowbottomsep=0pt
	\aboverulesep=.4ex
	\abovetopsep=0pt
	\cmidrulesep=\doublerulesep
	\cmidrulekern=.5em
	\defaultaddspace=.5em
	\renewcommand{\arraystretch}{1.8}

	\begin{tabular}{cSSS}
		
		\toprule
	
		\rowcolor[gray]{0.9} \textbf{Shape} & \multicolumn{1}{c}{\textbf{V-band}}  & \multicolumn{1}{c}{\textbf{W-band}}& \multicolumn{1}{c}{\textbf{V+W}}\\
		Local &  15.3\pm 5.3& 2.2\pm 7.1& 10.8\pm4.8\\
		\rowcolor[gray]{0.9} DBI &  37.4\pm 23.7& 6.6\pm 13.7& 7.5\pm11.1\\
		Equilateral &  44.4\pm 27.2& 8.0\pm 15.6&7.9 \pm 17.0\\
		\rowcolor[gray]{0.9} Constant &  27.4\pm 13.2& 4.6\pm 10.1&8.9 \pm 8.9\\
		Orthogonal & -47.3\pm 23.3& -6.8\pm 22.6& -29.2\pm 18.6\\
		\rowcolor[gray]{0.9} Flat & 5.8\pm 1.9& 0.9\pm 2.1& 2.7\pm 1.6\\

		\bottomrule
	\end{tabular}
\end{center}
\caption{Estimates for the point source contamination for each bispectrum model considered, for the V-, W- and coadded data, using the more realistic point source model. Comparison to Table~\ref{table:bispresults} reveals the consistency with the results obtained using the more simplistic constant flux model.\label{table:pointsource2}}
\end{table}

\para{Needlet scale dependence.} 
In order to assess the dependence of our results on the needlet scales, in Fig.~\ref{fig:NeedletScale} we present (in the left hand column) a plot of the best-fit value and the error bar of $\fnl$, as we include more needlet scales, up to the maximum scale used in our analysis. On the right hand column we plot the corresponding quantities as we increase the minimum needlet scale. For each case it is necessary to recompute the inverse covariance matrix for the needlet scales under consideration. We present the plots for the local, equilateral and flattened models, noting that the other models considered in this work may be represented as linear combinations of these. The equilateral and flattened models are largely insensitive to the maximum and minimum scale chosen, supporting the observation that we have achieved optimal error bars for each. In the case of the local model there is the possibility that the error bars may shrink slightly with an increased number of needlet scales, as detailed earlier in the section. Nevertheless, the results show the robustness of our results to the choice of scales used.

\subsection{Trispectrum Constraints}\label{subsec:trispecConstraints}

By comparison with the bispectrum,
obtaining constraints on the CMB trispectrum is numerically challenging.
Here we briefly review constraints which have appeared in the literature.
Desjacques and Seljak found the constraint
$\gnlhat^{\text{loc}} = [2.35 \pm 2.93] \times 10^5$ using the
scale-dependent bias of dark matter haloes in the local model~\cite{DS2010}.
Smidt et al. used a pseudo-$C_l$ estimator to obtain
$\gnlhat^{\text{loc}} = [0.4 \pm 3.9] \times 10^5$~\cite{Cooray2}.
Using a modal decomposition and a suboptimal trispectrum estimator,
working up to $\lmax=500$,
Regan et al. found
$\gnlhat^{\text{loc}} = [1.1 \pm 4.9] \times 10^5$~\cite{RSF1,FRS2}.%
    \footnote{In these papers, an extra factor of $\fsky$ accounting for the
    sky fraction was erroneously included.}
Recently, Sekiguchi \& Sugiyama, using $\lmax=1024$, established that the optimal error
bar for the local $\gnl$-mode is $2.2 \times 10^5$,
finding $\gnlhat^{\text{loc}} = [-3.3 \pm 2.2] \times 10^5$.
Their analysis implemented the optimal estimator developed in
Ref.~\cite{RSF1}, using the full pixel-by-pixel inverse covariance matrix.
In this paper we work up to $\lmax=1000$, with the pixel-based estimator replaced by a needlet
estimator.
This has the advantage that,
instead of inverting an
$\sim 10^6 \times 10^6$ matrix representing the pixel-by-pixel covariance,
we need only invert a
$\sim 10^3 \times 10^3$ matrix representing the covariance between needlets.
Nevertheless, this only results in a small loss of optimality.


\begin{itemize}
\item \textbf{Local $\gnl$ model.}
The local model was discussed in~{\S\ref{subsec:bispecConstraints}}.
It gives a trispectrum of the form~\eqref{eq:trispgnl}.
We find
\begin{equation}
	\gnl^{\text{loc}}= [-2.3\pm 2.3]\times 10^5
	\quad \text{and} \quad
	\Delta\gnl^{\text{loc}}= [1.8\pm 2.2]\times 10^5
\end{equation} 
where (estimated here for the first time)
$\Delta\gnl^{\text{loc}}$ represents the bias due to point sources.
Correcting for this bias, the constraint
$\gnl^{\text{loca}} = [-4.1 \pm 2.3] \times 10^5$ is within
$\lesssim 5\%$ of the optimal bound and is
consistent with the result reported by Sekiguchi \& Sugiyama~\cite{SekSug2013}.
This represents strong evidence in favour of the accuracy
and efficacy of the needlet-based estimator.

The point source constraints reported are calculated using the more accurate model described in the previous section. Using the simpler constant flux model, the point source constraint is calculated to be $\Delta\gnl^{\text{loc}}= [0.9\pm 2.2]\times 10^5$. Thus it appears that the constant flux model underestimates the bias due to point sources. However, as we shall see this simple model works well for the other trispectrum models considered, and we will not pursue this interesting issue further in this paper. It is worth noting, however, that the trispectrum constraints appear to show stronger dependency on the presence of point sources than those of the bispectrum. We also consider the dependency of the results on the maximum needlet scale considered. In Fig.~\ref{fig:gnlErrors} the best-fit and $1\sigma$ error bars are plotted as a function of the maximum needlet scale chosen, up until the maximum used in our data analysis. The results show the convergence to the reported values, indicating the robustness of the reported values on the needlet scales chosen.

\begin{figure}
\centering
\vspace{0.25cm}
\hspace{0.1cm}
{
\includegraphics[width=0.3\linewidth]{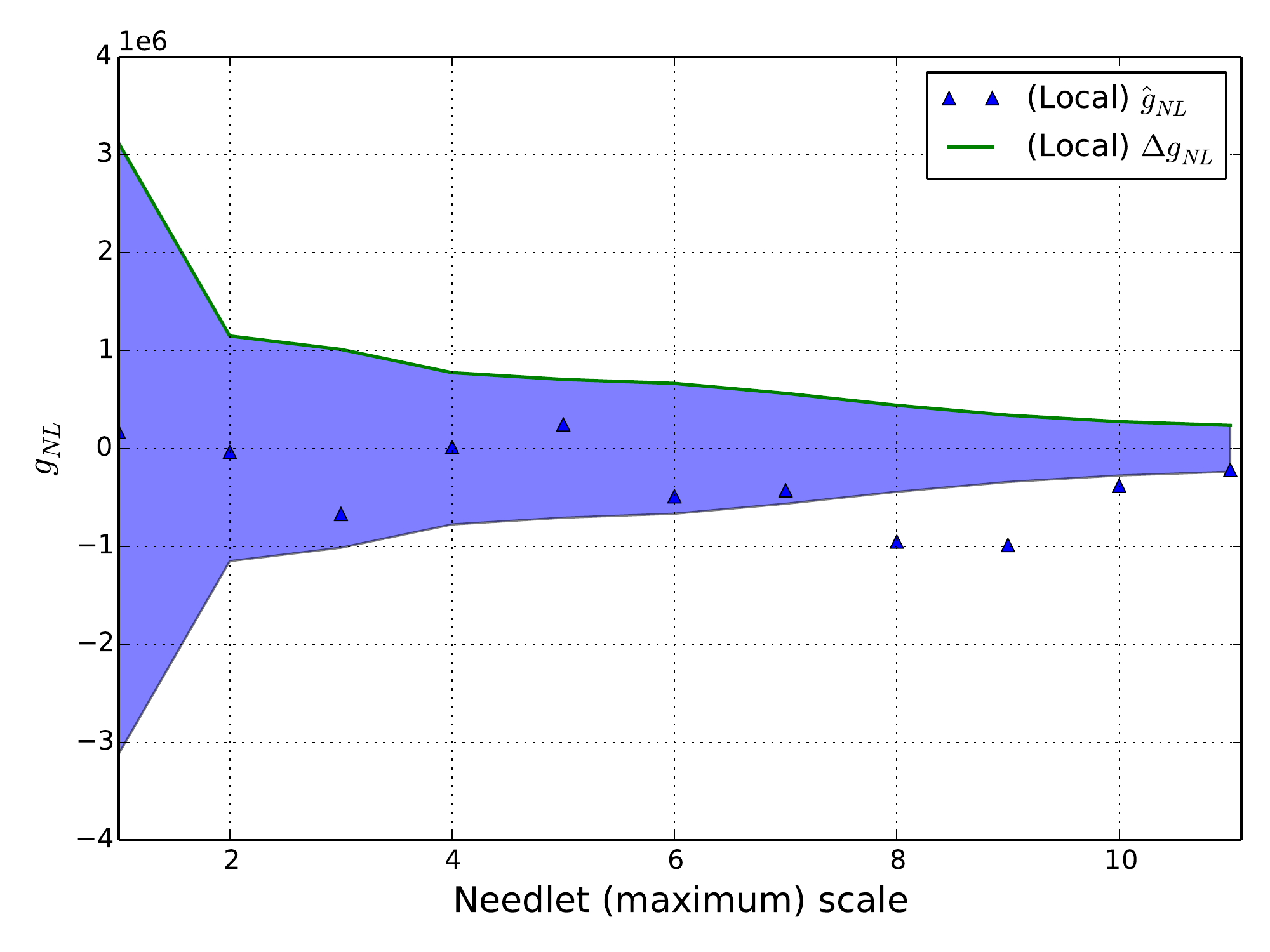}}
{
\includegraphics[width=0.3\linewidth]{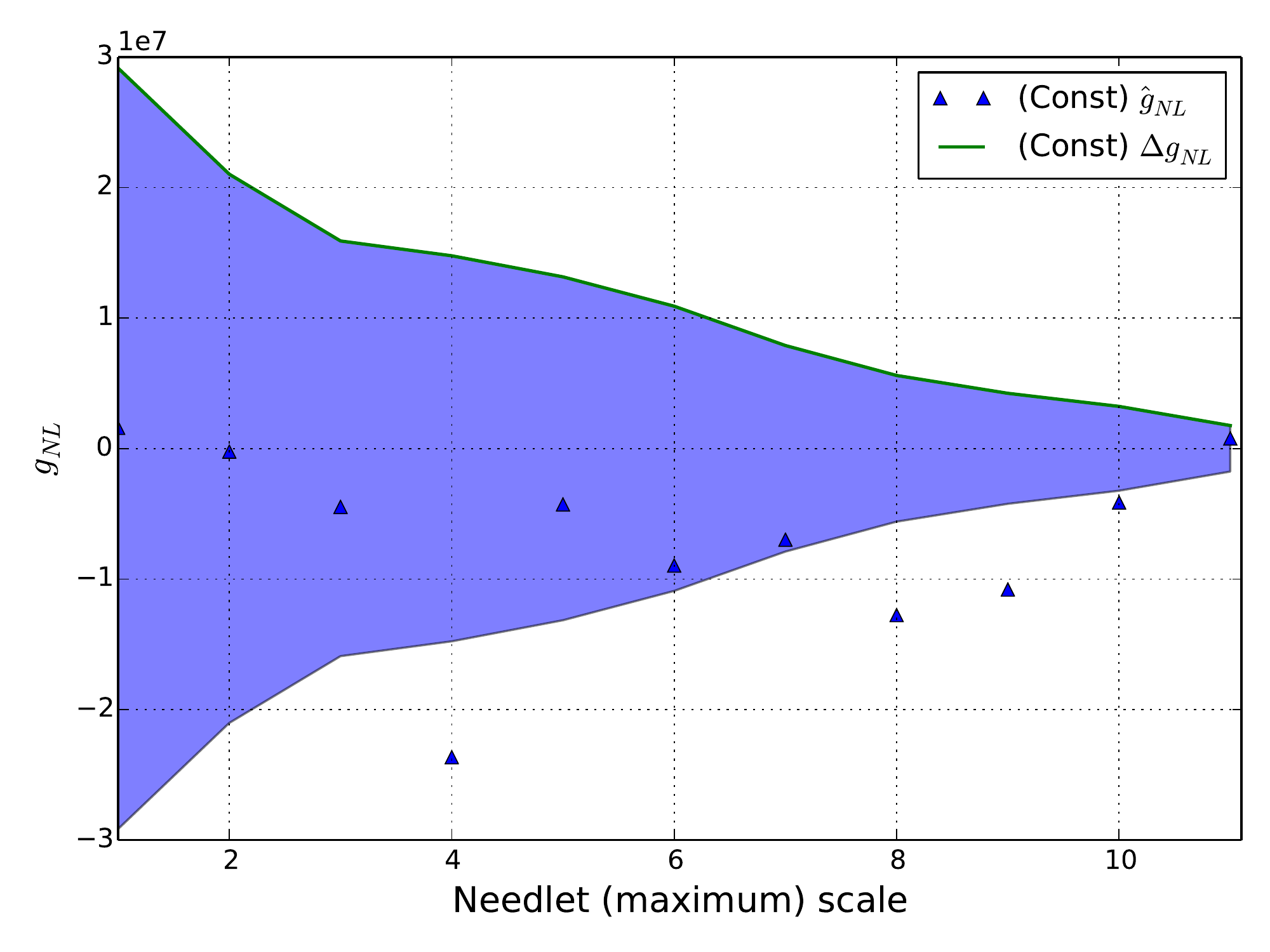}}
{
\includegraphics[width=0.3\linewidth]{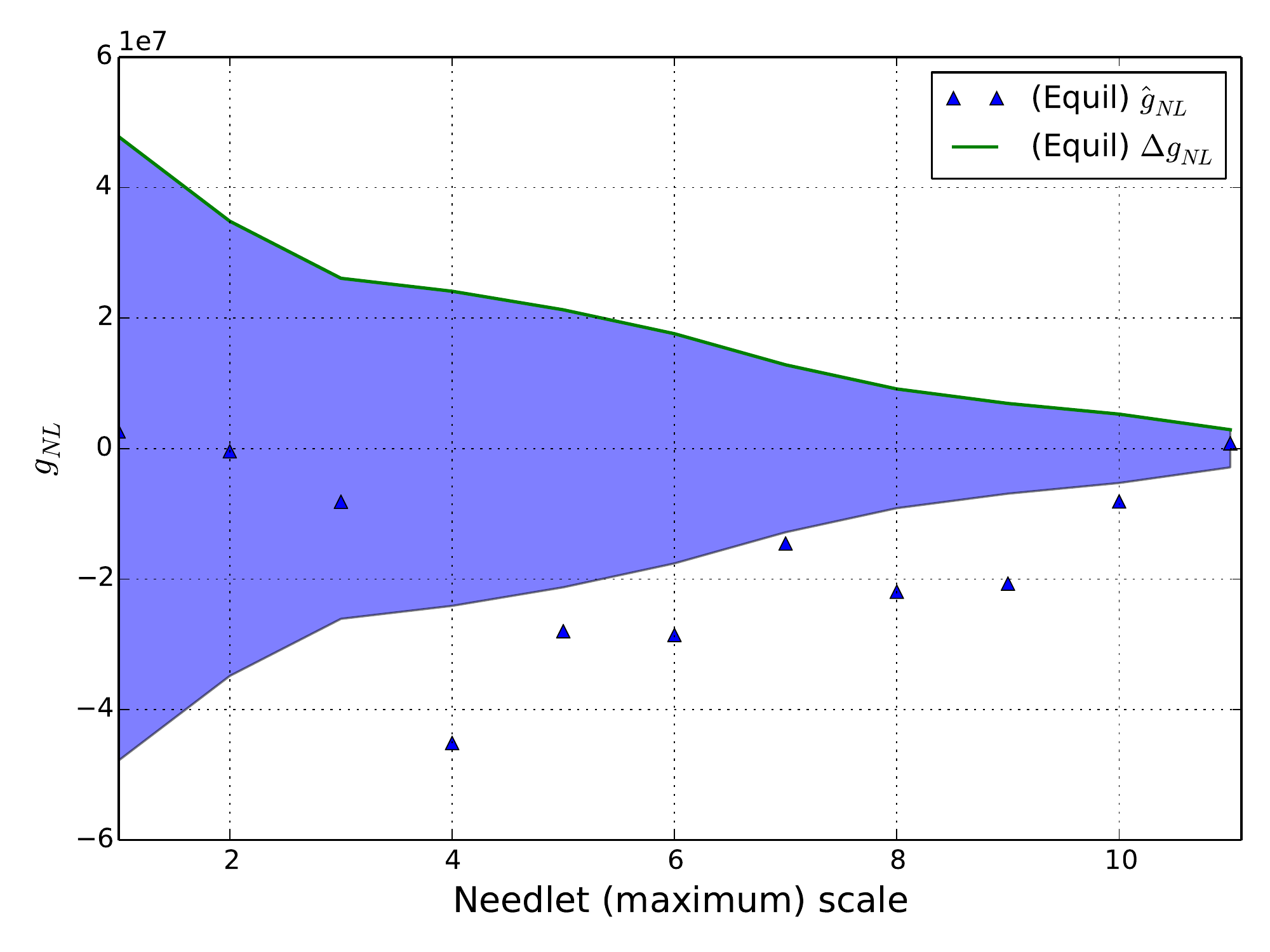}}
\caption{Represented left to right are the dependencies of best-fit (triangular glyphs) and $1\sigma$error bars on the maximum needlet scale used for the local $g_{NL}$ model, the constant trispectrum, and the equilateral, $c_1$, trispectrum, respectively. }
\label{fig:gnlErrors}
\end{figure}

\item \textbf{Constant model.}
The constant trispectrum was defined in Ref.~\cite{FRS2}
by analogy with the constant bispectrum. It gives
\begin{equation}
    T_{\Phi}(k_1,k_2,k_3,k_4)=24 \big( P_{\Phi}(k_1)P_{\Phi}(k_2)P_{\Phi}(k_3)P_{\Phi}(k_4) \big)^{3/4} .
\end{equation}
As with the constant bispectrum, features in this model are entirely due to the transfer functions.
We find

\begin{equation}
	\gnl^{\text{const}}=[ 0.8\pm 1.8] \times 10^6
	\quad \text{and} \quad
	\Delta\gnl^{\text{const}}= [1.0\pm 1.6] \times 10^6 .
\end{equation}
Correcting for the bias due to point sources this is equivalent to
$\gnl^{\text{const}} = [ -0.2\pm 1.8] \times 10^6$
and improves on the (corrected) constraint
$\gnl^{\text{const}}=[ 1.0\pm 2.8] \times 10^6$
presented in Ref.~\cite{FRS2}. The point source constraint uses the more accurate point source model but is very consistent with the constant flux model which gives $\Delta\gnl^{\text{const}}= [0.8\pm 1.6] \times 10^6$. In Fig.~\ref{fig:gnlErrors} we plot the dependency of the results on the (maximum) needlet scale used in the analysis. There appears to be a weak dependency on the maximum scale chosen, but the results clearly support the robust conclusion that the best-fit value lies within $\sim 1\sigma$ of zero.

\item \textbf{Equilateral $c_1$ model.}
In Refs.~\cite{aChen,XINGANG_t_NL,Arroja2009} it was
demonstrated that the trispectrum for single-field inflation models
with nontrivial kinetic terms
receives its dominant contribution from
a combination of three trispectra $T_{s_1,s_2,s_3}$
generated by scalar exchange,
and three trispectra $T_{c_1,c_2,c_3}$ generated by contact interactions.
As explained in Ref.~\cite{FRS2}, the $c_1$-type contact-interaction trispectrum
is strongly correlated with most of the other shapes. Helpfully, it can be
described by a simple diagonal-free formula, for which
\begin{equation}
    T_{c_1}(k_1,k_2,k_3,k_4)= \frac{24 A_{\Phi}^3}{k_1 k_2 k_3 k_4 (\sum_i k_i/4)^5} .
\end{equation}
We find the constraints
\begin{equation}
	\gnl^{c_1}= [0.8\pm 2.9]\times 10^6
	\quad \text{and} \quad
	\Delta\gnl^{c_1}= [1.6\pm 2.5] \times 10^6.
\end{equation} 
The constraint after correction for biasing
is $\gnl^{c_1} = [-0.8\pm 2.9]\times 10^6$. 
This is again consistent with
the (corrected) result of Ref.~\cite{FRS2},
which found $\gnl^{c_1}=[-2.4\pm 5.8]\times 10^6$. As for the constant model, the constraint on the point sources is largely insensitive to the model used with the constant flux model giving the constraint $\Delta\gnl^{c_1}= [1.2\pm 2.5] \times 10^6$. In addition, the plot of the dependency on the maximum needlet scale in Fig.~\ref{fig:gnlErrors} further supports the conclusion that the best-fit value lies within $\sim 1\sigma$ of zero.

\item \textbf{Local $\taunl$ model.}
The local model~\eqref{eq:localmodel} also generates a trispectrum
in the $\taunl$-mode,
which gives
$\taunl^{\text{loc}} = (6 \fnl^{\text{loc}} / 5)^2$.
In models with more contributions to the curvature perturabtion
this is softened to an inequality
$\taunl^{\text{loc}} \geq (6 \fnl^{\text{loc}} / 5)^2$~\cite{Suyama2008,Smith:2011if,Assassi:2012zq}.
Therefore, simultaneous detections of $\taunl^{\text{loc}}$
and $\fnl^{\text{loc}}$
would
provide an opportunity to probe the microphysics of the inflationary era.

Unfortunately, estimating $\taunllocal$ is a challenging undertaking.
Ref.~\cite{RSF1}
derived the estimator
$\taunlhat^{\text{loc}} = \mathcal{E}^{\taunl}/\langle \mathcal{E}^{\taunl}\rangle$, where
\begin{equation}
\label{eq:estimtrisp}
\begin{split}
    \mathcal{E}^{\taunl}
    =
    3 &
    \sum_{l_i m_i}\sum_{L M}(-1)^M p^{l_1 l_2}_{l_3 l_4}(L)\mathcal{G}^{l_1 l_2 L}_{m_1 m_2 M}\mathcal{G}^{l_3 l_4 L}_{m_3 m_4 -M}
    \\
    &
    \mbox{} \times
    \Bigg(\prod_{i=1}^4\tilde{a}_{l_i m_i}-\Big(\langle \tilde{a}^G_{l_1 m_1}\tilde{a}^G_{l_2 m_2}\rangle \tilde{a}_{l_3 m_3}\tilde{a}_{l_4 m_4}+\text{5 perms}\Big) + \langle \tilde{a}^G_{l_1 m_1}\tilde{a}^G_{l_2 m_2}\rangle \langle \tilde{a}^G_{l_3 m_3}\tilde{a}^G_{l_4 m_4}\rangle+\text{2 perms})\Bigg)
\end{split}
\end{equation}
\begin{equation}
\begin{split}
    \langle\mathcal{E}^{\taunl}\rangle
    \approx
    3
    \sum_{l_i,L}\frac{ p^{l_1 l_2}_{l_3 l_4}(L)h_{l_1 l_2 L}h_{l_3 l_4 L} }{C_{l_1}C_{l_2}C_{l_3}C_{l_4}}
    \Bigg( & \frac{p^{l_1 l_2}_{l_3 l_4}(L)h_{l_1 l_2 L}h_{l_3 l_4 L}}{2L+1} +\sum_{L'}(-1)^{l_2+l_3}\Big\{ \begin{array}{ccc}
l_1 & l_2 & L \\
l_4 & l_3 & L' \end{array}\Big\} p^{l_1 l_3}_{l_2 l_4}(L')h_{l_1 l_3 L'}h_{l_2 l_4 L'}\\
& \mbox{} +\sum_{L'}(-1)^{L+L'}\Big\{ \begin{array}{ccc} 
l_1 & l_2 & L \\
l_3 & l_4 & L' \end{array}\Big\} p^{l_1 l_4}_{l_3 l_2}(L')h_{l_1 l_4 L'}h_{l_3 l_2 L'}\Bigg)
,
\end{split}
\end{equation}
where $\tilde{a}_{l m}=(C^{-1} a)_{ lm}$ are the inverse-covariance weighted
spherical harmonics
and
\begin{equation}
    h_{l_1 l_2 l_3}
    \equiv
    \sqrt{\frac{\prod_{i=1}^3 (2l_i+1)}{4\pi}}\Big( \begin{array}{ccc} 
    l_1 & l_2 & l_3 \\
    0 & 0 &0 \end{array}\Big) .
\end{equation}
We have expressed $\langle \mathcal{E}^{\taunl}\rangle$ using the approximation of a
diagonal covariance matrix.
Unfortunately,
the presence of the $6j$ symbols $\{\dots\}$
means that the computation of $\langle\mathcal{E}^{\taunl}\rangle$ is prohibitive
and therefore some approximations are required.
For example, it is an accurate approximation to restrict the calculation to $L\lesssim 100$
and neglect the effect of the $6j$ symbols~\cite{0602099}.

However, the needlet estimator \eqref{eq:taunlT} obviates this
by removing the necessity to calculate the $6j$ symbols explicitly.
Nevertheless, the $\taunl$ model still represents a formidable numerical challenge
due to the presence of \emph{two} line-of-sight integrals in \eqref{eq:taunlfull}.

To make progress, we use the observation of
Pearson et al.~\cite{Pearson}
that
Eq.~\eqref{eq:ptrisp}
represents an accurate approximation to the $\taunl$ trispectrum shape.
In this paper we employ it for the calculation of
expectation values of the quartic needlet statistic, Eq.~\eqref{eq:taunlNJ}.
We note that a similar approximation could be applied to the
trispectrum generated by cosmic strings, or due to lensing.

Alternative approaches are possible.
A modulation-based estimator was developed by Pearson et al.~\cite{Pearson}
and was applied to Planck data for the Planck2013 data release~\cite{PlancknonG}.
This modulation-based estimator is similar
to Eq.~\eqref{eq:estimtrisp} with~\eqref{eq:ptrisp} used to approximate the
$\taunl$-mode trispectrum.
That is,
\begin{equation}
\mathcal{E}^{\taunl}_{\text{mod}}= 12\sum_L C_L^{\zeta_*}\sum_{M}\Big( |f_{LM}-\langle f_{LM}\rangle|^2 - \langle |f_{LM}^G-\langle f_{LM}^G\rangle|^2\rangle \Big) ,
\end{equation}
where
\begin{equation}
f_{LM}=\int \d^2 \hat{\bn} \; Y_{LM}^*(\hat{\bn}) \sum_{l_1 m_1} \tilde{a}_{l_1 m_1}Y_{l_1 m_1}(\hat{\bn})\sum_{l_2 m_2} C_{l_2}\tilde{a}_{l_2 m_2}Y_{l_2 m_2}(\hat{\bn}) .
\end{equation}

Applying the needlet estimator~\eqref{eq:taunlT}
gives constraints for the amplitude and bias due to point sources,
\begin{equation}
	\taunllocal = 3020,
	\quad \sigma(\taunllocal)=3910
	\quad \text{and} \quad
	\Delta\taunllocal =-230, \quad \sigma(\taunllocal)=1890.
\end{equation}
In the single-field case,
the 1-$\sigma$ error bar $\sigma(\taunl)$
would correspond to an error on $\fnllocal$ equal to $52$.
We do not express the error bars in the symmetric form
$\taunllocal = \taunlhat^{\text{loc}} \pm \sigma(\taunlhat^{\text{loc}})$
because this would assume a null hypothesis
of zero signal, and a Gaussian-distributed estimator.
Howevever, as explained by
Hanson \& Lewis and
Smith \& Kamionkowski~\cite{HansonLewis,SmithKam},
the distribution of $\taunlhat^{\text{loc}}$ is not symmetric:
it corresponds to a weighted sum of $\chi^2$ random variables.
Therefore, given a particular central value of $\taunllocal$,
it is necessary in general to evaluate the posterior distribution of
the error bar.
A suitable analysis was given in the Planck2013 data release~\cite{PlancknonG},
which we now briefly recapitulate.
Each mode of the modulation field $f_{LM}$ can be regarded as independently Gaussian distributed,
$f_{LM}\sim \mathcal{N}(\langle f_{LM}\rangle, \taunl C^{\zeta_*}_L+N_L\rangle)$,
where we have set
$N_L\equiv (2L+1)^{-1} \sum_L\langle |f_{LM}^G-\langle f^G_{LM}\rangle|^2\rangle$.
An estimate for $\taunl$ at each $L$ is
\begin{equation}
    \taunlhat(L)=\frac{1}{(2L+1)C_L^{\zeta_*}}\sum_{M}\Big( |f_{LM}-\langle f_{LM}\rangle|^2 - \langle |f_{LM}^G-\langle f_{LM}^G\rangle|^2\rangle \Big). 
\end{equation} 
Defining the quantity $x(L)$ as
\begin{equation}
    x(L)\equiv \frac{\taunlhat(L)+N_L/C^{\zeta_*}_L}{\taunl+N_L/C^{\zeta_*}_L}\,,
\end{equation}
and regarding the $\taunlhat(L)$
estimators as uncorrelated, the posterior distribution $P(\taunl;\{\taunlhat(L)\})$
is given by the product of inverse Gamma functions~\cite{PlancknonG,HamLewis2008}
\begin{equation}
    P(\taunl;\{\taunlhat(L)\})\propto \prod_{L=\lmin}^{L=\lmax} f\big[ x(L)^{-1}; (2L-1)/2, (2L+1)/2 \big] ,
\end{equation}
where the inverse Gamma distribution $f(x,\alpha,\beta)$
with shape parameter $\alpha$ and scale parameter $\beta$ is defined by
\begin{equation}
    f(x; \alpha, \beta) = \frac{\beta^\alpha}{\Gamma(\alpha)} x^{-\alpha-1}
    \e{-\beta/x} .
\end{equation}

In this paper we wish to work with the needlet-based estimator.
Therefore we approximate the quantity $\taunlhat(L)$
by our estimate $\taunlhat$ for all $L$.
Although not strictly correct, we expect that this will yield
qualitatively accurate constraints.
To deduce $N_L$ we utilise the expression $\sigma(\taunl)^{-2}\approx\sum_L (2L+1) {C_L^{\zeta_*}}^2/N_L^2$, and note that it represents white noise, and
therefore is independent of $L$.

The resulting constraint is
\begin{equation}
    \taunllocal < 22000
\end{equation}
which compares with
the Planck2013 error bar $\taunllocal < 2800$ \cite{PlancknonG}.
\end{itemize}

\section{Conclusions}\label{sec:conclusions}
In this paper we have coupled the successful `modal' or partial-wave
method for non-separable bi- and tri-spectra to
a needlet-based estimator.
This extends the approach of Ref.~\cite{RMS2013}
in which the partial-wave method was coupled to
a wavelet-based estimator.
The key step in this approach is the introduction
of `change-of-basis' matrices
$\langle N_{nJ} \rangle$
and $\langle N_{nJ}^T \rangle$.
In principle, a variant of this method can be used
to couple the partial-wave decomposition
to any desired estimator.

The needlet- and wavelet-based estimators
are efficient because they require
inversion of a covariance matrix of
order $\sim 10^3 \times 10^3$ rather than
the full pixel-by-pixel covariance matrix
of order $\sim 10^6 \times 10^6$.
Despite this reduced computational burden,
our comparison with the 9-year WMAP data demonstrates
that
these estimators are close
(within $\lesssim 5\%$--$10\%$)
to optimal.
Both the needlet- and wavelet-based estimators
are efficient detectors of point sources,
but our results suggest that the needlet-based
estimator is most sensitive.


We have used our approach to construct the first needlet-based estimator
for the trispectrum.
As a by-product, this estimator avoids the general (expensive) requirement
to explicitly calculate Wigner-6j symbols.
For the class of diagonal-free trispectra
(that is, those which depend only on the multipoles
$l_i$ in the harmonic decomposition of the CMB)
we employ the partial-wave expansion approach developed in Refs.~\cite{RSF1,FRS2}.
However, the estimator can equally well be applied
to trispectra which are not diagonal-free.
As an example, we have used it to contrain the local
$\taunl$-shape trispectrum.
Alternative uses could include searches for trispectra generated
by cosmic strings or lensing.


We have tabulated
constraints on the local, DBI, equilateral, constant, orthogonal and flattened bispectra.
For each of these models we provide estimates of the contamination due to point sources.
We have also studied the frequency-dependence of our results,
for which the constraints on the orthogonal model are particularly interesting.
While the WMAP team did not suggest a strong signal for this model 
in the 9-year daata ($\sim 2.5\sigma$), their analysis suggested a much stronger W-band signature ($\sim 2.9\sigma$)
comarped to the V-band ($\sim 1.1\sigma$).
Using the needlet-based estimator we have shown that all models,
including the orthogonal bispectrum, are essentially frequency independent.
The expected point-source contribution shows a mild frequency dependence, at the level of a fraction
of an error bar.

We also tabulate constraints
for a selection of trispectrum shapes, including
the local $\gnl$-shape, constant, and $c_1$ equilateral trispectra.
The $c_1$ model
is representative of certain inflationary models
with non-canonical kinetic terms. All three models are `diagonal-free', which
allows a decomposition into partial waves.
Our constraints on $\gnllocal$ are close to optimal.
Finally, we constrain the local $\taunl$-shape
trispectrum.
This is not diagonal-free, but can be modelled using an
accurate separable approximation similar to that
employed by Pearson et al.~\cite{Pearson}.
All of
these constraints show that the CMB does not deviate from the
standard paradigm of a Gaussian primordial fluctuation,
to both three-point and four-point order.
In addition we have computed, for the first time, the effect of point sources
on each of these trispectrum models. As with the bispectrum, each model
shows only mild bias due to the presence of unresolved point sources.


\section*{Acknowledgements}
It is a pleasure to thank
Antony Lewis for many helpful discussions.
DMR wishes to acknowledge work with
James Fergusson in developing many aspects of the modal methodology.

We acknowledge use of HEALPix (Hierarchical Equal Area isoLatitude Pixelization) software \cite{0409513} in computing many of the results presented in this paper.
Some of these numerical results were obtained using
the COSMOS supercomputer,
which is funded by STFC, HEFCE and SGI.
Other
numerical computations were carried out on the Sciama High Performance Compute (HPC)
cluster which is supported by
the ICG, SEPNet and the University of Portsmouth. 
We acknowledge support from the Science and Technology
Facilities Council [grant number ST/I000976/1].
The research leading to these results has received funding from
the European Research Council under the European Union's
Seventh Framework Programme (FP/2007--2013) / ERC Grant
Agreement No. [308082].
DS acknowledges support from the Leverhulme Trust.
MG thanks the Slovene Human Resources Development and Scholarship Fund for financial support.

\bibliography{NeedletsPaper}

\begin{thebibliography}{67}%
\makeatletter
\providecommand \@ifxundefined [1]{%
 \@ifx{#1\undefined}
}%
\providecommand \@ifnum [1]{%
 \ifnum #1\expandafter \@firstoftwo
 \else \expandafter \@secondoftwo
 \fi
}%
\providecommand \@ifx [1]{%
 \ifx #1\expandafter \@firstoftwo
 \else \expandafter \@secondoftwo
 \fi
}%
\providecommand \natexlab [1]{#1}%
\providecommand \enquote  [1]{``#1''}%
\providecommand \bibnamefont  [1]{#1}%
\providecommand \bibfnamefont [1]{#1}%
\providecommand \citenamefont [1]{#1}%
\providecommand \href@noop [0]{\@secondoftwo}%
\providecommand \href [0]{\begingroup \@sanitize@url \@href}%
\providecommand \@href[1]{\@@startlink{#1}\@@href}%
\providecommand \@@href[1]{\endgroup#1\@@endlink}%
\providecommand \@sanitize@url [0]{\catcode `\\12\catcode `\$12\catcode
  `\&12\catcode `\#12\catcode `\^12\catcode `\_12\catcode `\%12\relax}%
\providecommand \@@startlink[1]{}%
\providecommand \@@endlink[0]{}%
\providecommand \url  [0]{\begingroup\@sanitize@url \@url }%
\providecommand \@url [1]{\endgroup\@href {#1}{\urlprefix }}%
\providecommand \urlprefix  [0]{URL }%
\providecommand \Eprint [0]{\href }%
\providecommand \doibase [0]{http://dx.doi.org/}%
\providecommand \selectlanguage [0]{\@gobble}%
\providecommand \bibinfo  [0]{\@secondoftwo}%
\providecommand \bibfield  [0]{\@secondoftwo}%
\providecommand \translation [1]{[#1]}%
\providecommand \BibitemOpen [0]{}%
\providecommand \bibitemStop [0]{}%
\providecommand \bibitemNoStop [0]{.\EOS\space}%
\providecommand \EOS [0]{\spacefactor3000\relax}%
\providecommand \BibitemShut  [1]{\csname bibitem#1\endcsname}%
\let\auto@bib@innerbib\@empty
\bibitem [{\citenamefont {Bartolo}\ \emph {et~al.}(2004)\citenamefont
  {Bartolo}, \citenamefont {Komatsu}, \citenamefont {Matarrese},\ and\
  \citenamefont {Riotto}}]{0406398}%
  \BibitemOpen
  \bibfield  {author} {\bibinfo {author} {\bibfnamefont {N.}~\bibnamefont
  {Bartolo}}, \bibinfo {author} {\bibfnamefont {E.}~\bibnamefont {Komatsu}},
  \bibinfo {author} {\bibfnamefont {S.}~\bibnamefont {Matarrese}}, \ and\
  \bibinfo {author} {\bibfnamefont {A.}~\bibnamefont {Riotto}},\ }\href
  {\doibase 10.1016/j.physrep.2004.08.022} {\bibfield  {journal} {\bibinfo
  {journal} {Phys. Rept.}\ }\textbf {\bibinfo {volume} {402}},\ \bibinfo
  {pages} {103} (\bibinfo {year} {2004})},\ \Eprint
  {http://arxiv.org/abs/astro-ph/0406398} {arXiv:astro-ph/0406398} \BibitemShut
  {NoStop}%
\bibitem [{\citenamefont {{Yadav}}\ and\ \citenamefont
  {{Wandelt}}(2010)}]{2010AdAst2010E..71Y}%
  \BibitemOpen
  \bibfield  {author} {\bibinfo {author} {\bibfnamefont {A.~P.~S.}\
  \bibnamefont {{Yadav}}}\ and\ \bibinfo {author} {\bibfnamefont {B.~D.}\
  \bibnamefont {{Wandelt}}},\ }\href {\doibase 10.1155/2010/565248} {\bibfield
  {journal} {\bibinfo  {journal} {Advances in Astronomy}\ }\textbf {\bibinfo
  {volume} {2010}},\ \bibinfo {eid} {565248} (\bibinfo {year} {2010})},\
  \Eprint {http://arxiv.org/abs/1006.0275} {arXiv:1006.0275 [astro-ph.CO]}
  \BibitemShut {NoStop}%
\bibitem [{\citenamefont {{Fergusson}}\ \emph
  {et~al.}(2010{\natexlab{a}})\citenamefont {{Fergusson}}, \citenamefont
  {{Liguori}},\ and\ \citenamefont {{Shellard}}}]{FLS1}%
  \BibitemOpen
  \bibfield  {author} {\bibinfo {author} {\bibfnamefont {J.~R.}\ \bibnamefont
  {{Fergusson}}}, \bibinfo {author} {\bibfnamefont {M.}~\bibnamefont
  {{Liguori}}}, \ and\ \bibinfo {author} {\bibfnamefont {E.~P.~S.}\
  \bibnamefont {{Shellard}}},\ }\href {\doibase 10.1103/PhysRevD.82.023502}
  {\bibfield  {journal} {\bibinfo  {journal} {\prd}\ }\textbf {\bibinfo
  {volume} {82}},\ \bibinfo {pages} {023502} (\bibinfo {year}
  {2010}{\natexlab{a}})},\ \Eprint {http://arxiv.org/abs/0912.5516}
  {arXiv:0912.5516 [astro-ph.CO]} \BibitemShut {NoStop}%
\bibitem [{\citenamefont {{Regan}}\ \emph {et~al.}(2010)\citenamefont
  {{Regan}}, \citenamefont {{Shellard}},\ and\ \citenamefont
  {{Fergusson}}}]{RSF1}%
  \BibitemOpen
  \bibfield  {author} {\bibinfo {author} {\bibfnamefont {D.~M.}\ \bibnamefont
  {{Regan}}}, \bibinfo {author} {\bibfnamefont {E.~P.~S.}\ \bibnamefont
  {{Shellard}}}, \ and\ \bibinfo {author} {\bibfnamefont {J.~R.}\ \bibnamefont
  {{Fergusson}}},\ }\href {\doibase 10.1103/PhysRevD.82.023520} {\bibfield
  {journal} {\bibinfo  {journal} {\prd}\ }\textbf {\bibinfo {volume} {82}},\
  \bibinfo {pages} {023520} (\bibinfo {year} {2010})},\ \Eprint
  {http://arxiv.org/abs/1004.2915} {arXiv:1004.2915 [astro-ph.CO]} \BibitemShut
  {NoStop}%
\bibitem [{\citenamefont {{Fergusson}}\ \emph {et~al.}(2012)\citenamefont
  {{Fergusson}}, \citenamefont {{Liguori}},\ and\ \citenamefont
  {{Shellard}}}]{FLS2}%
  \BibitemOpen
  \bibfield  {author} {\bibinfo {author} {\bibfnamefont {J.~R.}\ \bibnamefont
  {{Fergusson}}}, \bibinfo {author} {\bibfnamefont {M.}~\bibnamefont
  {{Liguori}}}, \ and\ \bibinfo {author} {\bibfnamefont {E.~P.~S.}\
  \bibnamefont {{Shellard}}},\ }\href {\doibase 10.1088/1475-7516/2012/12/032}
  {\bibfield  {journal} {\bibinfo  {journal} {\jcap}\ }\textbf {\bibinfo
  {volume} {12}},\ \bibinfo {eid} {032} (\bibinfo {year} {2012})},\ \Eprint
  {http://arxiv.org/abs/1006.1642} {arXiv:1006.1642 [astro-ph.CO]} \BibitemShut
  {NoStop}%
\bibitem [{\citenamefont {{Fergusson}}\ \emph
  {et~al.}(2010{\natexlab{b}})\citenamefont {{Fergusson}}, \citenamefont
  {{Regan}},\ and\ \citenamefont {{Shellard}}}]{FRS2}%
  \BibitemOpen
  \bibfield  {author} {\bibinfo {author} {\bibfnamefont {J.~R.}\ \bibnamefont
  {{Fergusson}}}, \bibinfo {author} {\bibfnamefont {D.~M.}\ \bibnamefont
  {{Regan}}}, \ and\ \bibinfo {author} {\bibfnamefont {E.~P.~S.}\ \bibnamefont
  {{Shellard}}},\ }\href@noop {} {\bibfield  {journal} {\bibinfo  {journal}
  {ArXiv e-prints}\ } (\bibinfo {year} {2010}{\natexlab{b}})},\ \Eprint
  {http://arxiv.org/abs/1012.6039} {arXiv:1012.6039 [astro-ph.CO]} \BibitemShut
  {NoStop}%
\bibitem [{\citenamefont {Komatsu}\ \emph {et~al.}(2005)\citenamefont
  {Komatsu}, \citenamefont {Spergel},\ and\ \citenamefont {Wandelt}}]{KSW}%
  \BibitemOpen
  \bibfield  {author} {\bibinfo {author} {\bibfnamefont {E.}~\bibnamefont
  {Komatsu}}, \bibinfo {author} {\bibfnamefont {D.~N.}\ \bibnamefont
  {Spergel}}, \ and\ \bibinfo {author} {\bibfnamefont {B.~D.}\ \bibnamefont
  {Wandelt}},\ }\href {\doibase 10.1086/491724} {\bibfield  {journal} {\bibinfo
   {journal} {Astrophysical Journal}\ }\textbf {\bibinfo {volume} {634}},\
  \bibinfo {pages} {14} (\bibinfo {year} {2005})},\ \Eprint
  {http://arxiv.org/abs/astro-ph/0305189} {arXiv:astro-ph/0305189} \BibitemShut
  {NoStop}%
\bibitem [{\citenamefont {{Regan}}\ \emph {et~al.}(2013)\citenamefont
  {{Regan}}, \citenamefont {{Mukherjee}},\ and\ \citenamefont
  {{Seery}}}]{RMS2013}%
  \BibitemOpen
  \bibfield  {author} {\bibinfo {author} {\bibfnamefont {D.}~\bibnamefont
  {{Regan}}}, \bibinfo {author} {\bibfnamefont {P.}~\bibnamefont
  {{Mukherjee}}}, \ and\ \bibinfo {author} {\bibfnamefont {D.}~\bibnamefont
  {{Seery}}},\ }\href {\doibase 10.1103/PhysRevD.88.043512} {\bibfield
  {journal} {\bibinfo  {journal} {\prd}\ }\textbf {\bibinfo {volume} {88}},\
  \bibinfo {eid} {043512} (\bibinfo {year} {2013})},\ \Eprint
  {http://arxiv.org/abs/1302.5631} {arXiv:1302.5631 [astro-ph.CO]} \BibitemShut
  {NoStop}%
\bibitem [{\citenamefont {Curto}\ \emph {et~al.}(2009)\citenamefont {Curto},
  \citenamefont {Martinez-Gonzalez},\ and\ \citenamefont
  {Barreiro}}]{Curto:2009pv}%
  \BibitemOpen
  \bibfield  {author} {\bibinfo {author} {\bibfnamefont {A.}~\bibnamefont
  {Curto}}, \bibinfo {author} {\bibfnamefont {E.}~\bibnamefont
  {Martinez-Gonzalez}}, \ and\ \bibinfo {author} {\bibfnamefont {R.~B.}\
  \bibnamefont {Barreiro}},\ }\href {\doibase 10.1088/0004-637X/706/1/399}
  {\bibfield  {journal} {\bibinfo  {journal} {Astrophys. J.}\ }\textbf
  {\bibinfo {volume} {706}},\ \bibinfo {pages} {399} (\bibinfo {year}
  {2009})},\ \Eprint {http://arxiv.org/abs/0902.1523} {arXiv:0902.1523
  [astro-ph.CO]} \BibitemShut {NoStop}%
\bibitem [{\citenamefont {{Lan}}\ and\ \citenamefont
  {{Marinucci}}(2008)}]{Lan}%
  \BibitemOpen
  \bibfield  {author} {\bibinfo {author} {\bibfnamefont {X.}~\bibnamefont
  {{Lan}}}\ and\ \bibinfo {author} {\bibfnamefont {D.}~\bibnamefont
  {{Marinucci}}},\ }\href {\doibase 10.1214/08-EJS197} {\bibfield  {journal}
  {\bibinfo  {journal} {Electronic Journal of Statistics}\ }\textbf {\bibinfo
  {volume} {2}},\ \bibinfo {pages} {332} (\bibinfo {year} {2008})},\ \Eprint
  {http://arxiv.org/abs/0802.4020} {arXiv:0802.4020} \BibitemShut {NoStop}%
\bibitem [{\citenamefont {{Pietrobon}}\ \emph {et~al.}(2006)\citenamefont
  {{Pietrobon}}, \citenamefont {{Balbi}},\ and\ \citenamefont
  {{Marinucci}}}]{Pietrobon}%
  \BibitemOpen
  \bibfield  {author} {\bibinfo {author} {\bibfnamefont {D.}~\bibnamefont
  {{Pietrobon}}}, \bibinfo {author} {\bibfnamefont {A.}~\bibnamefont
  {{Balbi}}}, \ and\ \bibinfo {author} {\bibfnamefont {D.}~\bibnamefont
  {{Marinucci}}},\ }\href {\doibase 10.1103/PhysRevD.74.043524} {\bibfield
  {journal} {\bibinfo  {journal} {\prd}\ }\textbf {\bibinfo {volume} {74}},\
  \bibinfo {eid} {043524} (\bibinfo {year} {2006})},\ \Eprint
  {http://arxiv.org/abs/arXiv:astro-ph/0606475} {arXiv:astro-ph/0606475}
  \BibitemShut {NoStop}%
\bibitem [{\citenamefont {{Baldi}}\ \emph
  {et~al.}(2006{\natexlab{a}})\citenamefont {{Baldi}}, \citenamefont
  {{Kerkyacharian}}, \citenamefont {{Marinucci}},\ and\ \citenamefont
  {{Picard}}}]{Baldi}%
  \BibitemOpen
  \bibfield  {author} {\bibinfo {author} {\bibfnamefont {P.}~\bibnamefont
  {{Baldi}}}, \bibinfo {author} {\bibfnamefont {G.}~\bibnamefont
  {{Kerkyacharian}}}, \bibinfo {author} {\bibfnamefont {D.}~\bibnamefont
  {{Marinucci}}}, \ and\ \bibinfo {author} {\bibfnamefont {D.}~\bibnamefont
  {{Picard}}},\ }\href@noop {} {\bibfield  {journal} {\bibinfo  {journal}
  {ArXiv Mathematics e-prints}\ } (\bibinfo {year} {2006}{\natexlab{a}})},\
  \Eprint {http://arxiv.org/abs/arXiv:math/0606599} {arXiv:math/0606599}
  \BibitemShut {NoStop}%
\bibitem [{\citenamefont {{Baldi}}\ \emph
  {et~al.}(2006{\natexlab{b}})\citenamefont {{Baldi}}, \citenamefont
  {{Kerkyacharian}}, \citenamefont {{Marinucci}},\ and\ \citenamefont
  {{Picard}}}]{Baldi2}%
  \BibitemOpen
  \bibfield  {author} {\bibinfo {author} {\bibfnamefont {P.}~\bibnamefont
  {{Baldi}}}, \bibinfo {author} {\bibfnamefont {G.}~\bibnamefont
  {{Kerkyacharian}}}, \bibinfo {author} {\bibfnamefont {D.}~\bibnamefont
  {{Marinucci}}}, \ and\ \bibinfo {author} {\bibfnamefont {D.}~\bibnamefont
  {{Picard}}},\ }\href@noop {} {\bibfield  {journal} {\bibinfo  {journal}
  {ArXiv Mathematics e-prints}\ } (\bibinfo {year} {2006}{\natexlab{b}})},\
  \Eprint {http://arxiv.org/abs/arXiv:math/0606154} {arXiv:math/0606154}
  \BibitemShut {NoStop}%
\bibitem [{\citenamefont {{Pietrobon}}\ \emph
  {et~al.}(2009{\natexlab{a}})\citenamefont {{Pietrobon}}, \citenamefont
  {{Cabella}}, \citenamefont {{Balbi}}, \citenamefont {{de Gasperis}},\ and\
  \citenamefont {{Vittorio}}}]{Pietrobon2008}%
  \BibitemOpen
  \bibfield  {author} {\bibinfo {author} {\bibfnamefont {D.}~\bibnamefont
  {{Pietrobon}}}, \bibinfo {author} {\bibfnamefont {P.}~\bibnamefont
  {{Cabella}}}, \bibinfo {author} {\bibfnamefont {A.}~\bibnamefont {{Balbi}}},
  \bibinfo {author} {\bibfnamefont {G.}~\bibnamefont {{de Gasperis}}}, \ and\
  \bibinfo {author} {\bibfnamefont {N.}~\bibnamefont {{Vittorio}}},\ }\href
  {\doibase 10.1111/j.1365-2966.2009.14847.x} {\bibfield  {journal} {\bibinfo
  {journal} {\mnras}\ }\textbf {\bibinfo {volume} {396}},\ \bibinfo {pages}
  {1682} (\bibinfo {year} {2009}{\natexlab{a}})},\ \Eprint
  {http://arxiv.org/abs/0812.2478} {arXiv:0812.2478} \BibitemShut {NoStop}%
\bibitem [{\citenamefont {{Rudjord}}\ \emph {et~al.}(2009)\citenamefont
  {{Rudjord}}, \citenamefont {{Hansen}}, \citenamefont {{Lan}}, \citenamefont
  {{Liguori}}, \citenamefont {{Marinucci}},\ and\ \citenamefont
  {{Matarrese}}}]{Rudjord:2009mh}%
  \BibitemOpen
  \bibfield  {author} {\bibinfo {author} {\bibfnamefont {O.}~\bibnamefont
  {{Rudjord}}}, \bibinfo {author} {\bibfnamefont {F.~K.}\ \bibnamefont
  {{Hansen}}}, \bibinfo {author} {\bibfnamefont {X.}~\bibnamefont {{Lan}}},
  \bibinfo {author} {\bibfnamefont {M.}~\bibnamefont {{Liguori}}}, \bibinfo
  {author} {\bibfnamefont {D.}~\bibnamefont {{Marinucci}}}, \ and\ \bibinfo
  {author} {\bibfnamefont {S.}~\bibnamefont {{Matarrese}}},\ }\href@noop {}
  {\bibfield  {journal} {\bibinfo  {journal} {\apj}\ }\textbf {\bibinfo
  {volume} {701}},\ \bibinfo {pages} {369} (\bibinfo {year} {2009})},\ \Eprint
  {http://arxiv.org/abs/0901.3154} {arXiv:0901.3154 [astro-ph.CO]} \BibitemShut
  {NoStop}%
\bibitem [{\citenamefont {{Pietrobon}}\ \emph
  {et~al.}(2009{\natexlab{b}})\citenamefont {{Pietrobon}}, \citenamefont
  {{Cabella}}, \citenamefont {{Balbi}}, \citenamefont {{de Gasperis}},\ and\
  \citenamefont {{Vittorio}}}]{Pietrobon2009}%
  \BibitemOpen
  \bibfield  {author} {\bibinfo {author} {\bibfnamefont {D.}~\bibnamefont
  {{Pietrobon}}}, \bibinfo {author} {\bibfnamefont {P.}~\bibnamefont
  {{Cabella}}}, \bibinfo {author} {\bibfnamefont {A.}~\bibnamefont {{Balbi}}},
  \bibinfo {author} {\bibfnamefont {G.}~\bibnamefont {{de Gasperis}}}, \ and\
  \bibinfo {author} {\bibfnamefont {N.}~\bibnamefont {{Vittorio}}},\ }\href
  {\doibase 10.1111/j.1365-2966.2009.14847.x} {\bibfield  {journal} {\bibinfo
  {journal} {\mnras}\ }\textbf {\bibinfo {volume} {396}},\ \bibinfo {pages}
  {1682} (\bibinfo {year} {2009}{\natexlab{b}})},\ \Eprint
  {http://arxiv.org/abs/0812.2478} {arXiv:0812.2478} \BibitemShut {NoStop}%
\bibitem [{\citenamefont {Rudjord}\ \emph {et~al.}(2010)\citenamefont
  {Rudjord}, \citenamefont {{Groeneboom}}, \citenamefont {Hansen},\ and\
  \citenamefont {Cabella}}]{Rudjord2010}%
  \BibitemOpen
  \bibfield  {author} {\bibinfo {author} {\bibfnamefont {O.}~\bibnamefont
  {Rudjord}}, \bibinfo {author} {\bibfnamefont {N.~E.}\ \bibnamefont
  {{Groeneboom}}}, \bibinfo {author} {\bibfnamefont {F.~K.}\ \bibnamefont
  {Hansen}}, \ and\ \bibinfo {author} {\bibfnamefont {P.}~\bibnamefont
  {Cabella}},\ }\href@noop {} {\bibfield  {journal} {\bibinfo  {journal}
  {\apj}\ }\textbf {\bibinfo {volume} {718}},\ \bibinfo {pages} {66} (\bibinfo
  {year} {2010})},\ \Eprint {http://arxiv.org/abs/1002.1811} {arXiv:1002.1811
  [astro-ph.CO]} \BibitemShut {NoStop}%
\bibitem [{\citenamefont {{Geller}}\ and\ \citenamefont
  {{Mayeli}}(2007{\natexlab{a}})}]{2007arXiv0709.2452G}%
  \BibitemOpen
  \bibfield  {author} {\bibinfo {author} {\bibfnamefont {D.}~\bibnamefont
  {{Geller}}}\ and\ \bibinfo {author} {\bibfnamefont {A.}~\bibnamefont
  {{Mayeli}}},\ }\href@noop {} {\bibfield  {journal} {\bibinfo  {journal}
  {ArXiv e-prints}\ } (\bibinfo {year} {2007}{\natexlab{a}})},\ \Eprint
  {http://arxiv.org/abs/0709.2452} {arXiv:0709.2452 [math.FA]} \BibitemShut
  {NoStop}%
\bibitem [{\citenamefont {{Geller}}\ and\ \citenamefont
  {{Mayeli}}(2007{\natexlab{b}})}]{2007arXiv0706.3642G}%
  \BibitemOpen
  \bibfield  {author} {\bibinfo {author} {\bibfnamefont {D.}~\bibnamefont
  {{Geller}}}\ and\ \bibinfo {author} {\bibfnamefont {A.}~\bibnamefont
  {{Mayeli}}},\ }\href@noop {} {\bibfield  {journal} {\bibinfo  {journal}
  {ArXiv e-prints}\ } (\bibinfo {year} {2007}{\natexlab{b}})},\ \Eprint
  {http://arxiv.org/abs/0706.3642} {arXiv:0706.3642 [math.CA]} \BibitemShut
  {NoStop}%
\bibitem [{\citenamefont {{Geller}}\ and\ \citenamefont
  {{Mayeli}}(2009)}]{2009arXiv0907.3164G}%
  \BibitemOpen
  \bibfield  {author} {\bibinfo {author} {\bibfnamefont {D.}~\bibnamefont
  {{Geller}}}\ and\ \bibinfo {author} {\bibfnamefont {A.}~\bibnamefont
  {{Mayeli}}},\ }\href@noop {} {\bibfield  {journal} {\bibinfo  {journal}
  {ArXiv e-prints}\ } (\bibinfo {year} {2009})},\ \Eprint
  {http://arxiv.org/abs/0907.3164} {arXiv:0907.3164 [math.FA]} \BibitemShut
  {NoStop}%
\bibitem [{\citenamefont {{Scodeller}}\ \emph {et~al.}(2011)\citenamefont
  {{Scodeller}}, \citenamefont {{Rudjord}}, \citenamefont {{Hansen}},
  \citenamefont {{Marinucci}}, \citenamefont {{Geller}},\ and\ \citenamefont
  {{Mayeli}}}]{Scodeller2010}%
  \BibitemOpen
  \bibfield  {author} {\bibinfo {author} {\bibfnamefont {S.}~\bibnamefont
  {{Scodeller}}}, \bibinfo {author} {\bibfnamefont {O.}~\bibnamefont
  {{Rudjord}}}, \bibinfo {author} {\bibfnamefont {F.~K.}\ \bibnamefont
  {{Hansen}}}, \bibinfo {author} {\bibfnamefont {D.}~\bibnamefont
  {{Marinucci}}}, \bibinfo {author} {\bibfnamefont {D.}~\bibnamefont
  {{Geller}}}, \ and\ \bibinfo {author} {\bibfnamefont {A.}~\bibnamefont
  {{Mayeli}}},\ }\href@noop {} {\bibfield  {journal} {\bibinfo  {journal}
  {\apj}\ }\textbf {\bibinfo {volume} {733}},\ \bibinfo {eid} {121} (\bibinfo
  {year} {2011})},\ \Eprint {http://arxiv.org/abs/1004.5576} {arXiv:1004.5576
  [astro-ph.CO]} \BibitemShut {NoStop}%
\bibitem [{\citenamefont {{Regan}}\ and\ \citenamefont
  {{Shellard}}(2010)}]{RS2009}%
  \BibitemOpen
  \bibfield  {author} {\bibinfo {author} {\bibfnamefont {D.~M.}\ \bibnamefont
  {{Regan}}}\ and\ \bibinfo {author} {\bibfnamefont {E.~P.~S.}\ \bibnamefont
  {{Shellard}}},\ }\href {\doibase 10.1103/PhysRevD.82.063527} {\bibfield
  {journal} {\bibinfo  {journal} {\prd}\ }\textbf {\bibinfo {volume} {82}},\
  \bibinfo {eid} {063527} (\bibinfo {year} {2010})},\ \Eprint
  {http://arxiv.org/abs/0911.2491} {arXiv:0911.2491 [astro-ph.CO]} \BibitemShut
  {NoStop}%
\bibitem [{\citenamefont {{Sekiguchi}}\ and\ \citenamefont
  {{Sugiyama}}(2013)}]{SekSug2013}%
  \BibitemOpen
  \bibfield  {author} {\bibinfo {author} {\bibfnamefont {T.}~\bibnamefont
  {{Sekiguchi}}}\ and\ \bibinfo {author} {\bibfnamefont {N.}~\bibnamefont
  {{Sugiyama}}},\ }\href@noop {} {\bibfield  {journal} {\bibinfo  {journal}
  {ArXiv e-prints}\ } (\bibinfo {year} {2013})},\ \Eprint
  {http://arxiv.org/abs/1303.4626} {arXiv:1303.4626 [astro-ph.CO]} \BibitemShut
  {NoStop}%
\bibitem [{\citenamefont {Lewis}\ \emph {et~al.}(2000)\citenamefont {Lewis},
  \citenamefont {Challinor},\ and\ \citenamefont {Lasenby}}]{9911177}%
  \BibitemOpen
  \bibfield  {author} {\bibinfo {author} {\bibfnamefont {A.}~\bibnamefont
  {Lewis}}, \bibinfo {author} {\bibfnamefont {A.}~\bibnamefont {Challinor}}, \
  and\ \bibinfo {author} {\bibfnamefont {A.}~\bibnamefont {Lasenby}},\ }\href
  {\doibase 10.1086/309179} {\bibfield  {journal} {\bibinfo  {journal}
  {Astrophys. J.}\ }\textbf {\bibinfo {volume} {538}},\ \bibinfo {pages} {473}
  (\bibinfo {year} {2000})},\ \Eprint {http://arxiv.org/abs/astro-ph/9911177}
  {arXiv:astro-ph/9911177} \BibitemShut {NoStop}%
\bibitem [{\citenamefont {Komatsu}\ and\ \citenamefont
  {Spergel}(2001)}]{KomatsuSpergel2001}%
  \BibitemOpen
  \bibfield  {author} {\bibinfo {author} {\bibfnamefont {E.}~\bibnamefont
  {Komatsu}}\ and\ \bibinfo {author} {\bibfnamefont {D.~N.}\ \bibnamefont
  {Spergel}},\ }\href {\doibase 10.1103/PhysRevD.63.063002} {\bibfield
  {journal} {\bibinfo  {journal} {Physical Review}\ }\textbf {\bibinfo {volume}
  {D63}},\ \bibinfo {pages} {063002} (\bibinfo {year} {2001})},\ \Eprint
  {http://arxiv.org/abs/astro-ph/0005036} {arXiv:astro-ph/0005036} \BibitemShut
  {NoStop}%
\bibitem [{\citenamefont {{Donzelli}}\ \emph {et~al.}(2012)\citenamefont
  {{Donzelli}}, \citenamefont {{Hansen}}, \citenamefont {{Liguori}},
  \citenamefont {{Marinucci}},\ and\ \citenamefont
  {{Matarrese}}}]{Donzelli2012}%
  \BibitemOpen
  \bibfield  {author} {\bibinfo {author} {\bibfnamefont {S.}~\bibnamefont
  {{Donzelli}}}, \bibinfo {author} {\bibfnamefont {F.~K.}\ \bibnamefont
  {{Hansen}}}, \bibinfo {author} {\bibfnamefont {M.}~\bibnamefont {{Liguori}}},
  \bibinfo {author} {\bibfnamefont {D.}~\bibnamefont {{Marinucci}}}, \ and\
  \bibinfo {author} {\bibfnamefont {S.}~\bibnamefont {{Matarrese}}},\ }\href
  {\doibase 10.1088/0004-637X/755/1/19} {\bibfield  {journal} {\bibinfo
  {journal} {\apj}\ }\textbf {\bibinfo {volume} {755}},\ \bibinfo {eid} {19}
  (\bibinfo {year} {2012})},\ \Eprint {http://arxiv.org/abs/1202.1478}
  {arXiv:1202.1478 [astro-ph.CO]} \BibitemShut {NoStop}%
\bibitem [{\citenamefont {{Curto}}\ \emph {et~al.}(2012)\citenamefont
  {{Curto}}, \citenamefont {{Martinez-Gonzalez}},\ and\ \citenamefont
  {{Barreiro}}}]{Curto2011}%
  \BibitemOpen
  \bibfield  {author} {\bibinfo {author} {\bibfnamefont {A.}~\bibnamefont
  {{Curto}}}, \bibinfo {author} {\bibfnamefont {E.}~\bibnamefont
  {{Martinez-Gonzalez}}}, \ and\ \bibinfo {author} {\bibfnamefont {R.~B.}\
  \bibnamefont {{Barreiro}}},\ }\href {\doibase
  10.1111/j.1365-2966.2012.21805.x} {\bibfield  {journal} {\bibinfo  {journal}
  {\mnras}\ }\textbf {\bibinfo {volume} {426}},\ \bibinfo {pages} {1361}
  (\bibinfo {year} {2012})},\ \Eprint {http://arxiv.org/abs/1111.3390}
  {arXiv:1111.3390 [astro-ph.CO]} \BibitemShut {NoStop}%
\bibitem [{\citenamefont {Seery}\ \emph {et~al.}(2007)\citenamefont {Seery},
  \citenamefont {Lidsey},\ and\ \citenamefont {Sloth}}]{Seery:2006vu}%
  \BibitemOpen
  \bibfield  {author} {\bibinfo {author} {\bibfnamefont {D.}~\bibnamefont
  {Seery}}, \bibinfo {author} {\bibfnamefont {J.~E.}\ \bibnamefont {Lidsey}}, \
  and\ \bibinfo {author} {\bibfnamefont {M.~S.}\ \bibnamefont {Sloth}},\ }\href
  {\doibase 10.1088/1475-7516/2007/01/027} {\bibfield  {journal} {\bibinfo
  {journal} {JCAP}\ }\textbf {\bibinfo {volume} {0701}},\ \bibinfo {pages}
  {027} (\bibinfo {year} {2007})},\ \Eprint
  {http://arxiv.org/abs/astro-ph/0610210} {arXiv:astro-ph/0610210 [astro-ph]}
  \BibitemShut {NoStop}%
\bibitem [{\citenamefont {Seery}\ \emph {et~al.}(2009)\citenamefont {Seery},
  \citenamefont {Sloth},\ and\ \citenamefont {Vernizzi}}]{Seery:2008ax}%
  \BibitemOpen
  \bibfield  {author} {\bibinfo {author} {\bibfnamefont {D.}~\bibnamefont
  {Seery}}, \bibinfo {author} {\bibfnamefont {M.~S.}\ \bibnamefont {Sloth}}, \
  and\ \bibinfo {author} {\bibfnamefont {F.}~\bibnamefont {Vernizzi}},\ }\href
  {\doibase 10.1088/1475-7516/2009/03/018} {\bibfield  {journal} {\bibinfo
  {journal} {JCAP}\ }\textbf {\bibinfo {volume} {0903}},\ \bibinfo {pages}
  {018} (\bibinfo {year} {2009})},\ \Eprint {http://arxiv.org/abs/0811.3934}
  {arXiv:0811.3934 [astro-ph]} \BibitemShut {NoStop}%
\bibitem [{\citenamefont {Seery}\ and\ \citenamefont
  {Lidsey}(2007)}]{Seery:2006js}%
  \BibitemOpen
  \bibfield  {author} {\bibinfo {author} {\bibfnamefont {D.}~\bibnamefont
  {Seery}}\ and\ \bibinfo {author} {\bibfnamefont {J.~E.}\ \bibnamefont
  {Lidsey}},\ }\href {\doibase 10.1088/1475-7516/2007/01/008} {\bibfield
  {journal} {\bibinfo  {journal} {JCAP}\ }\textbf {\bibinfo {volume} {0701}},\
  \bibinfo {pages} {008} (\bibinfo {year} {2007})},\ \Eprint
  {http://arxiv.org/abs/astro-ph/0611034} {arXiv:astro-ph/0611034 [astro-ph]}
  \BibitemShut {NoStop}%
\bibitem [{\citenamefont {Chen}\ \emph {et~al.}(2009)\citenamefont {Chen},
  \citenamefont {Hu}, \citenamefont {Huang}, \citenamefont {Shiu},\ and\
  \citenamefont {Wang}}]{aChen}%
  \BibitemOpen
  \bibfield  {author} {\bibinfo {author} {\bibfnamefont {X.}~\bibnamefont
  {Chen}}, \bibinfo {author} {\bibfnamefont {B.}~\bibnamefont {Hu}}, \bibinfo
  {author} {\bibfnamefont {M.}~\bibnamefont {Huang}}, \bibinfo {author}
  {\bibfnamefont {G.}~\bibnamefont {Shiu}}, \ and\ \bibinfo {author}
  {\bibfnamefont {Y.}~\bibnamefont {Wang}},\ }\href
  {doi:10.1088/1475-7516/2009/08/008} {\bibfield  {journal} {\bibinfo
  {journal} {Journal of Cosmology and Astroparticle Physics}\ }\textbf
  {\bibinfo {volume} {0908}},\ \bibinfo {pages} {008} (\bibinfo {year}
  {2009})}\BibitemShut {NoStop}%
\bibitem [{\citenamefont {{Pearson}}\ \emph {et~al.}(2012)\citenamefont
  {{Pearson}}, \citenamefont {{Lewis}},\ and\ \citenamefont
  {{Regan}}}]{Pearson}%
  \BibitemOpen
  \bibfield  {author} {\bibinfo {author} {\bibfnamefont {R.}~\bibnamefont
  {{Pearson}}}, \bibinfo {author} {\bibfnamefont {A.}~\bibnamefont {{Lewis}}},
  \ and\ \bibinfo {author} {\bibfnamefont {D.}~\bibnamefont {{Regan}}},\ }\href
  {\doibase 10.1088/1475-7516/2012/03/011} {\bibfield  {journal} {\bibinfo
  {journal} {\jcap}\ }\textbf {\bibinfo {volume} {3}},\ \bibinfo {eid} {011}
  (\bibinfo {year} {2012})},\ \Eprint {http://arxiv.org/abs/1201.1010}
  {arXiv:1201.1010 [astro-ph.CO]} \BibitemShut {NoStop}%
\bibitem [{\citenamefont {Okamoto}\ and\ \citenamefont
  {Hu}(2002)}]{Okamoto:2002ik}%
  \BibitemOpen
  \bibfield  {author} {\bibinfo {author} {\bibfnamefont {T.}~\bibnamefont
  {Okamoto}}\ and\ \bibinfo {author} {\bibfnamefont {W.}~\bibnamefont {Hu}},\
  }\href {\doibase 10.1103/PhysRevD.66.063008} {\bibfield  {journal} {\bibinfo
  {journal} {Phys.Rev.}\ }\textbf {\bibinfo {volume} {D66}},\ \bibinfo {pages}
  {063008} (\bibinfo {year} {2002})},\ \Eprint
  {http://arxiv.org/abs/astro-ph/0206155} {arXiv:astro-ph/0206155 [astro-ph]}
  \BibitemShut {NoStop}%
\bibitem [{\citenamefont {Boubekeur}\ and\ \citenamefont
  {Lyth}(2006)}]{Boubekeur:2005fj}%
  \BibitemOpen
  \bibfield  {author} {\bibinfo {author} {\bibfnamefont {L.}~\bibnamefont
  {Boubekeur}}\ and\ \bibinfo {author} {\bibfnamefont {D.}~\bibnamefont
  {Lyth}},\ }\href {\doibase 10.1103/PhysRevD.73.021301} {\bibfield  {journal}
  {\bibinfo  {journal} {Phys.Rev.}\ }\textbf {\bibinfo {volume} {D73}},\
  \bibinfo {pages} {021301} (\bibinfo {year} {2006})},\ \Eprint
  {http://arxiv.org/abs/astro-ph/0504046} {arXiv:astro-ph/0504046 [astro-ph]}
  \BibitemShut {NoStop}%
\bibitem [{\citenamefont {Sasaki}\ \emph {et~al.}(2006)\citenamefont {Sasaki},
  \citenamefont {Valiviita},\ and\ \citenamefont {Wands}}]{Sasaki:2006kq}%
  \BibitemOpen
  \bibfield  {author} {\bibinfo {author} {\bibfnamefont {M.}~\bibnamefont
  {Sasaki}}, \bibinfo {author} {\bibfnamefont {J.}~\bibnamefont {Valiviita}}, \
  and\ \bibinfo {author} {\bibfnamefont {D.}~\bibnamefont {Wands}},\ }\href
  {\doibase 10.1103/PhysRevD.74.103003} {\bibfield  {journal} {\bibinfo
  {journal} {Phys.Rev.}\ }\textbf {\bibinfo {volume} {D74}},\ \bibinfo {pages}
  {103003} (\bibinfo {year} {2006})},\ \Eprint
  {http://arxiv.org/abs/astro-ph/0607627} {arXiv:astro-ph/0607627 [astro-ph]}
  \BibitemShut {NoStop}%
\bibitem [{\citenamefont {{Bennett}}\ \emph {et~al.}(2013)\citenamefont
  {{Bennett}}, \citenamefont {{Larson}}, \citenamefont {{Weiland}},
  \citenamefont {{Jarosik}}, \citenamefont {{Hinshaw}}, \citenamefont
  {{Odegard}}, \citenamefont {{Smith}}, \citenamefont {{Hill}}, \citenamefont
  {{Gold}}, \citenamefont {{Halpern}}, \citenamefont {{Komatsu}}, \citenamefont
  {{Nolta}}, \citenamefont {{Page}}, \citenamefont {{Spergel}}, \citenamefont
  {{Wollack}}, \citenamefont {{Dunkley}}, \citenamefont {{Kogut}},
  \citenamefont {{Limon}}, \citenamefont {{Meyer}}, \citenamefont {{Tucker}},\
  and\ \citenamefont {{Wright}}}]{WMAP9}%
  \BibitemOpen
  \bibfield  {author} {\bibinfo {author} {\bibfnamefont {C.~L.}\ \bibnamefont
  {{Bennett}}}, \bibinfo {author} {\bibfnamefont {D.}~\bibnamefont {{Larson}}},
  \bibinfo {author} {\bibfnamefont {J.~L.}\ \bibnamefont {{Weiland}}}, \bibinfo
  {author} {\bibfnamefont {N.}~\bibnamefont {{Jarosik}}}, \bibinfo {author}
  {\bibfnamefont {G.}~\bibnamefont {{Hinshaw}}}, \bibinfo {author}
  {\bibfnamefont {N.}~\bibnamefont {{Odegard}}}, \bibinfo {author}
  {\bibfnamefont {K.~M.}\ \bibnamefont {{Smith}}}, \bibinfo {author}
  {\bibfnamefont {R.~S.}\ \bibnamefont {{Hill}}}, \bibinfo {author}
  {\bibfnamefont {B.}~\bibnamefont {{Gold}}}, \bibinfo {author} {\bibfnamefont
  {M.}~\bibnamefont {{Halpern}}}, \bibinfo {author} {\bibfnamefont
  {E.}~\bibnamefont {{Komatsu}}}, \bibinfo {author} {\bibfnamefont {M.~R.}\
  \bibnamefont {{Nolta}}}, \bibinfo {author} {\bibfnamefont {L.}~\bibnamefont
  {{Page}}}, \bibinfo {author} {\bibfnamefont {D.~N.}\ \bibnamefont
  {{Spergel}}}, \bibinfo {author} {\bibfnamefont {E.}~\bibnamefont
  {{Wollack}}}, \bibinfo {author} {\bibfnamefont {J.}~\bibnamefont
  {{Dunkley}}}, \bibinfo {author} {\bibfnamefont {A.}~\bibnamefont {{Kogut}}},
  \bibinfo {author} {\bibfnamefont {M.}~\bibnamefont {{Limon}}}, \bibinfo
  {author} {\bibfnamefont {S.~S.}\ \bibnamefont {{Meyer}}}, \bibinfo {author}
  {\bibfnamefont {G.~S.}\ \bibnamefont {{Tucker}}}, \ and\ \bibinfo {author}
  {\bibfnamefont {E.~L.}\ \bibnamefont {{Wright}}},\ }\href {\doibase
  10.1088/0067-0049/208/2/20} {\bibfield  {journal} {\bibinfo  {journal}
  {\apjs}\ }\textbf {\bibinfo {volume} {208}},\ \bibinfo {eid} {20} (\bibinfo
  {year} {2013})},\ \Eprint {http://arxiv.org/abs/1212.5225} {arXiv:1212.5225
  [astro-ph.CO]} \BibitemShut {NoStop}%
\bibitem [{\citenamefont {Hanson}\ \emph {et~al.}(2009)\citenamefont {Hanson},
  \citenamefont {Smith}, \citenamefont {Challinor},\ and\ \citenamefont
  {Liguori}}]{Hanson:2009kg}%
  \BibitemOpen
  \bibfield  {author} {\bibinfo {author} {\bibfnamefont {D.}~\bibnamefont
  {Hanson}}, \bibinfo {author} {\bibfnamefont {K.~M.}\ \bibnamefont {Smith}},
  \bibinfo {author} {\bibfnamefont {A.}~\bibnamefont {Challinor}}, \ and\
  \bibinfo {author} {\bibfnamefont {M.}~\bibnamefont {Liguori}},\ }\href
  {\doibase 10.1103/PhysRevD.80.083004} {\bibfield  {journal} {\bibinfo
  {journal} {Physical Review}\ }\textbf {\bibinfo {volume} {D80}},\ \bibinfo
  {pages} {083004} (\bibinfo {year} {2009})},\ \Eprint
  {http://arxiv.org/abs/0905.4732} {arXiv:0905.4732 [astro-ph.CO]} \BibitemShut
  {NoStop}%
\bibitem [{\citenamefont {{Komatsu}}\ \emph {et~al.}(2009)\citenamefont
  {{Komatsu}}, \citenamefont {{Dunkley}}, \citenamefont {{Nolta}},
  \citenamefont {{Bennett}}, \citenamefont {{Gold}}, \citenamefont {{Hinshaw}},
  \citenamefont {{Jarosik}}, \citenamefont {{Larson}}, \citenamefont {{Limon}},
  \citenamefont {{Page}}, \citenamefont {{Spergel}}, \citenamefont {{Halpern}},
  \citenamefont {{Hill}}, \citenamefont {{Kogut}}, \citenamefont {{Meyer}},
  \citenamefont {{Tucker}}, \citenamefont {{Weiland}}, \citenamefont
  {{Wollack}},\ and\ \citenamefont {{Wright}}}]{WMAP5}%
  \BibitemOpen
  \bibfield  {author} {\bibinfo {author} {\bibfnamefont {E.}~\bibnamefont
  {{Komatsu}}}, \bibinfo {author} {\bibfnamefont {J.}~\bibnamefont
  {{Dunkley}}}, \bibinfo {author} {\bibfnamefont {M.~R.}\ \bibnamefont
  {{Nolta}}}, \bibinfo {author} {\bibfnamefont {C.~L.}\ \bibnamefont
  {{Bennett}}}, \bibinfo {author} {\bibfnamefont {B.}~\bibnamefont {{Gold}}},
  \bibinfo {author} {\bibfnamefont {G.}~\bibnamefont {{Hinshaw}}}, \bibinfo
  {author} {\bibfnamefont {N.}~\bibnamefont {{Jarosik}}}, \bibinfo {author}
  {\bibfnamefont {D.}~\bibnamefont {{Larson}}}, \bibinfo {author}
  {\bibfnamefont {M.}~\bibnamefont {{Limon}}}, \bibinfo {author} {\bibfnamefont
  {L.}~\bibnamefont {{Page}}}, \bibinfo {author} {\bibfnamefont {D.~N.}\
  \bibnamefont {{Spergel}}}, \bibinfo {author} {\bibfnamefont {M.}~\bibnamefont
  {{Halpern}}}, \bibinfo {author} {\bibfnamefont {R.~S.}\ \bibnamefont
  {{Hill}}}, \bibinfo {author} {\bibfnamefont {A.}~\bibnamefont {{Kogut}}},
  \bibinfo {author} {\bibfnamefont {S.~S.}\ \bibnamefont {{Meyer}}}, \bibinfo
  {author} {\bibfnamefont {G.~S.}\ \bibnamefont {{Tucker}}}, \bibinfo {author}
  {\bibfnamefont {J.~L.}\ \bibnamefont {{Weiland}}}, \bibinfo {author}
  {\bibfnamefont {E.}~\bibnamefont {{Wollack}}}, \ and\ \bibinfo {author}
  {\bibfnamefont {E.~L.}\ \bibnamefont {{Wright}}},\ }\href {\doibase
  10.1088/0067-0049/180/2/330} {\bibfield  {journal} {\bibinfo  {journal}
  {\apjs}\ }\textbf {\bibinfo {volume} {180}},\ \bibinfo {pages} {330}
  (\bibinfo {year} {2009})},\ \Eprint {http://arxiv.org/abs/0803.0547}
  {arXiv:0803.0547} \BibitemShut {NoStop}%
\bibitem [{\citenamefont {{Tegmark}}\ and\ \citenamefont
  {{Efstathiou}}(1996)}]{TegEfsth96}%
  \BibitemOpen
  \bibfield  {author} {\bibinfo {author} {\bibfnamefont {M.}~\bibnamefont
  {{Tegmark}}}\ and\ \bibinfo {author} {\bibfnamefont {G.}~\bibnamefont
  {{Efstathiou}}},\ }\href@noop {} {\bibfield  {journal} {\bibinfo  {journal}
  {\mnras}\ }\textbf {\bibinfo {volume} {281}},\ \bibinfo {pages} {1297}
  (\bibinfo {year} {1996})},\ \Eprint
  {http://arxiv.org/abs/arXiv:astro-ph/9507009} {arXiv:astro-ph/9507009}
  \BibitemShut {NoStop}%
\bibitem [{\citenamefont {{Nolta}}\ \emph {et~al.}(2009)\citenamefont
  {{Nolta}}, \citenamefont {{Dunkley}}, \citenamefont {{Hill}}, \citenamefont
  {{Hinshaw}}, \citenamefont {{Komatsu}}, \citenamefont {{Larson}},
  \citenamefont {{Page}}, \citenamefont {{Spergel}}, \citenamefont {{Bennett}},
  \citenamefont {{Gold}}, \citenamefont {{Jarosik}}, \citenamefont {{Odegard}},
  \citenamefont {{Weiland}}, \citenamefont {{Wollack}}, \citenamefont
  {{Halpern}}, \citenamefont {{Kogut}}, \citenamefont {{Limon}}, \citenamefont
  {{Meyer}}, \citenamefont {{Tucker}},\ and\ \citenamefont
  {{Wright}}}]{Nolta08}%
  \BibitemOpen
  \bibfield  {author} {\bibinfo {author} {\bibfnamefont {M.~R.}\ \bibnamefont
  {{Nolta}}}, \bibinfo {author} {\bibfnamefont {J.}~\bibnamefont {{Dunkley}}},
  \bibinfo {author} {\bibfnamefont {R.~S.}\ \bibnamefont {{Hill}}}, \bibinfo
  {author} {\bibfnamefont {G.}~\bibnamefont {{Hinshaw}}}, \bibinfo {author}
  {\bibfnamefont {E.}~\bibnamefont {{Komatsu}}}, \bibinfo {author}
  {\bibfnamefont {D.}~\bibnamefont {{Larson}}}, \bibinfo {author}
  {\bibfnamefont {L.}~\bibnamefont {{Page}}}, \bibinfo {author} {\bibfnamefont
  {D.~N.}\ \bibnamefont {{Spergel}}}, \bibinfo {author} {\bibfnamefont {C.~L.}\
  \bibnamefont {{Bennett}}}, \bibinfo {author} {\bibfnamefont {B.}~\bibnamefont
  {{Gold}}}, \bibinfo {author} {\bibfnamefont {N.}~\bibnamefont {{Jarosik}}},
  \bibinfo {author} {\bibfnamefont {N.}~\bibnamefont {{Odegard}}}, \bibinfo
  {author} {\bibfnamefont {J.~L.}\ \bibnamefont {{Weiland}}}, \bibinfo {author}
  {\bibfnamefont {E.}~\bibnamefont {{Wollack}}}, \bibinfo {author}
  {\bibfnamefont {M.}~\bibnamefont {{Halpern}}}, \bibinfo {author}
  {\bibfnamefont {A.}~\bibnamefont {{Kogut}}}, \bibinfo {author} {\bibfnamefont
  {M.}~\bibnamefont {{Limon}}}, \bibinfo {author} {\bibfnamefont {S.~S.}\
  \bibnamefont {{Meyer}}}, \bibinfo {author} {\bibfnamefont {G.~S.}\
  \bibnamefont {{Tucker}}}, \ and\ \bibinfo {author} {\bibfnamefont {E.~L.}\
  \bibnamefont {{Wright}}},\ }\href {\doibase 10.1088/0067-0049/180/2/296}
  {\bibfield  {journal} {\bibinfo  {journal} {\apjs}\ }\textbf {\bibinfo
  {volume} {180}},\ \bibinfo {pages} {296} (\bibinfo {year} {2009})},\ \Eprint
  {http://arxiv.org/abs/0803.0593} {arXiv:0803.0593} \BibitemShut {NoStop}%
\bibitem [{\citenamefont {{Lyth}}\ and\ \citenamefont
  {{Rodriguez}}(2005)}]{LythRod2005}%
  \BibitemOpen
  \bibfield  {author} {\bibinfo {author} {\bibfnamefont {D.~H.}\ \bibnamefont
  {{Lyth}}}\ and\ \bibinfo {author} {\bibfnamefont {Y.}~\bibnamefont
  {{Rodriguez}}},\ }\href {\doibase 10.1103/PhysRevLett.95.121302} {\bibfield
  {journal} {\bibinfo  {journal} {Physical Review Letters}\ }\textbf {\bibinfo
  {volume} {95}},\ \bibinfo {eid} {121302} (\bibinfo {year} {2005})},\ \Eprint
  {http://arxiv.org/abs/arXiv:astro-ph/0504045} {arXiv:astro-ph/0504045}
  \BibitemShut {NoStop}%
\bibitem [{\citenamefont {{Enqvist}}\ and\ \citenamefont
  {{Sloth}}(2002)}]{EnqvistSloth2001}%
  \BibitemOpen
  \bibfield  {author} {\bibinfo {author} {\bibfnamefont {K.}~\bibnamefont
  {{Enqvist}}}\ and\ \bibinfo {author} {\bibfnamefont {M.~S.}\ \bibnamefont
  {{Sloth}}},\ }\href {\doibase 10.1016/S0550-3213(02)00043-3} {\bibfield
  {journal} {\bibinfo  {journal} {Nuclear Physics B}\ }\textbf {\bibinfo
  {volume} {626}},\ \bibinfo {pages} {395} (\bibinfo {year} {2002})},\ \Eprint
  {http://arxiv.org/abs/arXiv:hep-ph/0109214} {arXiv:hep-ph/0109214}
  \BibitemShut {NoStop}%
\bibitem [{\citenamefont {{Lyth}}\ and\ \citenamefont
  {{Wands}}(2002)}]{LythWands2001}%
  \BibitemOpen
  \bibfield  {author} {\bibinfo {author} {\bibfnamefont {D.~H.}\ \bibnamefont
  {{Lyth}}}\ and\ \bibinfo {author} {\bibfnamefont {D.}~\bibnamefont
  {{Wands}}},\ }\href {\doibase 10.1016/S0370-2693(01)01366-1} {\bibfield
  {journal} {\bibinfo  {journal} {Physics Letters B}\ }\textbf {\bibinfo
  {volume} {524}},\ \bibinfo {pages} {5} (\bibinfo {year} {2002})},\ \Eprint
  {http://arxiv.org/abs/arXiv:hep-ph/0110002} {arXiv:hep-ph/0110002}
  \BibitemShut {NoStop}%
\bibitem [{\citenamefont {{Moroi}}\ and\ \citenamefont
  {{Takahashi}}(2001)}]{Moroi2001}%
  \BibitemOpen
  \bibfield  {author} {\bibinfo {author} {\bibfnamefont {T.}~\bibnamefont
  {{Moroi}}}\ and\ \bibinfo {author} {\bibfnamefont {T.}~\bibnamefont
  {{Takahashi}}},\ }\href {\doibase 10.1016/S0370-2693(01)01295-3} {\bibfield
  {journal} {\bibinfo  {journal} {Physics Letters B}\ }\textbf {\bibinfo
  {volume} {522}},\ \bibinfo {pages} {215} (\bibinfo {year} {2001})},\ \Eprint
  {http://arxiv.org/abs/arXiv:hep-ph/0110096} {arXiv:hep-ph/0110096}
  \BibitemShut {NoStop}%
\bibitem [{\citenamefont {Salopek}\ and\ \citenamefont {Bond}(1990)}]{Salopek}%
  \BibitemOpen
  \bibfield  {author} {\bibinfo {author} {\bibfnamefont {D.~S.}\ \bibnamefont
  {Salopek}}\ and\ \bibinfo {author} {\bibfnamefont {J.~R.}\ \bibnamefont
  {Bond}},\ }\href {\doibase 10.1103/PhysRevD.42.3936} {\bibfield  {journal}
  {\bibinfo  {journal} {\prd}\ }\textbf {\bibinfo {volume} {42}},\ \bibinfo
  {pages} {3936} (\bibinfo {year} {1990})}\BibitemShut {NoStop}%
\bibitem [{\citenamefont {Alishahiha}\ \emph {et~al.}(2004)\citenamefont
  {Alishahiha}, \citenamefont {Silverstein},\ and\ \citenamefont
  {Tong}}]{AlishahihaSilversteinTong2004}%
  \BibitemOpen
  \bibfield  {author} {\bibinfo {author} {\bibfnamefont {M.}~\bibnamefont
  {Alishahiha}}, \bibinfo {author} {\bibfnamefont {E.}~\bibnamefont
  {Silverstein}}, \ and\ \bibinfo {author} {\bibfnamefont {D.}~\bibnamefont
  {Tong}},\ }\href {\doibase 10.1103/PhysRevD.70.123505} {\bibfield  {journal}
  {\bibinfo  {journal} {Physical Review}\ }\textbf {\bibinfo {volume} {D70}},\
  \bibinfo {pages} {123505} (\bibinfo {year} {2004})},\ \Eprint
  {http://arxiv.org/abs/hep-th/0404084} {arXiv:hep-th/0404084} \BibitemShut
  {NoStop}%
\bibitem [{\citenamefont {Chen}\ \emph
  {et~al.}(2007{\natexlab{a}})\citenamefont {Chen}, \citenamefont {Easther},\
  and\ \citenamefont {Lim}}]{ChenetAl2007}%
  \BibitemOpen
  \bibfield  {author} {\bibinfo {author} {\bibfnamefont {X.}~\bibnamefont
  {Chen}}, \bibinfo {author} {\bibfnamefont {R.}~\bibnamefont {Easther}}, \
  and\ \bibinfo {author} {\bibfnamefont {E.~A.}\ \bibnamefont {Lim}},\
  }\href@noop {} {\bibfield  {journal} {\bibinfo  {journal} {Journal of
  Cosmology and Astroparticle Physics}\ }\textbf {\bibinfo {volume} {0706}},\
  \bibinfo {pages} {023} (\bibinfo {year} {2007}{\natexlab{a}})},\ \Eprint
  {http://arxiv.org/abs/astro-ph/0611645} {arXiv:astro-ph/0611645} \BibitemShut
  {NoStop}%
\bibitem [{\citenamefont {Chen}\ \emph
  {et~al.}(2007{\natexlab{b}})\citenamefont {Chen}, \citenamefont {Huang},
  \citenamefont {Kachru},\ and\ \citenamefont {Shiu}}]{0605045}%
  \BibitemOpen
  \bibfield  {author} {\bibinfo {author} {\bibfnamefont {X.}~\bibnamefont
  {Chen}}, \bibinfo {author} {\bibfnamefont {M.-x.}\ \bibnamefont {Huang}},
  \bibinfo {author} {\bibfnamefont {S.}~\bibnamefont {Kachru}}, \ and\ \bibinfo
  {author} {\bibfnamefont {G.}~\bibnamefont {Shiu}},\ }\href@noop {} {\bibfield
   {journal} {\bibinfo  {journal} {Journal of Cosmology and Astroparticle
  Physics}\ }\textbf {\bibinfo {volume} {0701}},\ \bibinfo {pages} {002}
  (\bibinfo {year} {2007}{\natexlab{b}})},\ \Eprint
  {http://arxiv.org/abs/hep-th/0605045} {arXiv:hep-th/0605045} \BibitemShut
  {NoStop}%
\bibitem [{\citenamefont {{Chen}}\ and\ \citenamefont
  {{Wang}}(2010)}]{ChenWang2009}%
  \BibitemOpen
  \bibfield  {author} {\bibinfo {author} {\bibfnamefont {X.}~\bibnamefont
  {{Chen}}}\ and\ \bibinfo {author} {\bibfnamefont {Y.}~\bibnamefont
  {{Wang}}},\ }\href {\doibase 10.1103/PhysRevD.81.063511} {\bibfield
  {journal} {\bibinfo  {journal} {\prd}\ }\textbf {\bibinfo {volume} {81}},\
  \bibinfo {eid} {063511} (\bibinfo {year} {2010})},\ \Eprint
  {http://arxiv.org/abs/0909.0496} {arXiv:0909.0496 [astro-ph.CO]} \BibitemShut
  {NoStop}%
\bibitem [{\citenamefont {Holman}\ and\ \citenamefont
  {Tolley}(2008)}]{HolmanTolley2008}%
  \BibitemOpen
  \bibfield  {author} {\bibinfo {author} {\bibfnamefont {R.}~\bibnamefont
  {Holman}}\ and\ \bibinfo {author} {\bibfnamefont {A.~J.}\ \bibnamefont
  {Tolley}},\ }\href {\doibase 10.1088/1475-7516/2008/05/001} {\bibfield
  {journal} {\bibinfo  {journal} {Journal of Cosmology and Astroparticle
  Physics}\ }\textbf {\bibinfo {volume} {0805}},\ \bibinfo {pages} {001}
  (\bibinfo {year} {2008})},\ \Eprint {http://arxiv.org/abs/0710.1302}
  {arXiv:0710.1302 [hep-th]} \BibitemShut {NoStop}%
\bibitem [{\citenamefont {{Ashoorioon}}\ and\ \citenamefont
  {{Shiu}}(2011)}]{Ashoorioon}%
  \BibitemOpen
  \bibfield  {author} {\bibinfo {author} {\bibfnamefont {A.}~\bibnamefont
  {{Ashoorioon}}}\ and\ \bibinfo {author} {\bibfnamefont {G.}~\bibnamefont
  {{Shiu}}},\ }\href {\doibase 10.1088/1475-7516/2011/03/025} {\bibfield
  {journal} {\bibinfo  {journal} {\jcap}\ }\textbf {\bibinfo {volume} {3}},\
  \bibinfo {eid} {025} (\bibinfo {year} {2011})},\ \Eprint
  {http://arxiv.org/abs/1012.3392} {arXiv:1012.3392 [astro-ph.CO]} \BibitemShut
  {NoStop}%
\bibitem [{\citenamefont {Argueso}\ \emph {et~al.}(2006)\citenamefont
  {Argueso}, \citenamefont {Sanz}, \citenamefont {Barreiro}, \citenamefont
  {Herranz},\ and\ \citenamefont {Gonzalez-Nuevo}}]{Argueso:2006gu}%
  \BibitemOpen
  \bibfield  {author} {\bibinfo {author} {\bibfnamefont {F.}~\bibnamefont
  {Argueso}}, \bibinfo {author} {\bibfnamefont {J.}~\bibnamefont {Sanz}},
  \bibinfo {author} {\bibfnamefont {R.}~\bibnamefont {Barreiro}}, \bibinfo
  {author} {\bibfnamefont {D.}~\bibnamefont {Herranz}}, \ and\ \bibinfo
  {author} {\bibfnamefont {J.}~\bibnamefont {Gonzalez-Nuevo}},\ }\href
  {\doibase 10.1111/j.1365-2966.2006.11041.x} {\bibfield  {journal} {\bibinfo
  {journal} {\mnras}\ }\textbf {\bibinfo {volume} {373}},\ \bibinfo {pages}
  {311} (\bibinfo {year} {2006})},\ \Eprint
  {http://arxiv.org/abs/astro-ph/0609348} {arXiv:astro-ph/0609348 [astro-ph]}
  \BibitemShut {NoStop}%
\bibitem [{\citenamefont {De~Zotti}\ \emph {et~al.}(2005)\citenamefont
  {De~Zotti}, \citenamefont {Ricci}, \citenamefont {Mesa}, \citenamefont
  {Silva}, \citenamefont {Mazzotta} \emph {et~al.}}]{DeZotti:2004mn}%
  \BibitemOpen
  \bibfield  {author} {\bibinfo {author} {\bibfnamefont {G.}~\bibnamefont
  {De~Zotti}}, \bibinfo {author} {\bibfnamefont {R.}~\bibnamefont {Ricci}},
  \bibinfo {author} {\bibfnamefont {D.}~\bibnamefont {Mesa}}, \bibinfo {author}
  {\bibfnamefont {L.}~\bibnamefont {Silva}}, \bibinfo {author} {\bibfnamefont
  {P.}~\bibnamefont {Mazzotta}},  \emph {et~al.},\ }\href {\doibase
  10.1051/0004-6361:20042108} {\bibfield  {journal} {\bibinfo  {journal}
  {Astron.Astrophys.}\ }\textbf {\bibinfo {volume} {431}},\ \bibinfo {pages}
  {893} (\bibinfo {year} {2005})},\ \Eprint
  {http://arxiv.org/abs/astro-ph/0410709} {arXiv:astro-ph/0410709 [astro-ph]}
  \BibitemShut {NoStop}%
\bibitem [{\citenamefont {Curto}\ \emph {et~al.}(2013)\citenamefont {Curto},
  \citenamefont {Tucci}, \citenamefont {Gonzalez-Nuevo}, \citenamefont
  {Toffolatti}, \citenamefont {Martinez-Gonzalez} \emph
  {et~al.}}]{Curto:2013hi}%
  \BibitemOpen
  \bibfield  {author} {\bibinfo {author} {\bibfnamefont {A.}~\bibnamefont
  {Curto}}, \bibinfo {author} {\bibfnamefont {M.}~\bibnamefont {Tucci}},
  \bibinfo {author} {\bibfnamefont {J.}~\bibnamefont {Gonzalez-Nuevo}},
  \bibinfo {author} {\bibfnamefont {L.}~\bibnamefont {Toffolatti}}, \bibinfo
  {author} {\bibfnamefont {E.}~\bibnamefont {Martinez-Gonzalez}},  \emph
  {et~al.},\ }\href {\doibase 10.1093/mnras/stt511} {\bibfield  {journal}
  {\bibinfo  {journal} {Mon.Not.Roy.Astron.Soc.}\ }\textbf {\bibinfo {volume}
  {432}},\ \bibinfo {pages} {728} (\bibinfo {year} {2013})},\ \Eprint
  {http://arxiv.org/abs/1301.1544} {arXiv:1301.1544 [astro-ph.CO]} \BibitemShut
  {NoStop}%
\bibitem [{\citenamefont {{Desjacques}}\ and\ \citenamefont
  {{Seljak}}(2010)}]{DS2010}%
  \BibitemOpen
  \bibfield  {author} {\bibinfo {author} {\bibfnamefont {V.}~\bibnamefont
  {{Desjacques}}}\ and\ \bibinfo {author} {\bibfnamefont {U.}~\bibnamefont
  {{Seljak}}},\ }\href {\doibase 10.1155/2010/908640} {\bibfield  {journal}
  {\bibinfo  {journal} {Advances in Astronomy}\ }\textbf {\bibinfo {volume}
  {2010}} (\bibinfo {year} {2010}),\ 10.1155/2010/908640},\ \Eprint
  {http://arxiv.org/abs/1006.4763} {arXiv:1006.4763 [astro-ph.CO]} \BibitemShut
  {NoStop}%
\bibitem [{\citenamefont {{Smidt}}\ \emph {et~al.}(2010)\citenamefont
  {{Smidt}}, \citenamefont {{Amblard}}, \citenamefont {{Byrnes}}, \citenamefont
  {{Cooray}}, \citenamefont {{Heavens}},\ and\ \citenamefont
  {{Munshi}}}]{Cooray2}%
  \BibitemOpen
  \bibfield  {author} {\bibinfo {author} {\bibfnamefont {J.}~\bibnamefont
  {{Smidt}}}, \bibinfo {author} {\bibfnamefont {A.}~\bibnamefont {{Amblard}}},
  \bibinfo {author} {\bibfnamefont {C.~T.}\ \bibnamefont {{Byrnes}}}, \bibinfo
  {author} {\bibfnamefont {A.}~\bibnamefont {{Cooray}}}, \bibinfo {author}
  {\bibfnamefont {A.}~\bibnamefont {{Heavens}}}, \ and\ \bibinfo {author}
  {\bibfnamefont {D.}~\bibnamefont {{Munshi}}},\ }\href {\doibase
  10.1103/PhysRevD.81.123007} {\bibfield  {journal} {\bibinfo  {journal}
  {\prd}\ }\textbf {\bibinfo {volume} {81}},\ \bibinfo {eid} {123007} (\bibinfo
  {year} {2010})},\ \Eprint {http://arxiv.org/abs/1004.1409} {arXiv:1004.1409
  [astro-ph.CO]} \BibitemShut {NoStop}%
\bibitem [{\citenamefont {Chen}\ \emph {et~al.}(2006)\citenamefont {Chen},
  \citenamefont {{Huang}},\ and\ \citenamefont {{Shiu}}}]{XINGANG_t_NL}%
  \BibitemOpen
  \bibfield  {author} {\bibinfo {author} {\bibfnamefont {X.}~\bibnamefont
  {Chen}}, \bibinfo {author} {\bibfnamefont {M.}~\bibnamefont {{Huang}}}, \
  and\ \bibinfo {author} {\bibfnamefont {G.}~\bibnamefont {{Shiu}}},\ }\href
  {\doibase 10.1103/PhysRevD.74.121301} {\bibfield  {journal} {\bibinfo
  {journal} {\prd}\ }\textbf {\bibinfo {volume} {74}},\ \bibinfo {pages}
  {121301} (\bibinfo {year} {2006})}\BibitemShut {NoStop}%
\bibitem [{\citenamefont {{Arroja}}\ \emph {et~al.}(2009)\citenamefont
  {{Arroja}}, \citenamefont {{Mizuno}}, \citenamefont {{Koyama}},\ and\
  \citenamefont {{Tanaka}}}]{Arroja2009}%
  \BibitemOpen
  \bibfield  {author} {\bibinfo {author} {\bibfnamefont {F.}~\bibnamefont
  {{Arroja}}}, \bibinfo {author} {\bibfnamefont {S.}~\bibnamefont {{Mizuno}}},
  \bibinfo {author} {\bibfnamefont {K.}~\bibnamefont {{Koyama}}}, \ and\
  \bibinfo {author} {\bibfnamefont {T.}~\bibnamefont {{Tanaka}}},\ }\href
  {\doibase 10.1103/PhysRevD.80.043527} {\bibfield  {journal} {\bibinfo
  {journal} {\prd}\ }\textbf {\bibinfo {volume} {80}},\ \bibinfo {eid} {043527}
  (\bibinfo {year} {2009})},\ \Eprint {http://arxiv.org/abs/0905.3641}
  {arXiv:0905.3641 [hep-th]} \BibitemShut {NoStop}%
\bibitem [{\citenamefont {{Suyama}}\ and\ \citenamefont
  {{Yamaguchi}}(2008)}]{Suyama2008}%
  \BibitemOpen
  \bibfield  {author} {\bibinfo {author} {\bibfnamefont {T.}~\bibnamefont
  {{Suyama}}}\ and\ \bibinfo {author} {\bibfnamefont {M.}~\bibnamefont
  {{Yamaguchi}}},\ }\href {\doibase 10.1103/PhysRevD.77.023505} {\bibfield
  {journal} {\bibinfo  {journal} {\prd}\ }\textbf {\bibinfo {volume} {77}},\
  \bibinfo {eid} {023505} (\bibinfo {year} {2008})},\ \Eprint
  {http://arxiv.org/abs/0709.2545} {arXiv:0709.2545} \BibitemShut {NoStop}%
\bibitem [{\citenamefont {Smith}\ \emph {et~al.}(2011)\citenamefont {Smith},
  \citenamefont {LoVerde},\ and\ \citenamefont {Zaldarriaga}}]{Smith:2011if}%
  \BibitemOpen
  \bibfield  {author} {\bibinfo {author} {\bibfnamefont {K.~M.}\ \bibnamefont
  {Smith}}, \bibinfo {author} {\bibfnamefont {M.}~\bibnamefont {LoVerde}}, \
  and\ \bibinfo {author} {\bibfnamefont {M.}~\bibnamefont {Zaldarriaga}},\
  }\href {\doibase 10.1103/PhysRevLett.107.191301} {\bibfield  {journal}
  {\bibinfo  {journal} {Phys.Rev.Lett.}\ }\textbf {\bibinfo {volume} {107}},\
  \bibinfo {pages} {191301} (\bibinfo {year} {2011})},\ \Eprint
  {http://arxiv.org/abs/1108.1805} {arXiv:1108.1805 [astro-ph.CO]} \BibitemShut
  {NoStop}%
\bibitem [{\citenamefont {Assassi}\ \emph {et~al.}(2012)\citenamefont
  {Assassi}, \citenamefont {Baumann},\ and\ \citenamefont
  {Green}}]{Assassi:2012zq}%
  \BibitemOpen
  \bibfield  {author} {\bibinfo {author} {\bibfnamefont {V.}~\bibnamefont
  {Assassi}}, \bibinfo {author} {\bibfnamefont {D.}~\bibnamefont {Baumann}}, \
  and\ \bibinfo {author} {\bibfnamefont {D.}~\bibnamefont {Green}},\ }\href
  {\doibase 10.1088/1475-7516/2012/11/047} {\bibfield  {journal} {\bibinfo
  {journal} {JCAP}\ }\textbf {\bibinfo {volume} {1211}},\ \bibinfo {pages}
  {047} (\bibinfo {year} {2012})},\ \Eprint {http://arxiv.org/abs/1204.4207}
  {arXiv:1204.4207 [hep-th]} \BibitemShut {NoStop}%
\bibitem [{\citenamefont {Kogo}\ and\ \citenamefont {Komatsu}(2006)}]{0602099}%
  \BibitemOpen
  \bibfield  {author} {\bibinfo {author} {\bibfnamefont {N.}~\bibnamefont
  {Kogo}}\ and\ \bibinfo {author} {\bibfnamefont {E.}~\bibnamefont {Komatsu}},\
  }\href {\doibase 10.1103/PhysRevD.73.083007} {\bibfield  {journal} {\bibinfo
  {journal} {\prd}\ }\textbf {\bibinfo {volume} {73}},\ \bibinfo {pages}
  {083007} (\bibinfo {year} {2006})},\ \Eprint
  {http://arxiv.org/abs/astro-ph/0602099} {arXiv:astro-ph/0602099} \BibitemShut
  {NoStop}%
\bibitem [{\citenamefont {{Planck Collaboration}}\ \emph
  {et~al.}(2013)\citenamefont {{Planck Collaboration}}, \citenamefont {{Ade}},
  \citenamefont {{Aghanim}}, \citenamefont {{Armitage-Caplan}}, \citenamefont
  {{Arnaud}}, \citenamefont {{Ashdown}}, \citenamefont {{Atrio-Barandela}},
  \citenamefont {{Aumont}}, \citenamefont {{Baccigalupi}}, \citenamefont
  {{Banday}},\ and\ \citenamefont {et~al.}}]{PlancknonG}%
  \BibitemOpen
  \bibfield  {author} {\bibinfo {author} {\bibnamefont {{Planck
  Collaboration}}}, \bibinfo {author} {\bibfnamefont {P.~A.~R.}\ \bibnamefont
  {{Ade}}}, \bibinfo {author} {\bibfnamefont {N.}~\bibnamefont {{Aghanim}}},
  \bibinfo {author} {\bibfnamefont {C.}~\bibnamefont {{Armitage-Caplan}}},
  \bibinfo {author} {\bibfnamefont {M.}~\bibnamefont {{Arnaud}}}, \bibinfo
  {author} {\bibfnamefont {M.}~\bibnamefont {{Ashdown}}}, \bibinfo {author}
  {\bibfnamefont {F.}~\bibnamefont {{Atrio-Barandela}}}, \bibinfo {author}
  {\bibfnamefont {J.}~\bibnamefont {{Aumont}}}, \bibinfo {author}
  {\bibfnamefont {C.}~\bibnamefont {{Baccigalupi}}}, \bibinfo {author}
  {\bibfnamefont {A.~J.}\ \bibnamefont {{Banday}}}, \ and\ \bibinfo {author}
  {\bibnamefont {et~al.}},\ }\href@noop {} {\bibfield  {journal} {\bibinfo
  {journal} {ArXiv e-prints}\ } (\bibinfo {year} {2013})},\ \Eprint
  {http://arxiv.org/abs/1303.5084} {arXiv:1303.5084 [astro-ph.CO]} \BibitemShut
  {NoStop}%
\bibitem [{\citenamefont {{Hanson}}\ and\ \citenamefont
  {{Lewis}}(2009)}]{HansonLewis}%
  \BibitemOpen
  \bibfield  {author} {\bibinfo {author} {\bibfnamefont {D.}~\bibnamefont
  {{Hanson}}}\ and\ \bibinfo {author} {\bibfnamefont {A.}~\bibnamefont
  {{Lewis}}},\ }\href {\doibase 10.1103/PhysRevD.80.063004} {\bibfield
  {journal} {\bibinfo  {journal} {\prd}\ }\textbf {\bibinfo {volume} {80}},\
  \bibinfo {eid} {063004} (\bibinfo {year} {2009})},\ \Eprint
  {http://arxiv.org/abs/0908.0963} {arXiv:0908.0963 [astro-ph.CO]} \BibitemShut
  {NoStop}%
\bibitem [{\citenamefont {{Smith}}\ and\ \citenamefont
  {{Kamionkowski}}(2012)}]{SmithKam}%
  \BibitemOpen
  \bibfield  {author} {\bibinfo {author} {\bibfnamefont {T.~L.}\ \bibnamefont
  {{Smith}}}\ and\ \bibinfo {author} {\bibfnamefont {M.}~\bibnamefont
  {{Kamionkowski}}},\ }\href {\doibase 10.1103/PhysRevD.86.063009} {\bibfield
  {journal} {\bibinfo  {journal} {\prd}\ }\textbf {\bibinfo {volume} {86}},\
  \bibinfo {eid} {063009} (\bibinfo {year} {2012})},\ \Eprint
  {http://arxiv.org/abs/1203.6654} {arXiv:1203.6654 [astro-ph.CO]} \BibitemShut
  {NoStop}%
\bibitem [{\citenamefont {{Hamimeche}}\ and\ \citenamefont
  {{Lewis}}(2008)}]{HamLewis2008}%
  \BibitemOpen
  \bibfield  {author} {\bibinfo {author} {\bibfnamefont {S.}~\bibnamefont
  {{Hamimeche}}}\ and\ \bibinfo {author} {\bibfnamefont {A.}~\bibnamefont
  {{Lewis}}},\ }\href {\doibase 10.1103/PhysRevD.77.103013} {\bibfield
  {journal} {\bibinfo  {journal} {\prd}\ }\textbf {\bibinfo {volume} {77}},\
  \bibinfo {eid} {103013} (\bibinfo {year} {2008})},\ \Eprint
  {http://arxiv.org/abs/0801.0554} {arXiv:0801.0554} \BibitemShut {NoStop}%
\bibitem [{\citenamefont {{Gorski}}\ \emph {et~al.}(2005)\citenamefont
  {{Gorski}}, \citenamefont {{Hivon}}, \citenamefont {{Banday}}, \citenamefont
  {{Wandelt}}, \citenamefont {{Hansen}}, \citenamefont {{Reinecke}},\ and\
  \citenamefont {{Bartelmann}}}]{0409513}%
  \BibitemOpen
  \bibfield  {author} {\bibinfo {author} {\bibfnamefont {K.~M.}\ \bibnamefont
  {{Gorski}}}, \bibinfo {author} {\bibfnamefont {E.}~\bibnamefont {{Hivon}}},
  \bibinfo {author} {\bibfnamefont {A.~J.}\ \bibnamefont {{Banday}}}, \bibinfo
  {author} {\bibfnamefont {B.~D.}\ \bibnamefont {{Wandelt}}}, \bibinfo {author}
  {\bibfnamefont {F.~K.}\ \bibnamefont {{Hansen}}}, \bibinfo {author}
  {\bibfnamefont {M.}~\bibnamefont {{Reinecke}}}, \ and\ \bibinfo {author}
  {\bibfnamefont {M.}~\bibnamefont {{Bartelmann}}},\ }\href {\doibase
  10.1086/427976} {\bibfield  {journal} {\bibinfo  {journal} {\apj}\ }\textbf
  {\bibinfo {volume} {622}},\ \bibinfo {pages} {759} (\bibinfo {year}
  {2005})},\ \Eprint {http://arxiv.org/abs/arXiv:astro-ph/0409513}
  {arXiv:astro-ph/0409513} \BibitemShut {NoStop}%
\end{thebibliography}%

\end{document}